\documentclass[preprint]{sigplanconf}
\usepackage{etex}

\usepackage[pdftex]{graphicx}
\usepackage[cmex10]{amsmath}
\usepackage{array}
\usepackage{fixltx2e}
\usepackage{stmaryrd}%
\usepackage{amsthm}%
\usepackage{amssymb}%
\usepackage{bm}%
\usepackage{color}%
\usepackage{url}%
\usepackage{soul}%
\usepackage{proof}%
\usepackage{mathtools}%
\usepackage[pdftex,all]{xy}%
\allowdisplaybreaks[1]
\theoremstyle{definition}
\newtheorem{mydefinition}{Definition}[section]
\theoremstyle{plain}
\newtheorem{mylemma}[mydefinition]{Lemma}
\newtheorem{myproposition}[mydefinition]{Proposition}
\newtheorem{mytheorem}[mydefinition]{Theorem}
\newtheorem{mycorollary}[mydefinition]{Corollary}

\theoremstyle{definition}
\newtheorem{myremark}[mydefinition]{Remark}

\newtheorem{myexample}[mydefinition]{Example}

\newtheorem{myassumption}[mydefinition]{Assumption}

\theoremstyle{remark}
\newtheorem*{myproof}{Proof}
\def\myqed{\qed}
\newif\ifignore % when set to true, additional text appears containing
\ignorefalse
\newcommand{\auxproof}[1]{
\ifignore\mbox{}\newline
\color{gray}
\textbf{BEGIN: AUX-PROOF} \dotfill\newline
{#1}\mbox{}\newline
\textbf{END: AUX-PROOF}\dotfill\newline
\color{black}
\fi}

\newif\ifblind % when set to true, additional text appears containing
\blindfalse
\newcommand{\blindAlt}[2]{\ifblind #1 \else #2\fi}

\xyoption{v2}
\xyoption{curve}
\xyoption{2cell}
\SelectTips{cm}{}  % Tips (of arrows) are in accordance with Computer Modern
\UseAllTwocells
\SilentMatrices

\newdir{ >}{{}*!/-8pt/@{>}}  
\newdir{|>}{%
!/1.6pt/@{|}*:(1,-.2)@^{>}*:(1,+.2)@_{>}}
\newdir{pb}{:(1,-1)@^{|-}}
  \def\pb#1{\save[]+<20 pt,0 pt>:a(#1)\ar@{pb{}}[]\restore}
\newcommand{\rar}{\ar|-*\dir{|}}

\usepackage{wrapfig}
\usepackage{algorithm}
\usepackage{algpseudocode}
\makeatletter
\algrenewcommand\ALG@beginalgorithmic{\footnotesize}
\makeatother

\usepackage{tikz}
\usepackage{tikz-qtree}
\usetikzlibrary{automata}
\usetikzlibrary{trees}
\usepackage{pgfplots}
\usepackage{pgfplotstable}

\newcommand{\co}{\mathrel{\circ}}

\newcommand{\id}{\mathrm{id}}
\newcommand{\Sets}{\mathbf{Sets}}
\newcommand{\Rel}{\mathbf{Rel}}

\newcommand{\Kleisli}[1]{\mathcal{K}{\kern-.2ex}\ell(#1)}
\newcommand{\EM}[1]{\mathcal{E}{\kern-.5ex}\mathcal{M}(#1)}

\newcommand{\sem}[1]{\llbracket #1 \rrbracket}

\newcommand{\C}{\mathbb{C}}
\newcommand{\bbP}{\mathbb{P}}

\newcommand{\place}{\underline{\phantom{n}}\,} % place holder
\newcommand{\tuple}[1]{\langle#1\rangle}

\DeclareMathOperator{\tr}{tr}
\newcommand{\pow}{\mathcal P}

\newcommand{\dist}{\mathcal D}

\newcommand{\iso}{\mathrel{\stackrel{
           \raisebox{.5ex}{$\scriptstyle\cong\,$}}{
           \raisebox{0ex}[0ex][0ex]{$\rightarrow$}}}}

\newcommand{\oF}{\overline{F}}

\newcommand{\ev}{\mathsf{ev}}

\DeclareMathOperator{\beh}{\mathsf{beh}}

\def\compsign{\mathrel>\kern-2pt\joinrel>\kern-2pt\joinrel>}

\newcommand{\seq}[2]{{#1}_{1},\dotsc,{#1}_{#2}}
\newcommand{\Bool}{\mathbf{2}}

\newcommand{\defiff}{\stackrel{\text{def.}}{\Longleftrightarrow}}

\newcommand{\dar}{\ar@{..>}}
\newcommand{\lar}{\ar@{-}}

\newcommand{\Kco}{\mathbin{\odot}} % for composition in Kleisli
\newcommand{\relto}{\mathrel{\ooalign{\hfil\raisebox{.3pt}{$\shortmid$}\hfil\crcr$\rightarrow$}}}
\newcommand{\longrelto}{\mathrel{\ooalign{\hfil\raisebox{.3pt}{$\shortmid$}\hfil\crcr$\longrightarrow$}}}
\newcommand{\ttrue}{\mathrm{t{\kern-1.5pt}t}}
\newcommand{\ffalse}{\mathrm{f{\kern-1.5pt}f}}

\newcommand{\str}{\mathrm{str}}

\newcommand{\ssub}{\mathrel{\subset{\kern-1.6ex}\subset}}

\newcommand{\PredKl}[1]{\bbP^{\mathcal{K}{\kern-.2ex}\ell}(#1)}
\newcommand{\PredEM}[1]{\bbP^{\mathcal{E}{\kern-.5ex}\mathcal{M}}(#1)}
 \newcommand{\Upcl}{\mathcal{U}{\kern-.2ex}p}

\newcommand{\upcl}{\mathop{\uparrow}\nolimits}
\newcommand{\dwcl}{\mathop{\downarrow}\nolimits}
\newcommand{\UP}{\mathcal{U{\kern-.3ex}P}}
\newcommand{\CD}{\mathcal{C{\kern-.3ex}D}}

\newcommand{\Meas}{\mathbf{Meas}}

\newcommand{\Var}{\mathbf{Var}}

\newcommand{\RC}{\mathcal{C}{\kern-.3ex}v}

\newcommand{\AP}{\mathsf{AP}}
\newcommand{\Giry}{\mathcal{G}}

\newcommand{\PT}{\mathsf{PT}}

\DeclareMathOperator\X{\mathsf{X}}

\newcommand{\pri}{\mathop{\mathsf{pr}}}

\newcommand{\Cmu}{\mathbf{C}\boldsymbol{\mu}}
\newcommand{\CmuGL}{\Cmu_{\Gamma,\Lambda}}
\newcommand{\mntn}[1]{[#1]_{\mathsf{m}}}
\newcommand{\unitInt}{{[0,1]}}

\newcommand{\SmythLeq}{\mathrel{\sqsubseteq_{\mathsf{S}}}}
\newcommand{\HoareLeq}{\mathrel{\sqsubseteq_{\mathsf{H}}}}

\newcommand{\sol}{\mathrm{sol}}
\newcommand{\approximant}{p}
\newcommand{\CompLat}{L}
\newcommand{\even}{\mathsf{even}}
\newcommand{\odd}{\mathsf{odd}}
\newcommand{\ascCL}{\mathsf{ascCL}}
\newcommand{\NoGood}{\spadesuit}
\newcommand{\sigalg}{\mathfrak{F}}
\newcommand{\pPM}{\mathsf{pPM}}
\newcommand{\pPMpa}{\pPM_{\varphi,\alpha}}
\newcommand{\POM}{\mathsf{POM}}
\newcommand{\PTMPM}{\PT^{\rm M}}
\newcommand{\nextPO}{\mathrm{succ}_{\preceq}}
\newcommand{\collapse}{\mathrm{col}}
\newcommand{\ESatisPhi}{E^{\rm sat}_{\varphi}}
\newcommand{\rk}{\mathrm{rk}}
\newcommand{\Ord}{\mathrm{Ord}}

\makeatletter
\def\underparen#1{\mathop{\vtop{\m@th\ialign{##\crcr
   $\hfil\displaystyle{#1}\hfil$\crcr
   \noalign{\kern3\p@\nointerlineskip}%
   \upparenfill\crcr\noalign{\kern3\p@}}}}\limits}
\def\upparenfill{$\m@th \setbox\z@\hbox{$\braceld$}%
  \bracelu\leaders\vrule \@height\ht\z@ \@depth\z@\hfill\braceru$}
\makeatother

\begin{document}
\toappear{}

\setlength{\pdfpageheight}{\paperheight}
\setlength{\pdfpagewidth}{\paperwidth}

\conferenceinfo{POPL 2016}{...} 
\copyrightyear{2016} 
\copyrightdata{???} 
\doi{nnnnnnn.nnnnnnn}
% \doi{nnnnnnn.nnnnnnn}

% Uncomment one of the following two, if you are not going for the 
% traditional copyright transfer agreement.

%\exclusivelicense                % ACM gets exclusive license to publish, 
                                  % you retain copyright

%\permissiontopublish             % ACM gets nonexclusive license to publish
                                  % (paid open-access papers, 
                                  % short abstracts)

\titlebanner{}        % These are ignored unless
\preprintfooter{}   % 'preprint' option specified.

\title{Lattice-Theoretic Progress Measures
 and Coalgebraic Model Checking
  % \\ {\LARGE From Coalgebraic Components to Algebraic Effects}
}
%\subtitle{From Coalgebraic Components to Algebraic Effects}

\blindAlt{
\authorinfo{Omitted for Submission}{}{}
}{
\authorinfo{Ichiro Hasuo \and Shunsuke Shimizu}
           {University of Tokyo, Japan}
           {
 \{ichiro,shunsuke\}@is.s.u-tokyo.ac.jp
}
\authorinfo{Corina C\^{\i}rstea}
           {University of Southampton, UK}
           {
 cc2@ecs.soton.ac.uk
}
}

\maketitle

  \begin{abstract}
     In the context of formal verification in general and model checking in
  particular, \emph{parity games} serve as a mighty vehicle: many
  problems are encoded as parity games, which are then solved by the
  seminal algorithm by Jurdzinski. In this paper we identify the essence
  of this workflow to be the notion of \emph{progress measure}, and
  formalize it in general, possibly infinitary, lattice-theoretic
  terms. Our view on progress measures is that they are to
  nested/alternating fixed points what \emph{invariants} are to
  safety/greatest fixed points, and what \emph{ranking functions} are to
  liveness/least fixed points. That is, progress measures are
  combination of the
  latter two notions (invariant and ranking function) that have been
  extensively studied in the context of (program) verification.

  We then apply our theory of progress measures to a general model-checking
  framework, where systems are categorically presented as coalgebras.
  The framework's
   theoretical robustness  is witnessed by a smooth
  transfer from the branching-time setting to the linear-time one.
  Although the framework can be used to 
  derive some decision procedures for finite settings, we also
  expect the proposed framework to form a basis for sound proof methods
  for some undecidable/infinitary problems.  
  \end{abstract}

% \category{Software and its engineering}{}{Model checking}
% \category{Theory of computation}{}{Categorical semantics}
% \category{Theory of computation}{}{Modal and temporal logics}
%\category{F.3.2}{Semantics of Programming Languages}{Process models}

% general terms are not compulsory anymore, 
% you may leave them out
%\terms
%Theory
% term1, term2

% (keywords seem compulsory)

\auxproof{
After finishing the paper, some thoughts...
\begin{itemize}
 \item In \S{}\ref{sec:progressMeasForCL}, why don't we say
       $u_{0},u_{2},\dotsc$ are $\nu$-variables and
       $u_{1},u_{3},\dotsc$ are $\mu$-variables? Justified by the Bekic
       lemma.
       \begin{itemize}
	\item This is OK in~\S{}\ref{sec:progressMeasForCL}, but for the
	      coalgebraic model checking the current presentation is
	      much more useful. Otherwise translation of formulas into
	      equational systems takes more efforts, grouping variables
	      of the same polarity
	      together.
       \end{itemize}
\end{itemize}
}

\section{Introduction}\label{sec:intro}
\subsection{Backgrounds}
\paragraph{Parity Games and Fixed-Point Logics}
  For the purpose of formal verification  where one aims at
establishing that a system satisfies a certain property (called a
\emph{specification}), it is common to express: a model of the system as
a state-based transition system such as an automaton or a Kripke structure;
and a specification as a formula in some modal logic. For the latter, in particular,
logics with \emph{fixed-point operators}---such as LTL and CTL---serve
well thanks to their remarkable expressivity~\cite{Pnueli77}. The
\emph{modal $\mu$-calculus} (see e.g.~\cite{Kozen83,BradfieldS06})
provides a clean syntax that incorporates the least and greatest fixed-point
operators ($\mu$ and $\nu$) in a systematic manner.

 Dealing with such fixed
points is however a nontrivial task---this is especially the case when $\mu$'s
and $\nu$'s are nested and they alternate.  Many engineers find it challenging
to express their intuition as a fixed-point formula; furthermore,  many
algorithms are first introduced for an alternation-free
fragment and then later extended to the full fragment (see
e.g.~\cite{CleavelandS93} and~\cite{CleavelandKS92}).

For the purpose of analyses of fixed-point logics and designing
algorithms for them, \emph{parity games} have emerged as a very useful
tool in the last decade or so. A parity game is played by two players
$\even$ and $\odd$, on a board each position $x$ of which has a natural
number $\pri(x)\in\omega$ called its \emph{priority}. Notably its
winning condition is the \emph{parity condition}: the player $\even$
wins if the largest priority that occurs infinitely often in a given
play (an infinite sequence of positions) is an even number. This
condition---that may seem ad-hoc at first sight---turns out to be
extremely useful for modeling nested and alternating $\mu$'s and
$\nu$'s. It is in a sense a \emph{combinatorial} presentation of
an alternation between
$\mu$'s and $\nu$'s.
 % One can say that the parity condition captures the
% ``essence'' of
% nested $\mu$'s
% and $\nu$'s.

The use of parity games has been boosted further by Jurdzinski's
algorithm that efficiently determines the winner at each position of a
parity game~\cite{Jurdzinski00}. It exhibits a practical complexity that
is exponential only in so-called the \emph{alternation depth} of a
parity game. It has then  become a norm, in the context of
fixed-point logics and algorithmic formal verification, to take the
following \emph{parity-game workflow}: it reduces a
problem in question to the decision problem of some parity game, and then
solves the latter by Jurdzinski's algorithm. A
notable example is the model-checking problem for the modal
$\mu$-calculus (see e.g.~\cite{Wilke01}).

The key ingredient of
Jurdzinski's algorithm is what is called a \emph{progress measure} (a
notion originally from~\cite{KlarlundK91})---it can be understood as an
extension of a \emph{ranking function} (used e.g.\ for
termination proofs) to a setting with nested  $\mu$'s and $\nu$'s.

  \paragraph{Coalgebras and Coalgebraic Modal Logics}
  On the other side of formal verification (namely system models),
  \emph{coalgebra} has attracted attention as a categorical abstraction
  of state-based systems~\cite{Rutten00a,Jacobs12CoalgBook} for more
  than a decade. An \emph{$F$-coalgebra} is an arrow $c\colon X\to FX$
  in some category $\C$, where $F\colon \C\to \C$ is an endofunctor.  By
  changing $\C$ and $F$ a coalgebra instantiates to a variety of
  transition systems, such as Kripke structures, LTSs, Markov chains,
  tree automata, processes in the $\pi$-calculus, and so on. Abstracting
  away from specific choices of $\C$ and $F$ allows us to develop a
  uniform theory  that applies to various systems.
  One notable success is a uniform definition of \emph{bisimulation}
  that is independent from $\C$ and
  $F$. See~\cite{Rutten00a,Jacobs12CoalgBook}. 

  Along with the development of the theory of coalgebras,
  \emph{coalgebraic modal logics} have been developed as
  languages suited for specifying about
  coalgebras (see e.g.~\cite{CirsteaKPSV11}).
  Besides the approaches with Moss' \emph{cover modality}~\cite{Moss99}
  and \emph{Stone-like dualities}~\cite{BonsangueK05}, the one with
  \emph{predicate liftings}~\cite{Pattinson03TCS} is widely adopted in
  the literature. The theory has since produced many uniform results
  about coalgebraic modal logics as specification languages. They
  are on:
  expressivity (i.e.\ that bisimilarity is
  captured)~\cite{Pattinson03TCS,Klin07}; sound and complete
  axiomatizations~\cite{Pattinson03TCS,SchroederP09};
   satisfiability complexity~\cite{SchroederP09}; cut elimination and
  interpolation~\cite{Pattinson13,PattinsonS08}; and so on.
  
  Fixed-point operators in coalgebraic modal logics have been actively
  studied
  too. See
  e.g.~\cite{Jacobs02d,Cirstea14,SchroederP09,CirsteaKP09,SchroderV10,Cirstea11calco},
  and also~\cite{Venema06,FontaineLV10} where \emph{coalgebraic
  automata} are studied as translations of $\mu$-calculus formulas.
  In particular, in~\cite{CirsteaKP09}, algorithms for the model-checking
  and satisfiability problems of a coalgebraic $\mu$-calculus
  are presented. These algorithms reduce the
  problems to parity games---this follows the common parity-game
  workflow that we already discussed. For
  satisfiability they also need a tableau system devised for this purpose.

\subsection{Contributions}
%\begin{itemize}
% \item
In this paper we scrutinize the aforementioned \emph{parity-game
workflow} of: reducing to a parity game, and solving by Jurdzinski's
algorithm. We identify its essence in \emph{progress measures}---a key
notion in Jurdzinski's algorithm~\cite{Jurdzinski00}---rather than in
parity games themselves. This leads us to a lattice-theoretic
\emph{transfinite} notion of progress measure that works without any
finiteness assumption, a restriction that is inevitable in the
combinatorial notion of parity game. We then go on to develop a generic
(and not necessarily finitary) framework for model checking, where
system models and specifications  also have generic presentations in the language of
coalgebras and coalgebraic modal logics.

More specifically, our technical contributions are as follows.

\paragraph{Lattice-Theoretic Progress Measure}
      Taking an arbitrary  complete lattice $L$ as a value domain
      (instead of a finite power $\Bool^{m}$ of $\Bool=\{\ttrue,\ffalse\}$),
      we present a lattice-theoretic characterization of solutions of
       recursive equations
       with (nested and alternating) greatest and least
      fixed-points. The characterization is by the notions of
       \emph{prioritized ordinal} and        \emph{progress
       measure}---notions that are essentially generalization of what
       are in Jurdzinski's work~\cite{Jurdzinski00}.
       % In particular, we find the
       % \emph{$i$-th truncated lexicographic order} $\preceq_{i}$ between
       % prioritized ordinals essential in nested fixed points.
       % Indeed we observe that our notion of progress measure is a
       % ``transfinite extension'' of Jurdzinski's parity progress
       % 	measure.
    Our general formalization allows one to use
    progress measures also in infinitary settings where we deal with infinite-state systems,
    quantitative verification (i.e.\ the set of truth values is
    infinite), or both. 

    One can also think of our progress measure as the combination of the
    common proof methods by: \emph{invariants} for safety/gfp
    properties, and \emph{ranking functions} for liveness/lfp
    properties (Table~\ref{table:progressMeasuresCombineInvAndRankFunc}).  
These methods have been extensively studied especially
    in the field of \emph{program verification}---where problems
   are inherently infinitary due to the $\mathtt{Integer}$ datatype---with an emphasis on
    automatic synthesis of invariants and ranking functions (see
    e.g.\ recent~\cite{GargLMN14,Ben-AmramG15}).  
\begin{table}[tbp]
      \begin{tabular}{c|c}
      properties& witnessed by \\\hline\hline
      safety, gfp& invariants \\\hline
      liveness, lfp& ranking functions\\\hline
      \emph{nested gfp's}&
	  winning strategies for parity games
	  (if finitary),
	  \\
      \emph{and lfp's}
	  &  and
      \emph{progress measures} (in general)
     \end{tabular}
\caption{Progress measures $=$ (invariants $+$ ranking functions)}
\label{table:progressMeasuresCombineInvAndRankFunc}
\end{table}
Our current results therefore
    open the way to combining these automated synthesis techniques, and to
    obtaining automated proof methods for \emph{nested lfp/gfp
    properties} (like the \emph{response formula} $\mathsf{G}(p\to
    \mathsf{F}q)$ but much, much more). Once done  its impact will be significant, since
    currently most automation attempts in the field focus on only safety
    or liveness, and not their combination.  

  We  note that these results (in~\S{}\ref{sec:progressMeasForCL})
  are formulated solely in (rather elementary) lattice-theoretic terms,
  without any category theory. While their principal use in the current
  paper is in coalgebraic model checking, their application areas are
  expected to be widespread, in quantitative verification, program
  verification, and so on---by model checking and deductive methods alike.

 \paragraph{Progress Measure for Coalgebraic $\mu$-Calculus
       Model Checking} We apply the notion of progress measure to model
       checking of a \emph{coalgebraic modal $\mu$-calculus}
       $\CmuGL$. Specifically, given a coalgebra $c\colon X\to FX$ (as a
       system model), a $\CmuGL$-formula $\varphi$ (as a specification)
       and the domain $\Omega$ of truth values, we characterize the
       semantics $\sem{\varphi}_{c}\colon X\to \Omega$ of $\varphi$ over $c$
       in terms of progress measures.  Unlike the original definition of
       the semantics $\sem{\varphi}_{c}$ (that is highly nonlocal due to
       fixed-point operators), it can be checked locally whether given
       data constitute a progress measure.

       The lattice-theoretic
       generality of our progress measure allows: a
       state space $X$ that is infinite; a domain $\Omega$ of truth
       values that is other than
       $\Bool=\{\ttrue,\ffalse\}$ (such as the unit interval $[0,1]$);
       and so on. Furthermore, for its finitary special case, we derive
       a model-checking algorithm that is based on progress measures.

       We expect our theoretical framework
       (general, possibly infinitary, in~\S{}\ref{subsec:branchingTimeProgressMeasure})
       to be a foundation on which various verification
       techniques---a candidate being an extension of the simulation-based
       method in~\cite{UrabeH14}---can be formulated and proved sound. 

       Besides, our generic model-checking algorithm
       (in~\S{}\ref{subsec:branchingTimeAlgo}, as a finitary
       special case of the framework
       in~\S{}\ref{subsec:branchingTimeProgressMeasure}) is a
       uniform algorithm that works for a variety of
       endofunctors $F$ and modalities over $F$ (normal modal logic over
       Kripke models, neighborhood frames, graded modal logic, coalition
       logic, and so on; see Example~\ref{ex:lambdaAndGamma}). Moreover,
       thanks to its concrete presentation with matrices, our algorithm
       should be easy to implement. 

       Currently it is not clear whether our algorithm in~\S{}\ref{subsec:branchingTimeAlgo} competes with
       tailor-developed ones for a specific modal logic. However we believe
       our generic algorithm is at least worthwhile---much like
       a big part of the  coalgebraic attempts towards abstraction and
       genericity, see~\S{}\ref{sec:intro}---for the following reasons: 1) among the examples covered
       by our generic algorithm, not all enjoy 
       tailor-developed algorithms; and 2) we believe our algorithm,
       though currently basic, can expose further ``handles'' for
        optimization. The latter means: in many parity game-based
       algorithms, the part of solving parity games is left as a
       blackbox; and in principle opening up a blackbox (like we do)
       should be good for optimization, possibly allowing for ``shortcut
       fusion''-like optimization.

       %\item
       %  \paragraph{Model-Checking Algorithms for coalgebraic $\mu$-calculus.}
       % For the Boolean setting we derive a \emph{model checking}
       % algorithm that: takes a coalgebra $c\colon X\to FX$ (as a system
       % model) and a formula $\varphi$ in a coalgebraic $\mu$-calculus
       % (as a specification); and returns $\{x\in X\mid \text{$x$
       % satisfies $\varphi$}\}$. The
       % algorithm
       % exhibits complexity that is exponential in the number of
       % $\mu$-operators; this is comparable to usual complexity results
       % in model-checking that are typically exponential to an
       % alternation depth.

       %      \item

       \paragraph{Coalgebraic $\mu$-Calculus as a Linear-Time Logic}
       In order to further demonstrate the theoretical robustness of our
        framework, we present an
       adaptation of the framework to \emph{linear-time model checking}.
       In this case a system is a coalgebra $c\colon X\to \pow FX$ (with additional nondeterministic
       branching
       represented by the powerset monad $\pow$); and the question is
       whether  there is an \emph{infinitary trace}
       $z$ of $c$ starting from $x$ such that $z$ satisfies
       a $\CmuGL$-formula $\varphi$.

       It turns out that the combination with coalgebraic theory of
       traces and simulations (developed e.g.\
       in~\cite{Jacobs04c,HasuoJS07b,UrabeH15CALCOtoAppear}) allows a
       smooth transfer from the previous ``branching-time'' setting to the
       current linear-time one. 
       The outcome is a uniform treatment of branching and
       (nondeterministic) linear-time logics---which does not seem to be
       achieved before  despite the obvious efforts by the coalgebra community. This venture also needs a technical piece, namely the ``pumping''-like result (Thm.~\ref{thm:smallExistentialProgMeas}) by Zorn's lemma.

       Our technical contributions are: a progress
       measure-based characterization of linear-time model checking
       (where, again, whether given data is a valid progress measure or not
       can be checked locally); and a decision procedure for linear-time
       model checking (with the restriction that the state space $X$ is
       finite and the truth values are Boolean). The former solves the
       challenge, presented in~\cite{Cirstea15CALCOtoAppear}, of a local characterization
       of linear-time semantics
       (called ``step-wise semantics'' in~\cite{Cirstea15CALCOtoAppear}) 
       for coalgebraic fixed-point logics.
       % (wait for Corina's reply)
       % This seems to be the
       % first decidability result for linear-time model checking of
       % coalgebraic $\mu$-calculi: 
       
       % We further exemplify the robustness of our lattice-theoretic and
       % categorical (or coalgebraic) approach, by presenting an
       % adaptation of the framework to \emph{linear-time model checking}.
       % Here the problem is: given a \emph{nondeterministic
       % $F$-coalgebra}
       % $c\colon X\to \pow FX$ (notice additional nondeterministic
       % branching
       % represented by the powerset monad $\pow$), and a formula
       % $\varphi$; to return if there exists an \emph{infinitary trace}
       % $z\in \tr(c)(x)$ of $c$ starting from $x$ such that $z$ satisfies
       % $\varphi$. It turns out that the combination with coalgebraic theory
       % of traces and simulations (developed e.g.\
       % in~\cite{Jacobs04c,HasuoJS07b,UrabeH15CALCOtoAppear}) allows a
       % smooth transfer from the previous branching-time setting to the
       % current linear-time one. Our main technical result is a decision
       % 	     procedure for linear-time model checking.
 %\end{itemize}

\subsection{Future Work}
\label{subsec:futureWork} There are a lot of further topics to study in
our current venture to coalgebraic $\mu$-calculus. They include:
implementation of our model-checking algorithms
in~\S{}\ref{subsec:branchingTimeAlgo} and \S{}\ref{subsec:linearTimeAlgo}
(the one in~\S{}\ref{subsec:branchingTimeAlgo} should
especially be easy because of the presentation by matrices);
experiments, comparison with tailor-made algorithms and further optimization;
satisfiability and small-model property; universal linear-time model
checking (in this paper we study the existential one); synthesis; and
$\CmuGL$ as linear-time logic for systems with \emph{probabilistic}
branching. In particular we expect the last to be not hard, given the
 lattice-theoretic generality of the current results. 
It should also
 help that 
the coalgebraic theory of traces and simulations has been recently
extended to the probabilistic setting~\cite{Cirstea11,UrabeH15CALCOtoAppear} (using the Giry monad over the
category of measurable spaces). We can say we understand the
mathematical structures therein fairly well:
 these studies  suggest that  the probabilistic setting is
better-behaved than the nondeterministic setting, from  a
coalgebraic point of view. See~\cite{UrabeH15CALCOtoAppear} for further details.

Besides, our lattice-theoretic theory of nested fixed points allows
progress measures (which we identify as the essence of parity games)
to be applied to infinitary settings.
% Our lattice-theoretic generalization of progress measures essentially
% extends Jurdzinski's
% algorithm to \emph{infinitary} settings---a departure from the
% finiteness requirement enforced by the combinatorial nature of parity
% games.
% Although our infinitary extension does not allow algorithmic
% search,
We believe it will be useful for the following purposes.
% , besides its
% application to coalgebraic model checking in the current paper.
Working out
these further applications is future work.
 \paragraph{Establishing an Alternating Fixed-Point in Theorem
       Proving}  In an infinitary setting (such as the state space
       $|X|$ and/or the truth domain $\Omega$ are infinite), the search
       space for our (infinitary) progress measures will be infinite,
       and hence is not amenable to algorithmic search. Even so, one could
       resort to human ingenuity to find one.

       An advantage of a progress measure-based characterization of the
       semantics $\sem{\varphi}_{c}$ is, as we mentioned earlier, the
       validity of a progress measure can be checked \emph{locally} in a
       straightforward manner. This is unlike the original definition of
       the semantics $\sem{\varphi}_{c}$ (see Def.~\ref{def:CmuFmlSem})
       that involves highly nonlocal information like $V\colon X\to
       \Omega$.  We believe this advantage will be especially useful
       when one works with fixed-point specifications in a \emph{proof
       assistant}.

       Due to the same advantage,
       our progress measure-based characterization might also form a
       basis of  sound (but not necessarily complete) model-checking algorithms
       that rely e.g.\ on mathematical programming. This is much like
       in~\cite{UrabeH14} where Kleisli simulations (whose existence is
       checked by linear programming and hence is PTIME) give a sound proof
       method
       for weighted language inclusion (an undecidable property).
       
       % It is then important that the ``witness''
       % thus discovered can be easily checked to be valid. We believe our
       % (lattice-theoretic and hence uniform) definition 
       % allows such checks, and thus helps establishing fixed points 
       % e.g.\ in a proof assistant.

       \paragraph{As a Tool in a Meta-Theory} In \emph{higher-order model
       checking} (see e.g.~\cite{Ong06,KobayashiO09,TsukadaO14}), a
       \emph{higher-order recursion scheme (HORS)}
       % (that is essentially
       % a typed $\lambda$-term with recursion)
       generates an infinite tree that is then model-checked against a
       modal $\mu$-formula.  The generated tree is in general
       irrational---hence cannot be identified with a finite-state
       automaton. However it is shown~\cite{Ong06,KobayashiO09} that the
       model-checking is decidable; an algorithm operates directly with
       the HORS that generates the tree, but not with the tree itself.
       In this setting (and similar ones), we expect our infinitary
       progress measure to be a useful tool on the level of
       meta-theory, e.g.\ for showing the correctness of
       an algorithm. 

       We also envisage the use of our current results
       in lifting \emph{(bi)simulation} notions for B\"{u}chi and parity
       automata (see e.g.~\cite{EtessamiWS05}) to the coalgebraic level
       of abstraction and generality. 
       In this direction we have obtained
       some preliminary results that characterize the accepted languages
       of B\"uchi/parity automata via coalgebras in a Kleisli
       category---results that will hopefully enable us to extend our
       coalgebraic theory of traces and simulations
       in~\cite{Jacobs04c,HasuoJS07b,UrabeH15CALCOtoAppear} to
       B\"uchi/parity acceptance conditions.
 We also intend to study the
       relationship between our current work and \emph{quantitative}
       extensions of parity games, a topic of extensive research efforts~\cite{ChatterjeeJH04,ChatterjeeD12}.

% \subsection{Related Work}
% \label{subsec:relatedWork}
% Such ``universality'' of parity games seems to arise from: their parity
% acceptance condition that allows  to encode nested and alternating
% fixed point operators; and their structure as games that allows to
% encode both universal and existential quantification.

\subsection{Notations}
Throughout the paper, the domain of truth
values is denoted by $\Omega$ and is assumed to be a complete lattice,
with its order denoted by $\sqsubseteq$, and its supremums and infimums
denoted by $\bigsqcup$ and $\bigsqcap$.
Typical examples of $\Omega$ are the set $\Bool=\{\ttrue,\ffalse\}$ of Boolean truth
values, and the unit interval $\unitInt$ for a quantitative notion of truth.  
In~\S{}\ref{sec:progressMeasForCL} we will use another complete lattice
$L$; this will be instantiated by $L=\Omega^{X}$---where $X$ is the
state space of the system in question---for the use in later sections.
Since $\Omega$ is a complete lattice, any monotonic
endofunction $f$ on $\Omega$ has the greatest and
least fixed points $\nu f,\mu f$. The same holds for $L$ in place of $\Omega$.

  We fix a countable set $\Var$ of
\emph{(fixed-point) variables}. It is ranged by $u,v,w,\dotsc$.
We let $\eta$ designate fixed-point operators in general; it is either
$\mu$ or $\nu$. Confusion with a monad unit is unlikely.

The set of natural numbers is identified with the smallest infinite
ordinal and denoted by $\omega$. 

% (do we need this?)
% A function space is  denoted by $[X,Y]$, and by $Y^{X}$. By $\mntn{
% X,Y
% }$ we designate the set of \emph{monotone} functions (with respect to
% suitable orders on $X,Y$).

\subsection{Organization of the Paper}
In~\S{}\ref{sec:progressMeasForCL}
we present our lattice-theoretic notion of progress measure and prove
that it characterizes the solution of a system of fixed-point equations.
In~\S{}\ref{sec:coalgMuCal} we introduce our logic $\CmuGL$---it is a
coalgebraic modal logic with both greatest and least fixed-point
operators ($\nu$, $\mu$); it is parametrized not only by the set
$\Lambda$ of predicate liftings (i.e.\ modalities) for a functor $F$, but also by the set
$\Gamma$ of propositional connectives.
In~\S{}\ref{sec:branchingTimeModelChecking} we adapt progress measures
in~\S{}\ref{sec:progressMeasForCL} to the purpose of $\CmuGL$ model
checking (against $F$-coalgebras), derive a model-checking algorithm and
analyze its complexity. This framework is further adapted
 in~\S{}\ref{sec:nondetLinearTime} to (existential) linear-time
model checking---where a system has additional nondeterministic
branching. We present a decision procedure there.

%===============================================================
\iffalse
%===============================================================
Appendices to the current paper are found in the extended
version~\cite{HasuoSC16POPLExtended}. Omitted proofs are there, too.
%===============================================================
\fi
%===============================================================
Omitted proofs are found in Appendix~\ref{appendix:proofs}.

\section{Progress Measures for Equational
 Systems}\label{sec:progressMeasForCL}
 \subsection{Prelude: (Unnested) Fixed Points, Invariants and Ranking Functions}
 \label{subsec:KnasterTarskiAndCousotCousot}
In general, there are two different ways for characterizing (not nested)
least/greatest fixed points (lfp's and gfp's). The first is the \emph{Knaster-Tarski} one:
the lfp is the least prefixed point; and the gfp is the greatest
postfixed point. The second is the \emph{Cousot-Cousot}
one~\cite{CousotC79}: the lfp $\mu f$ of a monotone function $f\colon
\CompLat\to \CompLat$ over a complete lattice $\CompLat$ is the (possibly
transfinite) supremum of the chain $\bot\sqsubseteq f(\bot)\sqsubseteq
f^{2}(\bot)\sqsubseteq\cdots$; similarly the gfp $\nu f$ is the infimum of
 $\top\sqsupseteq f(\top)\sqsupseteq\cdots$. Sometimes these chains 
 are guaranteed to stabilize
 after $\omega$ steps, for example when $f$ satisfies suitable continuity
 conditions (the \emph{Kleene} fixed-point theorem).

 In this paper our principal interests will be finding \emph{lower bounds} for
 fixed points; see Rem.~\ref{rem:progressMeasForEqSys} for system
 verification motivations.
Among the last four characterizations (Knaster-Tarski and Cousot-Cousot,
for each of lfp and gfp), what are suited for this  purpose of
ours are: the Cousot-Cousot one for lfp's; and the
Knaster-Tarski one for gfp's (the other two only give us \emph{upper bounds}). 
We explicitly note this fact for the record:
% These two indeed give us an \emph{under-approximation} $p$ such that
% $p\sqsubseteq \lfp(f)$ (or $p\sqsubseteq \gfp(f)$); the other two are
% suited for \emph{over-approximation}, in contrast.
% We are currently after methods of
% \emph{under-approximating} a fixed point, that is, finding a witness $p$ such
% that $p\sqsubseteq \lfp(f)$ (or $p\sqsubseteq \gfp(f)$). The two
% characterizations mentioned are used to find such an
% \emph{underapproximation} $w$ of a fixed point. In contrast, the remaining
% two
% will give us an \emph{overapproximation} $w$. 
%
% Our assumptions---especially that $X$ is finite---ensure that the Cousot-Cousot chains
% $\bot\sqsubseteq f(\bot)\sqsubseteq
% \cdots$ for the conventional semantics $\sem{\mu u.\, \varphi}$
% stabilizes after $\omega$ steps. 
\begin{mylemma}[lower bounds for fixed points]
 \label{lem:FPLowerApprox}
 Let $\CompLat$ be a complete lattice and $f\colon \CompLat\to\CompLat$
 be a monotone function.
 \begin{enumerate}
  \item\label{item:lemFPLowerApproxLFP}
       For each ordinal $\alpha$ we have $f^{\alpha}(\bot)\sqsubseteq
	\mu f$. Here $f^{\alpha}(\bot)$ is defined by obvious induction:
	$f^{\alpha+1}(\bot)=f(f^{\alpha}(\bot))$ for a successor
	ordinal; and
	$f^{\alpha}(\bot)=\bigsqcup_{\beta<\alpha}f^{\beta}(\bot)$
	for a limit ordinal.
  \item\label{item:lemFPLowerApproxGFP}
       For any $l\in \CompLat$, $l\sqsubseteq f(l)$ implies
	$l\sqsubseteq \nu f$.
	\myqed
 \end{enumerate}
\end{mylemma}
\noindent
 We emphasize that this simple theoretical  observation is what underlies the
 difference between the common proof methods for \emph{safety/gfp}
 properties and for \emph{liveness/lfp} properties (Table~\ref{table:progressMeasuresCombineInvAndRankFunc}). For the
 former (gfp's) one would seek for an \emph{invariant}, that is, a
 postfixed point $l$ such that $l\sqsubseteq f(l)$. For the latter
 (lfp's)  one would typically synthesize a 
 \emph{ranking function}, an $\omega$-valued function that strictly
 decreases in each step. We formulate---also for the sake of some intuitions---the general principle
 behind the latter, focusing on  $L=\Bool^{X}$. 
 \begin{mydefinition}\label{def:rankingFunc}
  Let $f\colon \Bool^{X}\to\Bool^{X}$ be a monotone function. 
A \emph{ranking
 function} for $f$ is an ordinal- (or $\NoGood$, indicating ``failure'')
  valued function $\rk\colon X\to \Ord\amalg\{\NoGood\}$
such that: 1) 
 $\rk(x)\neq 0$ for each $x\in X$;  2) for each ordinal $\alpha$, 
 $\{x\mid \rk(x)\le \alpha+1\}\subseteq
f\bigl(\{x\mid \rk(x)\le \alpha\}\bigr)
$; and 3) for each limit ordinal $\alpha$, 
  $\{x\mid \rk(x)\le \alpha\}= \bigcup_{\beta<\alpha} 
\{x\mid \rk(x)\le \beta\}
$.
 \end{mydefinition}
\begin{myexample}\label{ex:rankingFunc}
Assume that $X$ is equipped with a transition relation $R\subseteq
 X\times X$ and we are interested in reachability to a subset $U\subseteq
 X$. We would then define $f$ by: $f(X'):=U\cup \{x\mid \exists x'.\,
 xRx'\land x'\in X'\}$; this yields $f^{\alpha}(\bot)$ to be the set of
 states from which $U$ is reachable within $\alpha-1$ steps. 
 A prototypical ranking function is given by $\rk(x):=(\text{the distance
 from $x$ to $U$})+1$.
\end{myexample}
\begin{mylemma}\label{lem:soundnessRkFunc}
 In Def.~\ref{def:rankingFunc},
 a ranking function $\rk$ for $f$ witnesses $\mu f$, the least fixed
 point of $f$. That is,
 $\rk(x)\neq\NoGood$ implies $x\in \mu f$.
\end{mylemma}
\begin{myproof}
 The following is easily shown by induction on an ordinal $\alpha$:
  for any $x\in X$ such that $\rk(x)=\alpha$, we have
 $x\in f^{\alpha}(\bot)$. The claim then follows from Lem.~\ref{lem:FPLowerApprox}.\ref{item:lemFPLowerApproxLFP}.
 \myqed
\end{myproof}
\begin{myremark}\label{rem:stoneLikeDuality}
 Implicit in the above is a bijective correspondence---not unlike in \emph{Stone-like
 dualities}---between: 
 \begin{itemize}
 \item 
 a ranking function $\rk\colon X\to
 \Ord\amalg\{\NoGood\}$; and
 \item 
 an \emph{approximating sequence}
 $U_{0}\subseteq U_{1}\subseteq\cdots$ such that: 1)
       $U_{0}=\bot=\emptyset$, 2)
 $U_{\alpha+1}\subseteq f(U_{\alpha})$, and 3)
 $U_{\alpha}=\bigcup_{\beta<\alpha}U_{\beta}$ for any limit ordinal
 $\alpha$. 
 \end{itemize}
From the former to the latter we let $U_{\alpha}:=\{x\mid \rk(x)\le
 \alpha\}$; conversely we let $\rk(x):=\inf\{\alpha\mid x\in
 U_{\alpha}\}$. 
\end{myremark}

 % In formal verification, accordingly,  common methods for establishing
 % safety properties (gfp's) and liveness properties (lfp's) come in
 % quite different flavors. For the former it suffices to discover an
 % \emph{invariant}; for the latter typically one would synthesize a 
 % \emph{ranking function}, an $\omega$-valued function that strictly
 % decreases in each step.

\subsection{Equational Systems}\label{subsec:eqSys}
With the preparations
in~\S{}\ref{subsec:KnasterTarskiAndCousotCousot} for unnested fixed
points, we set out to study nested and alternating ones. 
As a formalism of expressing them
we prefer \emph{equational systems}, to the (probably more common)
modal $\mu$-calculus-like notations.
Here we shall follow the accounts of similar notions
in~\cite{CleavelandKS92} and~\cite[\S{}1.4]{ArnoldN01}.
% See
% e.g.~\cite{CleavelandKS92,ArnoldN01}; here we loosely follow the notations in~\cite[\S{}1.4]{ArnoldN01}.
\begin{mydefinition}[equational system]\label{def:eqSys}
Let $L$ be a complete lattice. An \emph{equational system} $E$
over $L$ is an expression of the form
\begin{equation}\label{eq:sysOfEq}
 \begin{array}{c}
  u_{1}=_{\eta_{1}}f_{1}(u_{1},\dotsc, u_{m}),\quad
  \dotsc,\quad
  u_{m}=_{\eta_{m}} f_{m}(u_{1},\dotsc, u_{m})
 \end{array}
\end{equation}
where: $\seq{u}{m}$ are \emph{variables},
 $\eta_{1},\dotsc,\eta_{m}\in\{\mu,\nu\}$, and $\seq{f}{m}\colon
 L^{m}\to L$ are monotone functions.

 A variable $u_{j}$ is said to be a \emph{$\mu$-variable} if
 $\eta_{j}=\mu$; it is a \emph{$\nu$-variable} if $\eta_{j}=\nu$.

 We say $u_{i}$ has a \emph{bigger priority} than $u_{j}$ if $j<i$.
\end{mydefinition}
\noindent
Note that, in the last definition,  we have been vague  about the distinction between 
a function $f_{i}$ as a semantical object and a syntactic symbol that denotes
it. 

It is straightforward to generalize
the definition and allow different variables to take values in different
complete lattices $\seq{L}{m}$, and extend accordingly our technical
developments below.
We  assume $L_{1}=\cdots=L_{m}=L$ for
ease of presentation.

 The order of equations \emph{matters} in an equational system
like~(\ref{eq:sysOfEq}).\footnote{Here we follow the ordering convention
in~\cite{ArnoldN01}. In~\cite{CleavelandKS92} the order is reversed, and
the rightmost equation is solved first.} Intuitively, the system~(\ref{eq:sysOfEq}) is
solved starting from the leftmost equation, where the remaining 
variables $u_{2},\dotsc, u_{m}$ are left as undetermined parameters. 
The \emph{interim} solution of the leftmost equation (for $u_{1}$, in terms of
$u_{2},\dotsc, u_{m}$)  is then used in the second equation
$u_{2}=_{\eta_{2}}f_{2}(u_{1},\dotsc, u_{m})$ to eliminate
the occurrences of $u_{1}$ in its right-hand side. We continue this way;
then solving the last
(rightmost)
equation would give us a \emph{closed} (i.e.\ without any
variables occurring in it) solution for $u_{m}$. Such closed solutions are then propagated
from right to left in~(\ref{eq:sysOfEq}), finally giving a closed
solution to each variable $u_{i}$. 
 
The above intuitions can be put  in the following precise terms. 
\begin{mydefinition}[solution]\label{def:solOfEqSys}
 The \emph{solution} of an equational system~(\ref{eq:sysOfEq}) is defined 
 as follows. 
For each $i\in[1,m]$ and $j\in[1,i]$, we define monotone functions
	\begin{displaymath}
	 f^{\ddagger}_i\colon L^{m-i+1} \rightarrow L
	 \quad\text{and}\quad
	 l^{(i)}_{j}\colon L^{m-i}\rightarrow L
	\end{displaymath}
	as follows, inductively on $i$. For the base case $i=1$:
	\begin{displaymath}
\begin{aligned}
 	 f^{\ddagger}_{1}(l_{1},\dotsc,l_{m})&:= f_{1}(l_{1},\dotsc,l_{m}),\quad
 \\
	 l^{(1)}_{1}(l_{2},\dotsc, l_{m})&:= 
	 \eta_{1}\bigl[f^{\ddagger}_{1}(\place,l_{2},\dotsc, l_{m})\colon L\to
 L\bigr].
\end{aligned}\end{displaymath}
In the last line we take the lfp or gfp (according to
 $\eta_{1}\in\{\mu,\nu\}$) of the (monotone) function 
$f^{\ddagger}_{1}(\place,l_{2},\dotsc, l_{m})\colon L\to
 L$. 

For the step case, the function $f^{\ddagger}_{i+1}$ makes use of the 
$i$-th interim solutions $l^{(i)}_{1},\dotsc,l^{(i)}_{i}$ for the variables
	$u_{1},\dotsc, u_{i}$ obtained so far:
	\begin{displaymath}
\begin{aligned}
  & f^{\ddagger}_{i+1}(l_{i+1},\dotsc, l_{m}):=
\\
&
    f_{i+1}\bigl(\,l^{(i)}_{1}(l_{i+1},\dotsc, l_{m}),\;\dotsc,\;
 l^{(i)}_{i}(l_{i+1},\dotsc, l_{m}), \;
l_{i+1},\dotsc, l_{m}
\,\bigr).
\end{aligned}
\end{displaymath}
We then let
	\begin{displaymath}
\begin{aligned}
&l^{(i+1)}_{i+1}(l_{i+2},\dotsc, l_{m})
:=
\eta_{i+1}
\bigl[\,
f^{\ddagger}_{i+1}(\place,l_{i+2},\dotsc, l_{m})\colon L\to L
\,\bigr]
\end{aligned}
\end{displaymath}
and use it to obtain the $(i+1)$-th interim solutions
 $l^{(i+1)}_{1},\dotsc,l^{(i+1)}_{i}$. That is, for each $j\in [1,i]$, 
	\begin{equation}\label{eq:defInterimSolutionForSmallerJ}
 l^{(i+1)}_{j}(l_{i+2},\dotsc, l_{m})
:=
 l^{(i)}_{j}\bigl(\,l^{(i+1)}_{i+1}(l_{i+2},\dotsc, l_{m}),\; l_{i+2},\dotsc,l_{m}\,\bigr)
\end{equation}
Finally, the \emph{solution} 
$(
l^{\sol}_{1},
\dotsc,
l^{\sol}_{m}
)
\in L^{m}$ of the equational system~(\ref{eq:sysOfEq}) is defined by
\begin{math}
(
l^{\sol}_{1},
\dotsc,
l^{\sol}_{m}
)
:=
(
l^{(m)}_{1},
\dotsc,
l^{(m)}_{m}
) 
\end{math}, where we identify a function $l^{(m)}_{j}\colon 1\to L$ with an element of $L$.

It is easy to see that all the functions $f^{\ddagger}_{i}$ and
 $l^{(i)}_{j}$ involved here are monotone. 
That the solution uniquely exists is then guaranteed by the
 Knaster-Tarski theorem.
\end{mydefinition}

\begin{myexample}\label{ex:solOfEqSys}
 As a simple example, consider an equational system
 \begin{math}
  u_{1}=_{\mu} u_{2},\,
  u_{2}=_{\nu} u_{1}.
 \end{math}
 Solving the first equation yields $u_{1}=u_{2}$ (i.e.\ $l^{(1)}_{1}(l_{2})=l_{2}$); using it
 to eliminate $u_{1}$ in the second equation, we obtain $u_{2}=_{\nu}
 u_{2}$ (i.e.\ $f^{\ddagger}_{2}(l_{2})=l_{2}$). We conclude
 $u_{1}=u_{2}=\top$
 is the solution. 

 It is not hard to see that, if we change the order of the equations,
 the resulting system
 \begin{math}
  u_{2}=_{\nu} u_{1},
 \,
  u_{1}=_{\mu} u_{2}
 \end{math}
 has a different solution $u_{1}=u_{2}=\bot$.
\end{myexample}

It is not hard to give a precise correspondence between equational
systems and their modal $\mu$-calculus-like presentations. 
Each equation $u_{j}=_{\eta_{j}} f_{j}(u_{1},\dotsc, u_{m})$ corresponds
to a fixed-point formula $\eta_{j} u_{j}.\, f_{j}(u_{1},\dotsc, u_{m})$; 
since an
equational
system like~(\ref{eq:sysOfEq}) is solved from left to right, the formula
that
corresponds to an equation on the left occurs \emph{inside} the formula
for an equation on the right. For example, if $m=2$, the equational
system~(\ref{eq:sysOfEq}) is presented as
\begin{math}
 \eta_{2} u_{2}.\, f_{2}\bigl(\,
 %u_{1},\;
 \bigl(\eta_{1} u_{1}.\, f_{1}(u_{1}, u_{2})\bigr)\,,
 \;
 u_{2}\,\bigr)
%\enspace.
\end{math}.
In the light of such a correspondence to $\mu$-calculus-like formulas, the definition
of bigger/smaller priorities in Def.~\ref{def:eqSys} coincides with what
is customary (an outside fixed-point operator has a
bigger priority).
A precise translation can be defined following~\cite{CleavelandKS92};
see also Def.~\ref{def:translationFromFmlToEqSys} later, in the special case of
coalgebraic fixed-point logic.

%\subsection{Lower-Bounds of Solutions of Equational Systems}

\begin{myremark}[aiming at lower bounds]
 \label{rem:progressMeasForEqSys}
 Assume that an equational system $E$ is given. For the purpose of 
 system \emph{verification}, one is typically not so much interested
 in its
 solution 
 % $(
 % l^{\sol}_{1},
 % \dotsc,
 % l^{\sol}_{m}
 % )\in L^{m}$
 itself, as in a suitable \emph{lower bound} of it.
 For a simple example consider the setting of
 Example~\ref{ex:rankingFunc}, and assume that $X$, $R$  and $U$ are
 given as follows.
 \begin{displaymath}
\entrymodifiers={+++[o][F-]}
\vcenter{  \xymatrix@1@C-1.5em{
  *{\cdots}
     \ar[r]
  &
  {x_{-2}}
          \ar[r]
  &
  {x_{-1}}
          \ar[r]
  &
  {x_{0}}
  &
  {x_{1}}
          \ar[l]
  &
  {x_{2}}
          \ar[l]
  &
  *{\cdots}
          \ar[l]   
}},
\; U:=\{x_{0}\}.
 \end{displaymath}
 A common question would be if $U$ is reachable from a specific state of
 our interest,
 say $x_{3}$. To \emph{verify} it the ranking function 
\begin{displaymath}
\begin{array}{l}
   \rk(x_{0})=1,\quad
  \rk(x_{i})=i+1 \;\text{for each $i\ge 1$},\; 
\\
  \rk(x_{i})=\NoGood\;\text{for each $i<0$}
\end{array}
\end{displaymath} 
suffices. This choice of a ranking function---while it gives a lower bound
 $\{x_{0},x_{1},\dotsc\}\subseteq\mu f$ of $\mu f$---does not witness e.g.\
$x_{-3}\in \mu f$ (that actually holds). This is not a problem 
because we are interested only in $x_{3}$.

 This phenomenon (of only
 giving a lower bound) is the case with
 verification algorithms in general:  they conduct ``directed'' searches from
 the states in question. Therefore in this paper we focus on
 characterizing \emph{lower bounds} of the solution of an equational
 system. \emph{Upper bounds}, in contrast, are useful in
 \emph{refuting} that certain states have certain properties.

\end{myremark}

 \subsection{Progress Measures}
 We shall now characterize lower bounds of (nested and alternating) fixed
points specified by an equational system. 
We use the technical notion of \emph{progress measure}; 
it is a
lattice-theoretic generalization of the notion of  \emph{parity progress measure}
in~\cite{Jurdzinski00}, and hence is seen as a generalization of
\emph{winning strategies} for parity games, too.
Roughly speaking, these are how one combines
\emph{invariants} (for gfp's) and \emph{ranking functions} (for lfp's,
 see Table~\ref{table:progressMeasuresCombineInvAndRankFunc} and \S{}\ref{subsec:KnasterTarskiAndCousotCousot})
in an intricate way so that  priorities in alternation are respected.

Following
Lem.~\ref{lem:FPLowerApprox} we approximate least fixed
points by transfinite sequences starting from $\bot$.  In general there are
multiple $\mu$-variables in an equational system---we have one ``counter''
for each of them, and use their tuple that we call a \emph{prioritized
ordinal}. 
In particular,  the definition of the preorder $\preceq_{i}$ between
prioritized ordinals---derived from the one in~\cite{Jurdzinski00} and
defined for each variable $u_{i}$---lies in the technical core.

\begin{mydefinition}[prioritized ordinal, $\preceq_{i}$]\label{def:prioritizedOrdinal}
Let $E$ be the equational system in~(\ref{eq:sysOfEq}), over a complete
 lattice $L$. Let us collect all those indices $i\in [1,m]$ for which $u_{i}$ is a
 $\mu$-variable in the equational system  $E$, and arrange them so that
 $i_{1}<\cdots <i_{k}$. That is,
\begin{displaymath}
 \{i_{1},\dotsc,i_{k}\}
 \;=\;
 \{i\in[1,m]\mid \eta_{i}=\mu \text{ in~(\ref{eq:sysOfEq})}\}.
\end{displaymath}
Then a \emph{prioritized ordinal} for $E$ is a $k$-tuple
 $(\seq{\alpha}{k})$ of ordinals. Note that $k$ is the number of
 $\mu$-variables in $E$.

 For each $i\in [1,m]$ we define a  preorder $\preceq_{i}$ between
 prioritized ordinals---we call $\preceq_{i}$ the \emph{$i$-th truncated
 lexicographic order}---as follows.
Let $a\in [1,k]$ be such that
	       \begin{displaymath}
		i_{1}<\cdots <i_{a-1}<i\le i_{a}<\cdots < i_{k},
	       \end{displaymath}
 that is, $u_{i_{a}}$ is the $\mu$-variable with the smallest priority 
 that is at least as big as that of $i$.
 Then we define
 \begin{displaymath}
(\alpha_{1},\dotsc,\alpha_{k})
  \preceq_{i}
  (\alpha'_{1},\dotsc,\alpha'_{k})
 \end{displaymath}
 if, between the \emph{$i$-truncations} 
$  (\alpha_{a},\dotsc,\alpha_{k})$
and
$  (\alpha'_{a},\dotsc,\alpha'_{k})$
of the prioritized ordinals, 
 we have
 \begin{math}
  (\alpha_{a},\dotsc,\alpha_{k})
  \preceq
  (\alpha'_{a},\dotsc,\alpha'_{k})
 \end{math}. Here the last $\preceq$ denotes the lexicographic extension of 
 the usual order $\le$ between ordinals, with the latter elements being
 the more significant. Note here that the $i$-truncation
 $(\alpha_{a},\dotsc,\alpha_{k})$ 
 of 
 $(\alpha_{1},\dotsc,\alpha_{k})$ 
 is obtained by dropping the first elements that correspond to the
 $\mu$-variables with priorities smaller than that of $u_{i}$.

In case $\preceq_{i}$ holds in both ways we write $=_{i}$. Note that
 $=_{i}$
is in general coarser than the equality between prioritized ordinals
(see Example~\ref{ex:prioritizedOrd}).
 We define
 \begin{math}
(\alpha_{1},\dotsc,\alpha_{k})
  \prec_{i}
  (\alpha'_{1},\dotsc,\alpha'_{k})
 \end{math}
if 
 \begin{math}
(\alpha_{a},\dotsc,\alpha_{k})
  \preceq_{i}
  (\alpha'_{a},\dotsc,\alpha'_{k})
 \end{math}
holds but
 \begin{math}
(\alpha_{a},\dotsc,\alpha_{k})
  =_{i}
  (\alpha'_{a},\dotsc,\alpha'_{k})
 \end{math} fails.
\end{mydefinition}

\begin{myexample}\label{ex:prioritizedOrd}
 Let us consider the following example $E_{0}$ of an equational system:
\begin{multline*}
 u_{1}=_{\mu} f_{1}(\vec{u}),
 \quad
 u_{2}=_{\nu} f_{2}(\vec{u}),
 \quad
 u_{3}=_{\mu} f_{3}(\vec{u}),
 \quad
\\
 u_{4}=_{\mu} f_{4}(\vec{u}),
 \quad
 u_{5}=_{\nu} f_{5}(\vec{u}),
\end{multline*}
where $\vec{u}$ stands for $u_{1},\dotsc,u_{5}$. A prioritized 
ordinal for this $E_{0}$ is a tuple $(\alpha_{1},\alpha_{2},\alpha_{3})$
 of ordinals, where the ordinals $\alpha_{1}$, $\alpha_{2}$ and $\alpha_{3}$
 correspond to
the $\mu$-variables $u_{1}$, $u_{3}$ and $u_{4}$, respectively.

It holds that $(\omega,2,2)\preceq_{1}(0,3,2)$. To see that, since $u_{1}$ is
 with the smallest priority, we have to check $(\omega,2,2)\preceq (0,3,2)$.
 This holds; recall that $\preceq$ is the lexicographic order with
  the latter  being the more significant. We can similarly see that:
\begin{align*}
 &(\omega,2,2)\prec_{2}(0,3,2),
\quad
 &&(\omega,2,2)\prec_{3}(0,3,2),
\\
&(\omega,2,2)=_{4}(0,3,2), \quad\text{and}
&&(\omega,2,2)=_{5}(0,3,2).
\end{align*}
Note here that  the $3$-, $4$- and $5$-truncations of $(\omega,2,2)$ and
 $(0,3,2)$ are: $(2,2)$ and $(3,2)$;  $(2)$ and $(2)$; and $()$ and
 $()$, respectively.
\end{myexample}

In the following definition, 
 the element
$\approximant_{i}(\alpha_{1},\dotsc,\alpha_{k})\in \CompLat$ 
 is understood as the ``$(\alpha_{1},\dotsc,\alpha_{k})$-th
 approximation''
 of the solution for the variable $u_{i}$ in the equational
 system~(\ref{eq:sysOfEq}).
 % Here $(\alpha_{1},\dotsc,\alpha_{k})$ is a $k$-tuple of
 % (possibly infinite) ordinals---we call it a \emph{progress
 % measure}. In a progress measure we  associate one counter $\alpha_{a}$ to
 % each $\mu$-variable $u_{i_{a}}$. Hence the length $k$ of a progress
 % measure  is the number of $\mu$-variables in
 % the equational system.

\begin{mydefinition}[progress measure for an equational system]
\label{def:progressMeasForEqSys}
Assume the same setting as in Def.~\ref{def:prioritizedOrdinal}, with
 $E$ being the equational system~(\ref{eq:sysOfEq}) and  $i_{1}<\cdots
 <i_{k}$
 enumerating the indices of all the $\mu$-variables.

 A \emph{progress measure} $p$ for $E$ is given by a tuple
 \begin{displaymath}
  p\;=\;
  \bigl(\,
  (
  \overline{\alpha_{1}},\dotsc,
  \overline{\alpha_{k}}),
  \,
  \bigl(\,\approximant_{i}(\alpha_{1},\dotsc,\alpha_{k})\,\bigr)_{i,\seq{\alpha}{k}}
\,\bigr)
 \end{displaymath}
that
consists of:
 \begin{itemize}
  \item the \emph{maximum prioritized ordinal}
 $(\overline{\alpha_{1}},\dotsc, \overline{\alpha_{k}})$; and 
  \item the \emph{approximants} $\approximant_{i}(\alpha_{1},\dotsc,\alpha_{k})\in \CompLat$, defined for
	each
	$i\in[1,m]$ and each
	prioritized ordinal
 $(\alpha_{1},\dotsc,\alpha_{k})$
 such that
	$
	\alpha_{1}\le\overline{\alpha_{1}},\dotsc,
	\alpha_{k}\le\overline{\alpha_{k}}
	$. 
% Such a $k$-tuple $(\alpha_{1},\dotsc,\alpha_{k})$
% 	of ordinals is called a \emph{progress measure}.
 \end{itemize}
 The approximants $\approximant_{i}(\alpha_{1},\dotsc,\alpha_{k})$
 are subject to:
	\begin{enumerate}
	 \item \label{item:progressMeasDefMonotonicity}
	      \textbf{(Monotonicity)}
	       Let $i\in[1,m]$ (hence $u_{i}$ is either a $\mu$- or
	       $\nu$-variable). Then 
	       \begin{multline*}
		(\alpha_{1},\dotsc,\alpha_{k})
		\preceq_{i}
		(\alpha'_{1},\dotsc,\alpha'_{k})
		\;\text{implies}
		\\
		\approximant_{i}(\alpha_{1},\dotsc,\alpha_{k})
		\sqsubseteq
		\approximant_{i}(\alpha'_{1},\dotsc,\alpha'_{k}).
	       \end{multline*}
	 \item\label{item:progressMeasDefMuVarBaseCase}
	      \textbf{($\mu$-variables, base case)}
	       Let $a\in [1,k]$. Then $\alpha_{a}=0$ implies
%      $\approximant_{i_{a}}(\alpha_{1},\dotsc,
%      \overset{\underparen{a}}{0},\dotsc,\alpha_{k})$
	       $\approximant_{i_{a}}(\alpha_{1},\dotsc, \alpha_{a},\dotsc,\alpha_{k})=\bot$.
	      (Note the correspondence between: the  subscript $i_{a}$ of
	      $\approximant_{i_{a}}$; and the counter $\alpha_{a}$ that
	      is assumed to be $0$.)
	 \item\label{item:progressMeasDefMuVarStepCase}
	      \textbf{($\mu$-variables, step case)}
	      Let $a\in [1,k]$, and let 
	       $(\alpha_{1},\dotsc,
	       \alpha_{a}+1,\dotsc,\alpha_{k})$ be
	       a prioritized ordinal
	       such that its $a$-th counter
	       $\alpha_{a}+1$ is a successor ordinal. Then, regarding
	       the approximant $\approximant_{i_{a}} (\alpha_{1},\dotsc,\alpha_{a-1},
		\alpha_{a}+1,\alpha_{a+1},\dotsc,\alpha_{k})$, 
	       there exist ordinals
	       $\beta_{1},\dotsc,
	       \beta_{a-1}$ such that
	       \begin{equation}\label{eq:progressMeasDefMuVarStepCase}
 	       \begin{aligned}
		&
		\approximant_{i_{a}} (\alpha_{1},\dotsc,\alpha_{a-1},
		\alpha_{a}+1,\alpha_{a+1},\dotsc,\alpha_{k})
		\\
		&\quad\sqsubseteq\;
		f_{i_{a}}
		\left(\,
		\begin{array}{c}
		\approximant_{1} (\beta_{1},\dotsc,\beta_{a-1},
		 \alpha_{a},\alpha_{a+1},\dotsc,\alpha_{k}),
		 \\
		 \dotsc,
		  \\
		\approximant_{m} (\beta_{1},\dotsc,\beta_{a-1},
		\alpha_{a},\alpha_{a+1},\dotsc,\alpha_{k})
		\end{array}
		\,\right)
	       \end{aligned}
	       \end{equation}	       
and $\beta_{1}\le\overline{\alpha_{1}},\dotsc,
	       \beta_{a-1}\le \overline{\alpha_{a-1}}$.
	       Recall here that $f_{i_{a}}$ is a function in the
	       system~(\ref{eq:sysOfEq}).

	       Cond.~(\ref{eq:progressMeasDefMuVarStepCase}) originates
	       from the definition
	       $f^{\alpha+1}(\bot)=f(f^{\alpha}(\bot))$
	       in Lem.~\ref{lem:FPLowerApprox}.\ref{item:lemFPLowerApproxLFP}; a notable difference
	       here is
	       that the counters with smaller priorities (i.e.\ from the
	       first to the $(a-1)$-th) can be modified arbitrarily.
	 \item\label{item:progressMeasDefMuVarLimitCase}
	      \textbf{($\mu$-variables, limit case)}
	       Let $a\in [1,k]$, and
	       let
	       $(\alpha_{1},\dotsc,
	      %\alpha_{a},\dotsc,
	      \alpha_{k})$
	       be a prioritized ordinal
	       such that its $a$-th counter
	       $\alpha_{a}$
	       is a limit ordinal. Then, regarding the approximant
	       $\approximant_{i_{a}} (\alpha_{1},\dotsc,
	      %\alpha_{a},\dotsc,
	      \alpha_{k})$, we have
	       \begin{equation}\label{eq:progressMeasDefMuVarLimitCase}
		\approximant_{i_{a}} (\alpha_{1},\dotsc,
		\alpha_{a},\dotsc,\alpha_{k})
		% \\
		% &\qquad\sqsubseteq\;
		\sqsubseteq 
		\bigsqcup_{\beta<\alpha_{a}}
				\approximant_{i_{a}} (\alpha_{1},\dotsc,
		\beta,\dotsc,\alpha_{k})\enspace.
	       \end{equation}	       
	 \item\label{item:progressMeasDefNuVar}
	      \textbf{($\nu$-variables)}
	       Let $i\in [1,m]\setminus\{i_{1},\dotsc, i_{k}\}$ (i.e.\
	       $u_{i}$ is a $\nu$-variable in the
	       system~(\ref{eq:sysOfEq})); let $a\in [1,k]$ such that
	       \begin{displaymath}
		i_{1}<\cdots <i_{a-1}<i<i_{a}<\cdots < i_{k}.
	       \end{displaymath}
	       Let $(\alpha_{1},\dotsc,\alpha_{k})$ be a prioritized ordinal.
	       Then, regarding the approximant
	       $\approximant_{i} (\alpha_{1},\dotsc,\alpha_{k})$, there exist
	       ordinals
	       $\beta_{1},\dotsc,\beta_{a-1}$ such that
	       \begin{equation}\label{eq:progressMeasDefNuVar}
 	       \begin{aligned}
		&
		\approximant_{i} (\alpha_{1},\dotsc,
		\alpha_{a-1},\alpha_{a},\dotsc,\alpha_{k})
		\\
		&\qquad\sqsubseteq\;
		f_{i}
		\left(\,
		\begin{array}{c}
		\approximant_{1} (\beta_{1},\dotsc,\beta_{a-1},
		 \alpha_{a},\dotsc,\alpha_{k}),
		 \\
		 \dotsc,
		  \\
		\approximant_{m} (\beta_{1},\dotsc,\beta_{a-1},
		\alpha_{a},\dotsc,\alpha_{k})
		\end{array}
		\,\right)
	       \end{aligned}
	       \end{equation}	       
and $\beta_{1}\le \overline{\alpha_{1}},\dotsc,
	       \beta_{a-1}\le \overline{\alpha_{a-1}}$.

	       This condition is somewhat similar to 
	      Cond.~\ref{item:progressMeasDefMuVarStepCase} above: it
	      comes from the condition $l\sqsubseteq f(l)$ in
	      Lem.~\ref{lem:FPLowerApprox}.\ref{item:lemFPLowerApproxGFP};
	      and much like in 
	      Cond.~\ref{item:progressMeasDefMuVarStepCase}, the
	      counters with smaller priorities can be modified arbitrarily.
	\end{enumerate}
\end{mydefinition}

\begin{mytheorem}[correctness of progress measures]
\label{thm:correctnessOfProgMeasEqSys}
Let $E$ be the equational system in~(\ref{eq:sysOfEq}) over $L$,
       and $(
l^{\sol}_{1},
\dotsc,
l^{\sol}_{m}
)$ be its solution (Def.~\ref{def:solOfEqSys}). 
\begin{enumerate}
 \item\label{item:soundnessProgressMeas} \textbf{(Soundness)}
      A progress measure gives a lower bound of the solution. That is,
       assume 
 \begin{math}
  \bigl(\,
  (
  \overline{\alpha_{1}},\dotsc,
  \overline{\alpha_{k}}),
  \,
  \bigl(\,\approximant_{i}(\alpha_{1},\dotsc,\alpha_{k})\,\bigr)_{i,\overrightarrow{\alpha}}
\,\bigr)
 \end{math}
is a progress measure. 
Then for each $i\in
       [1,m]$ we have 
 \begin{displaymath}
  \approximant_{i}(
  \overline{\alpha_{1}},\dotsc,
  \overline{\alpha_{k}})
  \;\sqsubseteq\;
  l^{\sol}_{i}.
 \end{displaymath}
 \item\label{item:completenessProgressMeas} 
  \textbf{(Completeness)}
  There exists a progress measure
 that achieves the
       optimal, that is, 
 \begin{math}
  \bigl(\,
  (
  \overline{\alpha_{1}},\dotsc,
  \overline{\alpha_{k}}),
  \,
  \bigl(\,\approximant_{i}(\alpha_{1},\dotsc,\alpha_{k})\,\bigr)_{i,\overrightarrow{\alpha}}
\,\bigr)
 \end{math}
such that
 \begin{displaymath}
  \approximant_{i}
  (
  \overline{\alpha_{1}},\dotsc,
  \overline{\alpha_{k}})
  \;=\;
  l^{\sol}_{i}
 \end{displaymath}
 for each $i\in[1,m]$.

\auxproof{I have to be precise about the
      correspondence between cardinals and ordinals.}

 Moreover, such an ``optimal'' progress measure can be chosen
 so that the ordinals in its maximum prioritized ordinal
\begin{math}
   (
  \overline{\alpha_{1}},\dotsc,
  \overline{\alpha_{k}})
\end{math}
 are suitably bounded, in the following sense.
 Let $\ascCL(L)$ be the ordinal defined by the  supremum of the length
 of any (possibly transfinite) strictly ascending chain in $L$. Then
 $\overline{\alpha_{a}}\le \ascCL(L)$ for
      each $a\in [1,k]$.
      \myqed
\end{enumerate}
 \end{mytheorem}
\noindent
In the item~\ref{item:completenessProgressMeas}, the bound $\ascCL(L)$
is generally better than the bound by the size $|L|$ of the complete lattice
$L$. For example, in case $L=\Bool^{X}$ (where $\Bool=\{\ttrue,\ffalse\}$
and $X$ is a set), $\ascCL(L)=|X|$ while $|L|=2^{|X|}$.

We will need the following relaxation in establishing a correspondence
to Jurdzinski's notion of parity progress measure~\cite{Jurdzinski00}.
\begin{mydefinition}[extended progress measure for equational systems]
\label{def:extendedProgressMeasForEqSys}
 Assume the setting of Def.~\ref{def:progressMeasForEqSys}. An
 \emph{extended progress measure} $p$ for $E$ is the same as a progress
 measure, except that Cond.~\ref{item:progressMeasDefMuVarBaseCase} of
 Def.~\ref{def:progressMeasForEqSys} is replaced by the following:
\begin{itemize}
 \item[\ref{item:progressMeasDefMuVarBaseCase}'.] 
 Let $a\in [1,k]$. Then $\alpha_{a}=0$ implies either
 $p_{i_{a}}(\alpha_{1},\dotsc,\alpha_{k}) =\bot$, or
 there exists a
	      prioritized
 ordinal $(\alpha'_{1},\dotsc,\alpha'_{k})$ such that
 $(\alpha'_{1},\dotsc,\alpha'_{k})\prec_{i_{a}}
 (\alpha_{1},\dotsc,\alpha_{k})$ and 
 $p_{i_{a}}(\alpha_{1},\dotsc,\alpha_{k}) \sqsubseteq
  p_{i_{a}}(\alpha'_{1},\dotsc,\alpha'_{k})$.
\end{itemize}
\end{mydefinition}

\begin{myproposition}\label{prop:soundnessOfExtProgMeas}
 An extended progress measure is still sound in the sense of
 Thm.~\ref{thm:correctnessOfProgMeasEqSys}.\ref{item:soundnessProgressMeas}. \myqed
\end{myproposition}

In Appendix~\ref{appendix:parityProgressMeasure},
%===============================================================
\iffalse
%===============================================================
in the extended version~\cite{HasuoSC16POPLExtended},
%===============================================================
\fi
%===============================================================
as a sanity check, we 
present a
correspondence between our notion of progress measure
 (Def.~\ref{def:progressMeasForEqSys}) and 
Jurdzinski's parity progress measure~\cite{Jurdzinski00}.
Jurdzinski's formalization follows that of ranking functions,
while ours here is based on approximation sequences $p(0)\sqsubseteq
p(1)\sqsubseteq \cdots$ in the lattice $L=\Bool^{X}$. The relationship
between the two is much like in
Rem.~\ref{rem:stoneLikeDuality}. 

\begin{myexample}\label{ex:progressMeasure}
 For a simple example following the spirit of
 Example~\ref{ex:rankingFunc}, let us consider a set $X$ and a
 transition relation $R\subseteq X\times X$, and introduce
 a ``modal operator'' $\Box\colon \Bool^{X}\to\Bool^{X}$ by
 $\Box(X'):=\{x\in X\mid \forall x'.\, xRx' \Rightarrow x'\in X'\}$.

 We now
 fix a subset $F\subseteq X$, and consider the following equational
 system
 over $L=\Bool^{X}$.
 \begin{equation}\label{eq:10272034}
  u_{1}
  \;=_{\mu}\;
  (F\cap u_{2})\cup 
  \Box u_{1},
  \qquad
  u_{2}
  \;=_{\nu}\;
  u_{1}.
 \end{equation}
 The system corresponds to the $\mu$-calculus formula 
 $\nu u_{2}. \mu u_{1}.\, (F\cap u_{2})\cup \Box u_{1}$, and
 it is not hard to see---possibly relying on the Knaster-Tarski and
 Cousot-Cousot characterizations,
 see~\S{}\ref{subsec:KnasterTarskiAndCousotCousot}---that the solution
 for $u_{2}$ is the set of states \emph{any infinite path from which visits $F$ infinitely often}.

 For this specific system~(\ref{eq:10272034}), a progress measure
 (Def.~\ref{def:progressMeasForEqSys}) is given by data
 \begin{math}
  \bigl(\,\overline{\alpha},\,
  \bigl(p_{1}(\alpha)\bigr)_{\alpha\le\overline{\alpha}}, \,
  \bigl(p_{2}(\alpha)\bigr)_{\alpha\le\overline{\alpha}} \,\bigr)
 \end{math}
 subject to suitable conditions. 
 Some simplifications are possible, exploiting that
 in~(\ref{eq:progressMeasDefNuVar}) (and elsewhere) counters with smaller
 priorities can be modified arbitrarily. We see, after this
 simplification,  that
 a \emph{progress measure} for the equational system~(\ref{eq:10272034})
 is given by
 \begin{displaymath}
  p_{1}(0)\subseteq p_{1}(1)\subseteq \cdots\subseteq p_{1}(\overline{\alpha}) 
  \quad\text{and}\quad p_{2},
 \end{displaymath}
 all being subsets of $X$, such that: 1) $p_{1}(0)=\emptyset$; 2) 
 $p_{1}(\alpha+1)\subseteq (F\cap p_{2})\cup \Box p_{1}(\alpha)$; 
 3) $p_{1}(\alpha)=\bigcup_{\beta<\alpha}p_{1}(\beta)$ for a limit
 ordinal $\alpha$; and 4) $p_{2}\subseteq p_{1}(\overline{\alpha})$. 
 This ``witnesses'' the solution of~(\ref{eq:10272034}),
 i.e.\ $x\in p_{2}$ implies that any infinite path from $x$ visits $F$
 infinitely often.
\end{myexample}

\section{Coalgebraic $\mu$-Calculus $\Cmu_{\Gamma,\Lambda}$}
\label{sec:coalgMuCal}
From this section on we apply the theory developed
in~\S{}\ref{sec:progressMeasForCL} to a coalgebraic $\mu$-calculus
$\CmuGL$.
In the current section, as a
preparation,  we introduce the logic $\CmuGL$: its syntax, semantics,
and a translation to equational systems (so that the results
in~\S{}\ref{sec:progressMeasForCL}  apply).

\subsection{Coalgebraic Preliminaries}
\label{subsec:coalgPrelim}
We start with a minimal set of coalgebraic preliminaries. For further backgrounds on coalgebras
 see e.g.~\cite{Rutten00a,Jacobs12CoalgBook}; and see
 e.g.~\cite{MacLane71,Awodey06} for categorical preliminaries.
 From now to~\S{}\ref{sec:branchingTimeModelChecking} we fix the base category to be the one $\Sets$ of
sets and functions. 

Let $F$ be an  endofunctor on $\Sets$. An \emph{$F$-coalgebra} is a
function $c\colon X\to FX$, where $X$, $F$ and $c$ are intuitively
understood as a state space, a behavior type and a transition structure,
respectively. Therefore an $F$-coalgebra is ``a transition system of the
behavior type $F$.'' Some examples are presented later in
Example~\ref{ex:lambdaAndGamma}.

\begin{wrapfigure}[3]{r}{0pt}
\begin{minipage}{.2\textwidth}
\vspace{-2em}
  \begin{equation}\label{eq:coalgHom}
\!\!\!\!\!\!\!\!\!\!\!\!
\!\!\!
%\!\!\!
       \vcenter{\xymatrix@R=.4em@C+1em{
     {FX}
     \ar@{->}[r]^-{Ff}
     &
     {FY}
     \\
     {X}
     \ar@{->}[r]^-{f}
     \ar[u]^{c}
     &
     {Y}\mathrlap{\enspace}
     \ar[u]^{d}
     }}
 \end{equation}
\end{minipage}
\end{wrapfigure}
Given two coalgebras $c\colon X\to FX$ and $d\colon Y\to FY$ for the
same functor, a \emph{coalgebra homomorphism} from $c$ to $d$ is a function
$f\colon X\to Y$ such that the above diagram~(\ref{eq:coalgHom}) commutes.
In many examples of $F$, the notion of homomorphism expresses a natural
definition of \emph{behavior-preserving map}. Conversely, it is common
in
the theory of coalgebras that the notion of \emph{$F$-behavioral
equivalence} is defined using homomorphism (namely via cospans,
see~\cite{Jacobs12CoalgBook}).

\begin{wrapfigure}[4]{r}{0pt}
\begin{minipage}{.2\textwidth}
\vspace{-2em}
    \begin{equation}\label{eq:behq}
    \hspace{-3em}
     \vcenter{\xymatrix@R=.8em@C+2em{
     {FX}
     \ar@{-->}[r]^-{F(\beh(c))}
     &
     {FZ}
     \\
     {X}
     \ar@{-->}[r]_-{\beh(c)}
     \ar[u]^{c}
     &
     {Z}
     \ar[u]^{\text{final}}_{\zeta}
     }}
%     \quad\text{in $\Sets$.}
         \end{equation}
\end{minipage}
\end{wrapfigure}
Furthermore, many functors $F$ allow a ``classifying coalgebra''---one
that contains every possible $F$-behavior without
redundancy. This is categorically captured by \emph{finality} of a
coalgebra. Precisely, a coalgebra $\zeta\colon Z\to FZ$ is \emph{final}
if, for any coalgebra $c\colon X\to FX$ there exists a unique
homomorphism $\beh(c)\colon X\to Z$, as shown in~(\ref{eq:behq}). 
 This way we understand the carrier $Z$ of a final coalgebra to be \emph{the
 set of all $F$-behaviors}; and the map $\beh(c)$ induced by finality (as
 in~(\ref{eq:behq})) as the \emph{behavior map}.

  \subsection{$\Cmu_{\Gamma,\Lambda}$: Syntax}
\label{subsec:coalgMuCalSyntax}
It is common in the study of coalgebraic modal logics that the set of
modalities is parametrized. 
In our logic $\Cmu_{\Gamma,\Lambda}$, moreover, we parametrize
propositional connectives too. This allows us to accommodate
unconventional connectives that occur in quantitative setting, like the \emph{truncated
sum}  in the {\L}ukasiewicz
	      $\mu$-calculus~\cite{Mio14} and the
	      \emph{average} operator in e.g.~\cite{AlmagorBK13}.

\begin{mydefinition}[$\Lambda$, $\Gamma$]
\label{def:modalSigAndPropSig}
 A \emph{modal signature} $\Lambda$ over $F$ is a ranked alphabet
 $\Lambda=(\Lambda_{n})_{n\in\omega}$. An element
       $\lambda\in\Lambda_{n}$ is  an \emph{$n$-ary modality},
 and we write $|\lambda|$ for its arity $n$.
% ---with each
%  element $\lambda\in\Lambda$ assigned its \emph{arity}
%  $|\lambda|$---and each element $\lambda\in\Lambda$ is called a
%  \emph{modality}. 

 We assume that a modal signature comes with its \emph{interpretation}.
Assigned to each $\lambda\in\Lambda$  is a natural transformation
 $\sem{\lambda}$, whose components are functions 
%of the form
 \begin{displaymath}
\begin{array}{l}
   \sem{\lambda}_{X}\;\colon\;
  \bigl(\Omega^{
  %(\place)
  X}\bigr)^{|\lambda|}\longrightarrow
  \Omega^{F
 %(\place)
  X
 }  
  % \;:\;
  % \Sets^{\op}\longrightarrow\Sets
  \enspace,
  \quad\text{natural in $X$,}
\end{array} 
\end{displaymath}
% (*** or:
%  \begin{displaymath}
% \begin{array}{l}
%    \sem{\lambda}_{X}\;\colon\;
%   \C(X, \Omega^{|\lambda|})
%   \longrightarrow
%   \C(FX, \Omega)
%   % \;:\;
%   % \Sets^{\op}\longrightarrow\Sets
%   \enspace,
%   \quad\text{natural in $X$,}
% \end{array} 
% \end{displaymath}
% ***)
 and a component $\sem{\lambda}_{X}$ must be monotone with respect to
  (pointwise extensions of)
 the order $\sqsubseteq$ of the domain $\Omega$ of truth values. Such $\sem{\lambda}$  is commonly called
 a 
 \emph{(monotone) predicate lifting}~\cite{HermidaJ98,Pattinson03TCS}.
 \auxproof{ (Do we need this?)
 We also identify $\Lambda$ with a polynomial functor, i.e.\ $\Lambda
 X=\coprod_{\lambda\in\Lambda} X^{|\lambda|}$.
 }

 Similarly,
a \emph{propositional signature} is a ranked alphabet $\Gamma$
%, much like a modal signature,
where each $\gamma\in\Gamma$ is called a \emph{propositional
 connective}. Unlike a modal signature, each $\gamma\in\Gamma$
 is interpreted by a function $\sem{\gamma}\colon
 \Omega^{|\gamma|}\to\Omega$; we require that $\sem{\gamma}$ be monotone.
\auxproof{That, in turn, induces
a natural transformation
 \begin{displaymath}
  \sem{\gamma}_{X}
  \;\colon\;
  \bigl(\,\mntn{\Omega^{X},\Omega}\,\bigr)^{|\gamma|}
  \longrightarrow
 \mntn{\Omega^{X},\Omega}
 \end{displaymath}
 by $\lambda f_{1}.\,\dotsc\lambda f_{|\gamma|}.\; \lambda v.\;
 \sem{\gamma}(f_{1}v,\dotsc, f_{|\gamma|v})$.

In the case $\Omega=\unitInt$ of continuous truth values, we 
 additionally assume that $\sem{\lambda}$ and $\sem{\gamma}$ are
\emph{continuous}, in the sense that they preserve the supremum $\bigsqcup_{i\in\omega}t_{i}$ of
an ascending $\omega$-chain $a_{0}\sqsubseteq a_{1}\sqsubseteq\cdots$.
 }

In what follows we will often write $\lambda$ and $\gamma$ for
 $\sem{\lambda}$ and $\sem{\gamma}$. 
\end{mydefinition}

% We parametrize propositional connectives, too, so
% that we accommodate  connectives like the \emph{truncated
% sum} operator $\oplus$ in {\L}ukasiewicz
% 	      $\mu$-calculus~\cite{MioS13fics,Mio14}.

% \begin{mydefinition}[propositional signature $\Gamma$]
% \label{def:propSig}
% A \emph{propositional signature} is a ranked alphabet
% %, much like a modal signature,
% where each $\gamma\in\Gamma$ is called a \emph{propositional
%  connective}. Differently from a modal signature, each $\gamma\in\Gamma$
%  is interpreted by a function $\sem{\gamma}\colon
%  \Omega^{|\gamma|}\to\Omega$; we require that $\sem{\gamma}$ be monotone.
% \auxproof{That, in turn, induces
% a natural transformation
%  \begin{displaymath}
%   \sem{\gamma}_{X}
%   \;\colon\;
%   \bigl(\,\mntn{\Omega^{X},\Omega}\,\bigr)^{|\gamma|}
%   \longrightarrow
%  \mntn{\Omega^{X},\Omega}
%  \end{displaymath}
%  by $\lambda f_{1}.\,\dotsc\lambda f_{|\gamma|}.\; \lambda v.\;
%  \sem{\gamma}(f_{1}v,\dotsc, f_{|\gamma|v})$.
% }
 
% \end{mydefinition}

% \begin{mynotation}
%   We shall simply write $\lambda$ and $\gamma$ for $\sem{\lambda}$ and 
%  $\sem{\gamma}$, respectively, when no confusion
%  is likely.
% \end{mynotation}

\begin{mydefinition}[$\Cmu_{\Gamma,\Lambda}$]
%  Let $\Gamma$ be a ranked alphabet of \emph{propositional connectives}; 
%  and $\Lambda$ be  a ranked alphabet of \emph{modalities}. Each
%  $\gamma\in\Gamma$
% or $\lambda\in\Lambda$ comes with its \emph{arity} $|\gamma|,
%  |\lambda|\in\nat$.
 The language of our \emph{coalgebraic modal logic}
 $\Cmu_{\Gamma,\Lambda}$ over $\Gamma$ and $\Lambda$ is given by the
 following set of formulas.
 \begin{align*}
  \varphi,\varphi_{i}
  \,::=\,
  &
  u
  \mid
  \boxdot_{\gamma}(\varphi_{1},\dotsc,\varphi_{|\gamma|})
  \mid
  \heartsuit_{\lambda}(\varphi_{1},\dotsc,\varphi_{|\lambda|})
  \mid
  \\
  &
  \mu u.\, \varphi
  \mid
  \nu u.\, \varphi
 \end{align*}
 Here $u\in\Var$ is a (fixed-point) variable. The notations
 $\boxdot_{\gamma}$  (for $\gamma\in\Gamma$) and $\heartsuit_{\lambda}$ 
 (for $\lambda\in\Lambda$) are to distinguish propositional connectives
 (the former) from modalities (the latter).
%; the two classes are handled differently in defining semantics. 
% although we do not require each $u_{i}$ necessarily occur freely in
%  $\varphi$. 
%The set of free variables in $\varphi$ is denoted by $\FV(\varphi)$.
\end{mydefinition}

   \begin{myexample}\label{ex:lambdaAndGamma}
 Examples of predicate lifting-based coalgebraic logics abound.
 \begin{enumerate}
 \item \emph{Standard (normal) modal logic} is obtained by taking
       $F=\pow(\AP)\times\pow(\place)$ (with $\pow$ the covariant
       powerset functor and $\AP$ a set of atomic propositions), $\Omega
       = \Bool$, $\Gamma=\{\ttrue,\ffalse,\land,\lor\}$ with the usual
       interpretations, and $\Lambda=\AP\cup\{\Box,\Diamond\}$ with
       %associated predicate liftings
      \begin{align*}
     &\sem{p}_{X}\colon 1\to \Bool^{FX},~~ *\mapsto \bigl\{\,(U,Y) \mid
       p \in U\,\bigr\}\;\text{(where $p\in \AP$),}
     \\
     &\sem{\Box}_{X}\colon \Bool^{X}\to \Bool^{FX},~~ 
     (V\subseteq X)\mapsto  \bigl\{\,(U,Y)
     \,\bigl|\bigr.\, Y \subseteq V \,\bigr\},
          \\
     &\sem{\Diamond}_{X}\colon \Bool^{X}\to \Bool^{FX},~~ 
     (V\subseteq X)\mapsto  \bigl\{\,(U,Y)
     \,\bigl|\bigr.\, Y \cap V \ne \emptyset \,\bigr\}.
    \end{align*} 
Atomic propositions are thus identified with $0$-ary modalities, as is standard in coalgebraic modal logic (see e.g.~\cite{SchroederP09}).
 \item  \emph{Hennessy-Milner logic} is obtained by taking $F = (\pow(\place))^A$ (with $A$ a set of labels), $\Omega$ and $\Gamma$ as before, and $\Lambda = \{[a],\langle a \rangle\}$ with associated predicate liftings
       \begin{align*}
	&\sem{[a]}_X : \Bool^X \to \Bool^{F X},~~
	%(Y \subseteq X)
	Y\mapsto \{f \in \pow(X)^A \mid f(a) \subseteq Y\},
     \\
	& \sem{\langle a\rangle}_X : \Bool^X \to \Bool^{F X},~~
	% \\
	% &\qquad\qquad\qquad
	%	(Y \subseteq X)
	Y\mapsto \{f \in \pow(X)^A \mid f(a) \cap Y \ne \emptyset \}.
    \end{align*} 
 \item \emph{Monotone neighborhood logic} \cite{Chellas1980ML} is
       obtained by taking $F X = \{ Y \in \pow (\pow (X)) \mid Y
       \mbox{ is upward-closed} \}$, $\Omega$ and $\Gamma$ as before and
       $\Lambda = \{\Box \}$, with an associated predicate lifting 
        \begin{align*}
        &\sem{\Box} : \Bool^X \to \Bool^{ FX},~~(U \subseteq X) \mapsto \{ Y \in \pow(\pow (X)) \mid U \in Y\}.
        \end{align*}
 \item \emph{Graded modal logic} \cite{Fine1972MPW} is obtained by taking $F X =  (\omega + 1)^X$, $\Omega$ and $\Gamma$ as before, and $\Lambda$ consisting of graded modalities $\Box_k$ ("for all but $k$ successors") and $\Diamond_k$ ("for more than $k$ successors") for $k \in \omega$, with associated predicate liftings
        \begin{align*}
	 &\sem{\Box_k} : \Bool^X \to \Bool^{ FX},~~
	 %(Y \subseteq X)
	 Y\mapsto \{f \in F X \mid\textstyle \sum_{x \notin Y} f(x) \le k\},\\
	 &\sem{\Diamond_k} : \Bool^X \to \Bool^{ FX} ,~~
	 %(Y \subseteq X)
	 Y\mapsto \{f \in F X \mid \textstyle\sum_{x \in Y} f(x) > k\}.
        \end{align*}
\item Our approach also covers the \emph{coalition logic}~\cite{Pauly2002MLC}, interpreted over \emph{game frames}---these
      are coalgebras of the (class-valued) functor
\begin{align*}
F X = \{(S_1,\ldots,S_N,f) \mid \emptyset \ne S_i \in \Sets, f : \prod_{i \in N}S_i \to X\}
\end{align*}
with $N$ a set of agents, tuples $(S_1,\ldots,S_N)$ capturing agent strategies, and functions $f : \prod_{i \in N}S_i \to X$ modeling the outcomes of strategy choices for the agents. The modalities $[C]$, with $C \subseteq N$ a \emph{coalition}, arise from predicate liftings $\sem{[C]}_X : \Bool^X \to \Bool^{F X}$ given by
$\sem{[C]}_X(Y) = \{ (S_1,\ldots,S_N,f) \in F X \mid \exists \sigma_C \in S_C.\forall \sigma_{\overline{C}} \in S_{\overline{C}}. f(\sigma_C,\sigma_{\overline{C}}) \in Y\}$, 
where $S_C = \prod_{i \in C}S_i$,\, $\overline{C} = N \setminus C$, and
      $(\sigma_C,\sigma_{\overline{C}})$ is defined as expected.
  \item Here is an example that would yield the usual ``linear-time
	logic'' like LTL (i.e.\ formulas are interpreted over infinite words),
	in the setting of~\S{}\ref{sec:nondetLinearTime}.
	Take
	$F=\pow(\AP)\times(\place)$, $\Omega = \Bool$, $\Gamma=\{\ttrue,
	\ffalse, \land,\lor\}$ and $\Lambda=\AP\cup\{\X\}$, with the
	predicate liftings
    % Furthermore, Section~\ref{sec:nondetLinearTime} considers
    % 	(non-deterministic) \emph{linear-time} logics; this time, the
    % 	coalgebras of interest are of type $\pow F$, with $F : \Sets \to
    % 	\Sets$ specifying linear-time behaviour. In this setting, taking
    % 	$F=\pow(\AP)\times(\place)$, $\Omega = \Bool$, $\Gamma=\{\ttrue, \ffalse, \land,\lor\}$ and $\Lambda=\AP\cup\{\X\}$ recovers a variant of (the modal fragment of) the temporal logic LTL. The required predicate liftings are given by
\begin{displaymath}
      \begin{aligned}
     &\sem{p}_{X}\colon 1\to \Bool^{FX},\;~~*\mapsto \bigl\{\,(U,x)\in
     \pow(\AP)\times X\,\bigl|\bigr.\, p\in U\,\bigr\},
     \\
     &\sem{\X}_{X}\colon \Bool^{X}\to \Bool^{FX},
     ~~
     (V\subseteq X)\mapsto  \bigl\{\,(U,x)
     %\in
     %\pow(\AP)\times X
     \,\bigl|\bigr.\, x\in V\,\bigr\}.
    \end{aligned}
\end{displaymath}
% This time, an $F$-coalgebra is a deterministic transition system, with each state equipped with a set of atomic propositions that hold there. This corresponds to the use of computation paths in the semantics of LTL. However, a difference w.r.t.~standard LTL is in the treatment of atomic propositions: in LTL, they are associated to states of a transition system, whereas here, they are part of the linear-time behaviour (and hence a state can satisfy different atomic propositions along different branches). Thus, our models ($\pow F$-coalgebras) are slightly more general; however, any transition system can be encoded into a $\pow F$-coalgebra using the observation that $\pow$ comes with a \emph{swapped strength map}, specifically $\pow (\AP) \times \pow X \to \pow(\pow(\AP) \times X)$.
	
\end{enumerate}
%  { \bf Corina: what is below can be kept; but feel free to modify. }
In addition to the above $\Bool$-valued logics---that are also accounted
    for by other coalgebraic approaches to modal logic (see
    e.g.~\cite{CirsteaKP09})---our approach additionally covers
    many-valued logics. For example, the {\L}ukasiewicz logic of
    \cite{Mio14} can be recovered by taking $F = \pow \dist$, where
    $\dist : \Sets \to \Sets$ is the \emph{probability distribution
    functor} 
    defined by $\dist X = \{\mu : X \to \unitInt \mid \sum_{x} \mu(x) = 1\}$, $\Omega = \unitInt$, $\Gamma = \{\sqcup,\oplus\}$ and $\Lambda = \{ \Diamond \}$, with interpretations $\sem{\sqcup}, \sem{\oplus}: \unitInt \times \unitInt \to \unitInt$ given by
        \begin{align*}
        & \sem{\sqcup}_X(p,q) = \max(p,q),    &
        &  \sem{\oplus}_X(p,q) = \min(1,p+q)
\end{align*}
and predicate lifting $\sem{\Diamond}_X : \unitInt^X \to \unitInt^{FX}$ given by
\begin{align*}
& \sem{\Diamond}_X(p)(f) = \max_{\mu \in f}\sum_{x}\mu(x) p(x).
\end{align*}

    Another many-valued logic that is covered by our framework is 
    logics with \emph{future discounting}~\cite{AlfaroHM03,AlmagorBK13,NakagawaH16ToAppear}.
    A basic fragment is given as follows:
	take
	$F=\pow(\AP)\times(\place)$, $\Omega = [0,1]$, $\Gamma=\{\ttrue,
	\ffalse, \land,\lor\}$ and $\Lambda=\AP\cup\{\X\}$, with the
	predicate liftings
\begin{displaymath}
      \begin{aligned}
       &\sem{p}_{X}\colon 1\to [0,1]^{FX},\;
       \sem{p}_{X}(*)(U,x) =
       \begin{cases}
	1 &\text{if $p\in U$}
	\\
	0 &\text{otherwise,}
       \end{cases}
     \\
       &\sem{\X}_{X}\colon [0,1]^{X}\to [0,1]^{FX},\;       
      d\mapsto  \bigl[\,(U,x)\mapsto \frac{1}{2}\cdot d(x)\,\bigr].
      \end{aligned}
\end{displaymath}
      Note the factor $\frac{1}{2}$ that discounts the value of truth in the next step.

%     { \bf Corina: what is below can be kept; but feel free to modify. }
%    In what follows we often use the combination of
%     $F=\pow(\AP)\times(\place)$,   $\Gamma=\{\land,\lor\}$  and
%     $\Lambda=\AP\cup\{\X\}$ as an example. An $F$-coalgebra is a
%     deterministic transition system, with each state equipped with a set
%     of atomic propositions that hold there. An atomic proposition $p\in
%     \AP$ is also identified as a $0$-ary modality (as is standard in
%     coalgebraic modal logic; see e.g.~\cite{SchroederP09}). The predicate lifting
%     for each modality is described below, for the record.
%     \begin{align*}
%      &\sem{p}_{X}\colon 1\to \Bool^{FX},\; &&*\mapsto \bigl\{\,(U,x)\in
%      \pow(\AP)\times X\,\bigl|\bigr.\, p\in U\,\bigr\}
%      \\
%      &\sem{\X}_{X}\colon \Bool^{X}\to \Bool^{FX},
%      &&
%      (V\subseteq X)\mapsto  \bigl\{\,(U,x)
%      %\in
%      %\pow(\AP)\times X
%      \,\bigl|\bigr.\, x\in V\,\bigr\}.
%     \end{align*}

%     (*** I have to expand this! ***)
%     In our general framework we allow
%  $\Omega$ can be $\unitInt$, in which case there are
%  other examples that are covered. The {\L}ukasiewicz
% 	      $\mu$-calculus~\cite{Mio14};
% % (which we interpret over
% %  nondeterministic systems, however);
  \end{myexample}

\subsection{Equational Presentation}
\label{subsec:formulasAsEqSys}
In this paper we
 %follow~\cite{CleavelandKS92} and
 favor working with equational
 presentations of $\mu$-calculus formulas.
 % , over their original
% presentation
% in~\S{}\ref{subsec:coalgMuCalSyntax}.
Furthermore, for simplicity,
we shall present $\mu$-calculus formulas as \emph{simple} equational
systems, meaning that each right-hand side is of depth at most $1$.
 \begin{mydefinition}[simple $\CmuGL$-equational system]
  \label{def:simpleCmuGLEqSys}
  A \emph{simple $\CmuGL$-equational system} 
  is an expression of the form
  \begin{equation}\label{eq:CmuGLEqSys}
     u_{1}=_{\eta_{1}}\varphi_{1},\quad
  \dotsc,\quad
  u_{m}=_{\eta_{m}} \varphi_{m}
  \end{equation}
  where: $\eta_{i}\in\{\mu,\nu\}$;
  $\seq{u}{m}\in\Var$  are fixed-point variables;  and
  $\seq{\varphi}{m}$ are \emph{simple $\CmuGL$-formulas}
  of the form
  \begin{displaymath}
   % \varphi_{i}
   % \;::=\;
   u_{i},\quad
     \boxdot_{\gamma}(u_{i_{1}},\dotsc,u_{i_{|\gamma|}})
  \quad\text{or}\quad
  \heartsuit_{\lambda}(u_{i_{1}},\dotsc,u_{i_{|\lambda|}}).
  \end{displaymath}
  We make a further requirement that, in case $\eta_{i}=\mu$,
  the corresponding equation is of the form $u_{i}=_{\mu} u_{j}$ for
  some $j\in [1,m]$. This inessential requirement simplifies our
  subsequent exposition.

  A simple $\CmuGL$-equational system~(\ref{eq:CmuGLEqSys}) is
  \emph{closed} if
  all the variables that occur in $\seq{\varphi}{m}$
  are among $u_{1},\dotsc,u_{m}$.
\end{mydefinition}
Note that, much like in~\S{}\ref{sec:progressMeasForCL}, 
the order of equations in~(\ref{eq:CmuGLEqSys}) matters---the equations
are solved from left to right, i.e.\ priorities increases as one goes from left
to right.

 Translation of $\mu$-calculus formulas into equational systems is
 standard; so is translation in the  other direction.
 \begin{mydefinition}[translation]
  \label{def:translationFromFmlToEqSys}
  For each $\CmuGL$-formula $\varphi$, its \emph{equational presentation}
  $E_{\varphi}$  is defined by the following
  induction. Here $u_{\varphi}\in\Var$ denotes the variable on the
  left-hand side of the last equation
  in the equational system $E_{\varphi}$,
  $E_{1};\dotsc;E_{k}$ denotes the concatenation of equational systems
  $E_{1},\dotsc, E_{k}$,
  and the variable $v$ in each
  clause is chosen to be a fresh one.
  \begin{align*}
   E_{u}
   \;&:=\; \bigl(\,v=_{\nu} u\,\bigr)
   \\
      E_{\boxdot_{\gamma}(\varphi_{1},\dotsc,\varphi_{n})}
   \;&:=\;
   \bigl(\,
   E_{\varphi_{1}};\,\dotsc;\,E_{\varphi_{n}};\,
   v=_{\nu} \boxdot_{\gamma}(u_{\varphi_{1}},\dotsc, u_{\varphi_{n}})
   \,\bigr)
   \\
      E_{\heartsuit_{\lambda}(\varphi_{1},\dotsc,\varphi_{n})}
   \;&:=\;
   \bigl(\,
   E_{\varphi_{1}};\,\dotsc;\,E_{\varphi_{n}};\,
   v=_{\nu} \heartsuit_{\lambda}(u_{\varphi_{1}},\dotsc, u_{\varphi_{n}})
   \,\bigr)
   \\
      E_{\eta u.\,\varphi}
   \;&:=\;
   \bigl(\,
   E_{\varphi};\,
   u =_{\eta} u_{\varphi}
      \,\bigr)
   \qquad\text{where $\eta\in\{\mu,\nu\}$}
  \end{align*}
  The choice of $=_{\nu}$ in the first three clauses is arbitrary from
  the semantical viewpoint:
  changing it into $=_{\mu}$ yields the same semantics. It is however
  beneficial from the algorithmic and presentational viewpoints---in
  particular the resulting  system enjoys the 
  requirement in Def.~\ref{def:simpleCmuGLEqSys} (that a $\mu$-equation
  is of the form $u_{i}=_{\mu}u_{j}$).
  
  It is straightforward to see that a closed formula $\varphi$ yields a
  closed equational system $E_{\varphi}$.

  Conversely, given a simple $\CmuGL$-equational system $E$ like
  in~(\ref{eq:CmuGLEqSys}), we define its \emph{formulaic presentation}
  $\varphi_{E}$ by induction on the number $m$ of equations. If $m=1$
  then an equation $u_{1}=_{\eta_{1}}\varphi_{1}$ becomes the formula
  $\eta_{1}u_{1}.\, \varphi_{1}$. For the step case,
  let $E'$ be obtained by dropping the first equation, that is,
  \begin{displaymath}
   E'
   \;=\;
   \bigl(\,
     u_{2}=_{\eta_{2}}\varphi_{2},\quad
  \dotsc,\quad
  u_{m}=_{\eta_{m}} \varphi_{m}
   \,\bigr).
  \end{displaymath}
  Then we define $\varphi_{E}$ to be the result of replacing $u_{1}$ in
  $\varphi_{E'}$ with $\eta_{1}u_{1}.\, \varphi_{1}$. That is,
  \begin{displaymath}
   \varphi_{E}
   \;:=\;
   \varphi_{E'}\bigl[\,\eta_{1}u_{1}.\varphi_{1}\,/\,u_{1}\,\bigr].
  \end{displaymath}
 \end{mydefinition}

 The two translations are mutually inverse---not necessarily
 syntactically, but the semantics is preserved. See
 Prop.~\ref{prop:translationIsCorrect} later. Therefore, in what
 follows,
 we do not distinguish a $\CmuGL$-formula $\varphi$
 and its equational presentation $E_{\varphi}$. Both will be denoted by
 $\varphi$.

 \begin{myexample}
  Let $\Gamma=\{\land,\lor\}$  and $\Lambda=\AP\cup\{\X\}$ (from
  Example~\ref{ex:lambdaAndGamma}). The $\CmuGL$-formula
  \begin{math}
   \nu u.\, \mu v.\, ((p \lor \X v)\land \X u)
  \end{math}
  gets translated into the simple $\CmuGL$-equational system
  \begin{multline*}
   u_{1}=_{\nu} p,
   \;\;
   u_{2}=_{\nu} v,
   \;\;
   u_{3}=_{\nu} \X u_{2},
   \;\;
   u_{4}=_{\nu} u_{1}\lor u_{3},
   \;\;
      \\
   u_{5}=_{\nu} u,
   \;\;
   u_{6}=_{\nu} \X u_{5},
   \;\;
   u_{7}=_{\nu} u_{4}\land u_{6},
   \;\;
   v=_{\mu} u_{7},
   \;\;
   u=_{\nu} v,
  \end{multline*}
  under 
  Def.~\ref{def:translationFromFmlToEqSys}. 
  The translation in the other direction
  gives rise to a complicated formula which, however,
  is easily seen to be equivalent to the original formula
  under (obviously sound) simplifications like
  \begin{math}
   \nu u_{1}.\, p
  \end{math}
  into $p$.
 \end{myexample}
 
 \subsection{$\Cmu_{\Gamma,\Lambda}$: Semantics}
 \label{section:CmuGLSemantics}
 Formulas of $\Cmu_{\Gamma,\Lambda}$ are interpreted over 
 $F$-coalgebras (see~\S{}\ref{subsec:coalgPrelim}). The following 
 inductive interpretation is a standard one; it follows the tradition of
 coalgebraic modal
 logic~\cite{Cirstea14,SchroederP09,FontaineLV10,CirsteaKP09}
 as well as that of fixed-point logics~\cite{Kozen83}.

 \begin{mydefinition}[semantics of $\Cmu_{\Gamma,\Lambda}$ formulas]
\label{def:CmuFmlSem}
 Let $\Gamma$ and $\Lambda$ be propositional and modal signatures in
  Def.~\ref{def:modalSigAndPropSig}, and 
Let $c\colon X\to FX$  be a coalgebra. A formula
 $\varphi$ of $\CmuGL$---with free variables $\seq{u}{m}$---is
 assigned its \emph{denotation} over $c$; it is given by a function
\begin{displaymath}
\begin{array}{l}
  \sem{\varphi}_{c}
 \;\colon\;
 \bigl(\Omega^{X}\bigr)^{m}\longrightarrow \Omega^{X}
\end{array}
%\enspace.
\end{displaymath}
that is defined inductively in the following way. Here $\vec{V}$ is short for
 $V_{1},\dotsc, V_{m}$, where $V_{i}\colon X\to \Omega$.
\begin{align*}
 \sem{u_{i}}_{c}
 (\vec{V})(x)
%(\seq{V}{m})
&:= V_{i}(x),
\\
 \sem{
 \boxdot_{\gamma}\bigl(\seq{\varphi}{n}\bigr)}_{c}
(\vec{V})(x)
&:=
\\ 
\gamma
\bigl(\,
&\sem{\varphi_{1}}_{c}(\vec{V})(x),
\,\dotsc,\,
\sem{\varphi_{n}}_{c}(\vec{V})(x)\,\bigr),
\\
 \sem{
 \heartsuit_{\lambda}\bigl(\seq{\varphi}{n}\bigr)}_{c}
(\vec{V})(x)
&:= \\
\Bigl(\,\lambda_{X}
\bigl(\,
&\sem{\varphi_{1}}_{c}(\vec{V}),\,\dotsc,\,
\sem{\varphi_{n}}_{c}(\vec{V})\,\bigr)
\,\Bigr)\bigl(c(x)\bigr),
\\
 \sem{
  \mu u.\, \varphi
}_{c}
(\vec{V})(x)
:=&
 \bigl(\,\mu\bigl(\,\sem{ \varphi}_{c}(\vec{V}, \place)\colon 
   \Omega^{X}\to \Omega^{X}\,\bigr)\,\bigr)(x),
\\
 \sem{
  \nu u.\, \varphi
}_{c}
(\vec{V})(x)
:=&
 \bigl(\,\nu\bigl(\,\sem{ \varphi}_{c}(\vec{V}, \place)\colon 
   \Omega^{X}\to \Omega^{X}\,\bigr)\,\bigr)(x).
\end{align*}

% (*** or:
% \begin{displaymath}
%    \sem{\lambda}_{X}\;\colon\;
%   \C(X, \Omega^{m})
%   \longrightarrow
%   \C(X, \Omega)
% \end{displaymath}
% that is defined inductively as follows.
% \begin{align*}
%  \sem{u_{i}}_{c}
%  (V)(x)
% %(\seq{V}{m})
% &:= (\pi_{i}\co V)(x) 
% \\
% &\quad\text{where $\pi_{i}\colon \Omega^{m}\to \Omega$ is a projection,}
% \\
%  \sem{
%  \boxdot_{\gamma}\bigl(\seq{\varphi}{k}\bigr)}_{c}
% (V)(x)
% &:=
% \\ 
% \gamma
% \bigl(\,
% &\sem{\varphi_{1}}_{c}(V)(x),
% \,\dotsc,\,
% \sem{\varphi_{k}}_{c}(V)(x)\,\bigr),
% \\
%  \sem{
%  \heartsuit_{\lambda}\bigl(\seq{\varphi}{k}\bigr)}_{c}
% (V)(x)
% &:= \\
% \Bigl(\,\lambda_{X}
% \bigl(\,
% &\sem{\varphi_{1}}_{c}(V),\,\dotsc,\,
% \sem{\varphi_{k}}_{c}(V)\,\bigr)
% \,\Bigr)\bigl(c(x)\bigr),
% \\
%  \sem{
%   \mu u.\, \varphi
% }_{c}
% (V)(x)
% &:=
%  \bigl(\,\mu\bigl[\,\sem{ \varphi}_{c}(V, \place)\colon 
%    \Omega^{X}\to \Omega^{X}\,\bigr]\,\bigr)(x),
% \\
%  \sem{
%   \nu u.\, \varphi
% }_{c}
% (V)(x)
% &:=
%  \bigl(\,\nu\bigl[\,\sem{ \varphi}_{c}(V, \place)\colon 
%    \Omega^{X}\to \Omega^{X}\,\bigr]\,\bigr)(x).
% \end{align*}
% ***)

Recall that $\gamma\colon \Omega^{n}\to\Omega$ and 
$\lambda_{X}\colon (\Omega^{X})^{n}\to\Omega^{FX}$
are assumed to be given  (Def.~\ref{def:modalSigAndPropSig}).
In the last two clauses it is assumed, by suitably rearranging
  variables,
 that the bound variable $u$
is the last one $u_{m}$ among the free variables $\seq{u}{m}$ of
  $\varphi$. The necessary fixed points of the function 
$\sem{ \varphi}_{c}(\vec{V}, \place)\colon 
   \Omega^{X}\to \Omega^{X}$ are guaranteed by the Knaster-Tarski
  theorem, since $\Omega$ (and hence $\Omega^{X}$) is a complete lattice
and the function $\sem{ \varphi}_{c}(\vec{V}, \place)$ is easily seen to
  be monotone.
 % In the last two lines, the functions with respect to which we take
 % fixed points are of the type $\Omega^{X}\to\Omega^{X}$ and their
 % monotonicity is an immediate consequence of that of $\gamma$ and
 % $\lambda_{X}$. Noting that $\Omega^{X}$ is a complete lattice, the Knaster-Tarski theorem guarantees that the
 % fixed points exist. 
 \end{mydefinition}

% \begin{wrapfigure}[3]{r}{0pt}
% \begin{math}
%        \vcenter{\xymatrix@R=.8em@C+2em{
%      {FX}
%      \ar@{->}[r]^-{Ff}
%      &
%      {FY}
%      \\
%      {X}
%      \ar@{->}[r]_-{f}
%      \ar[u]^{c}
%      &
%      {Y}\mathrlap{\enspace.}
%      \ar[u]^{d}
%      }}
% \end{math}
% \end{wrapfigure}
 % The following lemma is standard one in coalgebraic modal logic. Its
 % inductive proof relies on the naturality of $\lambda$.

\begin{mylemma}\label{lem:invariantUnderCoalgMor}
 Let $f$ be a coalgebra homomorphism
 from $c\colon X\to FX$ to $d\colon Y\to FY$, as in~(\ref{eq:coalgHom}).
For each closed $\CmuGL$-formula $\varphi$ and each $x\in X$, we have
 $\sem{\varphi}_{c}(x)=\sem{\varphi}_{d}(f(x))$.
 \myqed
\end{mylemma}

 As discussed in~\S{}\ref{subsec:formulasAsEqSys} we favor working with
 equational presentation of formulas. We shall therefore define their
 semantics, too.
 \begin{mydefinition}[semantics of simple
  $\Cmu_{\Gamma,\Lambda}$-equational systems]
 \label{def:CmuSimpleEqSysSem}
 Let $E$ be
a simple
  $\Cmu_{\Gamma,\Lambda}$-equational system
 \begin{equation}\label{eq:201506191457}
       u_{1}=_{\eta_{1}}\varphi_{1},\;
  \dotsc,\;
  u_{m}=_{\eta_{m}} \varphi_{m}
 \end{equation}
  from Def.~\ref{def:simpleCmuGLEqSys}; assume that it is closed.
  Let $c\colon X\to FX$ be an $F$-coalgebra.

 Then  $E$ and $c$  together induce an equational system $E_{c}$
 (in the sense of Def.~\ref{def:eqSys})
  over the complete lattice $L=\Omega^{X}$---this is by identifying 
  a simple formula $\varphi_{i}$ on a right-hand side with 
  the function $\sem{\varphi_{i}}_{c}\colon (\Omega^{X})^{m}\to \Omega^{X}$
  defined in Def.~\ref{def:CmuFmlSem}. 
  
 Finally, solving $E_{c}$ as in Def.~\ref{def:solOfEqSys} yields a
  solution $(
l^{\sol}_{1},
\dotsc,
l^{\sol}_{m}
)$ that is an element of  $(\Omega^{X})^{m}$. The last component
$l^{\sol}_{m}$ is referred to as the \emph{semantics}
of the simple
  $\Cmu_{\Gamma,\Lambda}$-equational system $E$ over the coalgebra $c$.
 \end{mydefinition}
  
 The two semantics---the direct one, and the one via equational
 presentation---coincide, as expected. 
 \begin{myproposition}\label{prop:translationIsCorrect}
 Let $\varphi$ be a closed $\CmuGL$-formula, and $c\colon X\to FX$
 be a coalgebra. Consider its equational presentation $E_{\varphi}$
  (a simple
  $\Cmu_{\Gamma,\Lambda}$-equational system,
  Def.~\ref{def:translationFromFmlToEqSys}).
 Then the semantics of $E_{\varphi}$ over $c$---in the sense of
  Def.~\ref{def:CmuSimpleEqSysSem}, i.e.\ the solution of
  the equational system $E_{\varphi,c}$ over $\Omega^{X}$---coincides with 
  $\sem{\varphi}_{c}$ from Def.~\ref{def:CmuFmlSem}.
 \end{myproposition}
  \begin{myproof}
   Straightforward by induction. \myqed
  \end{myproof}

\section{$\Cmu_{\Gamma,\Lambda}$ Model Checking against $F$-Coalgebras
 }
 \label{sec:branchingTimeModelChecking}
 Let us turn to the model-checking problem of the modal logic
 $\CmuGL$ against $F$-coalgebras. Later
 in~\S{}\ref{sec:nondetLinearTime} we study
 model checking against coalgebras with additional nondeterministic
 branching---i.e.\ there the logic $\CmuGL$ is thought of as a
 ``linear-time'' logic. In contrast, here $\CmuGL$ is 
 a ``branching-time'' logic, in the sense that there is no additional branching to
 be abstracted away.

 Prop.~\ref{prop:translationIsCorrect}, together
 with Thm.~\ref{thm:correctnessOfProgMeasEqSys}, already
 gives us a characterization of the semantics $\sem{\varphi}_{c}$ in
 terms of progress measures. In this section
 we shall  rephrase it to yet another form, called \emph{MC progress
 measure}, that is easier to manipulate. Using it we present our main
 technical results, namely a generic model-checking algorithm
 (Algorithm~\ref{algo:branchingTimeMC}) and its complexity (Thm.~\ref{thm:complexityBranchingTimeMC}).

 % this
 % characterization in such a way that is more amenable to algorithmic
 % search, and derive a model-checking algorithm.

 The following  correspondence for (polyadic) modalities---that is not unlike in the Yoneda lemma---will
be used in the following developments.
 \begin{mylemma}[$\lambda^{\tuple{\seq{j}{n}}},\tilde{\lambda}$]
  \label{lem:yonedaLikeCorForPolyadicModality}
  Let $\lambda$ be a natural transformation,
  given by arrows
  $\lambda_{X}\colon (\Omega^{X})^{n}\to
 \Omega^{FX}$ that are natural in $X$. (This is the setting in
 Def.~\ref{def:modalSigAndPropSig}, where $\lambda$ is an $n$-ary
 modality).  Let $m\in \omega$ and $j_{1},\dotsc, j_{n}\in [1,m]$.
These data induce an arrow
 \begin{displaymath}
  \lambda^{\tuple{j_{1},\dotsc, j_{n}}}\colon F(\Omega^{m})\rightarrow \Omega
 \end{displaymath}
 by
$\lambda^{\tuple{j_{1},\dotsc,
 j_{n}}}:=\lambda_{\Omega^{m}}(\pi_{j_{1}},\dotsc,\pi_{j_{n}})$. Recall
 that
 $\lambda_{\Omega^{m}}$ is of type $(\Omega^{\Omega^{m}})^{n}\to
 \Omega^{F(\Omega^{m})}$, and $\pi_{j}\colon \Omega^{m}\to \Omega$.
 % , $(\pi_{j_{1}},\dotsc,\pi_{j_{n}})\mapsto
 % \lambda_{\Omega^{m}}$.

 Moreover, let us define $\tilde{\lambda}\colon F(\Omega^{n})\to \Omega$
  by $\tilde{\lambda}:=\lambda^{\tuple{1,2,\dotsc,n}}=\lambda_{\Omega^{n}}(\pi_{1},\dotsc,\pi_{n})$,
  where $\pi_{1},\dotsc,\pi_{n}\colon \Omega^{n}\to \Omega$. Then we have
  \begin{displaymath}
   \vcenter{\xymatrix@R=.6em@C+3em{
   {F(\Omega^{m})}
   \ar[rd]^-{\quad\lambda^{\tuple{j_{1},\dotsc,j_{n}}}}
   \ar[d]_{F\tuple{\pi_{j_{1}},\dotsc,\pi_{j_{n}}}}
   &
   \\
   {F(\Omega^{n})}
   \ar[r]^(.3){\tilde{\lambda}}
   &
      {\Omega\mathrlap{\enspace.}}
      }}
%      \tag*{\myqed}
  \end{displaymath}

  \vspace{-2em}{\myqed}
 \end{mylemma}

 \subsection{MC Progress Measure}
 \label{subsec:branchingTimeProgressMeasure}
 We start with customizing the lattice-theoretic notion of progress
 measure (Def.~\ref{def:progressMeasForEqSys}) to one that is
 tailored to $\CmuGL$ model checking.  For reuse in
 later sections, the definition is separated into the
 transition-irrelevant part (which we call \emph{pre-progress measure}),
 and the full definition.

 \begin{mydefinition}[pre-progress measure, pPM]\label{def:preProgMeas}
  Let $F\colon \Sets\to\Sets$ be a functor.
  % as described in
  % Def.~\ref{def:nondetCoalgebra}.
  Let $\varphi$ be a
  $\CmuGL$-formula---where $\Lambda$ is a modal signature over
  $F$---that is identified with a simple equational system
  \begin{math}
     u_{1}=_{\eta_{1}}\varphi_{1},
  \dotsc,
  u_{m}=_{\eta_{m}} \varphi_{m}
  \end{math}
  as
  in~\S{}\ref{subsec:formulasAsEqSys}.
   Let $i_{1}<\cdots
 <i_{k}$
 enumerate the indices of all the $\mu$-variables.

  A \emph{pre-progress measure (pPM)} $p$ for $\varphi$
  is given by a tuple
 \begin{displaymath}
  p\;=\;
  \bigl(\,
  (
  \overline{\alpha_{1}},\dotsc,
  \overline{\alpha_{k}}),
  \,
  \bigl(\,\approximant_{i}(\alpha_{1},\dotsc,\alpha_{k})\,\bigr)_{i,\seq{\alpha}{k}}
\,\bigr)
 \end{displaymath}
that
consists of:
 \begin{itemize}
  \item the \emph{maximum prioritized ordinal}
 $(\overline{\alpha_{1}},\dotsc, \overline{\alpha_{k}})$; and 
  \item the \emph{approximants} $\approximant_{i}(\alpha_{1},\dotsc,\alpha_{k})\in \Omega$, defined for
	each
	$i\in[1,m]$ and each
	prioritized ordinal
 $(\alpha_{1},\dotsc,\alpha_{k})$
 such that
	$
	\alpha_{1}\le\overline{\alpha_{1}},\dotsc,
	\alpha_{k}\le\overline{\alpha_{k}}
	$. 
% Such a $k$-tuple $(\alpha_{1},\dotsc,\alpha_{k})$
% 	of ordinals is called a \emph{progress measure}.
 \end{itemize}
 The approximants $\approximant_{i}(\alpha_{1},\dotsc,\alpha_{k})$
 are subject to:
 	\begin{enumerate}
	 \item \label{item:preprogressMeasDefMonotonicity}
	      \textbf{(Monotonicity)}
	       Let $i\in[1,m]$ (hence $u_{i}$ is either a $\mu$- or
	       $\nu$-variable). Then 
	       \begin{math}
		(\alpha_{1},\dotsc,\alpha_{k})
		\preceq_{i}
		(\alpha'_{1},\dotsc,\alpha'_{k})
	       \end{math}
	       implies
	       	       \begin{math}
		\approximant_{i}(\alpha_{1},\dotsc,\alpha_{k})
		\sqsubseteq
		\approximant_{i}(\alpha'_{1},\dotsc,\alpha'_{k})
	       \end{math}.
	 \item\label{item:preprogressMeasDefMuVarBaseCase}
	      \textbf{($\mu$-variables, base case)}
	      Let $a\in [1,k]$. Then $\alpha_{a}=0$ implies
	      either:
	      $p_{i_{a}}(\alpha_{1},\dotsc,\alpha_{k}) =\bot$, or
	       there exists a
	      prioritized
 ordinal $(\alpha'_{1},\dotsc,\alpha'_{k})$ such that
 $(\alpha'_{1},\dotsc,\alpha'_{k})\prec_{i_{a}}
 (\alpha_{1},\dotsc,\alpha_{k})$ and 
 $p_{i_{a}}(\alpha_{1},\dotsc,\alpha_{k}) \sqsubseteq
	      p_{i_{a}}(\alpha'_{1},\dotsc,\alpha'_{k})$.
	      (Note here that this condition mirrors Cond.~2' of
	      Def.~\ref{def:extendedProgressMeasForEqSys}, rather than
	      Cond.~2 of
	      Def.~\ref{def:progressMeasForEqSys}. Prop.~\ref{prop:soundnessOfExtProgMeas}
	      justifies doing so.)
%      $\approximant_{i_{a}}(\alpha_{1},\dotsc,
%      \overset{\underparen{a}}{0},\dotsc,\alpha_{k})$
	       % $\approximant_{i_{a}}(\alpha_{1},\dotsc,
	       % 	  \alpha_{a},\dotsc,\alpha_{k})=\bot$.
	 \item\label{item:preprogressMeasDefMuVarStepCase}
	   \textbf{($\mu$-variables, step case)}
	   	      Let $a\in [1,k]$, and let 
	       $(\alpha_{1},\dotsc,
	       \alpha_{a}+1,\dotsc,\alpha_{k})$ be
	       a prioritized ordinal
	       such that its $a$-th counter
	      $\alpha_{a}+1$ is a successor ordinal.
	      Consider
	       the approximant $\approximant_{i_{a}} (\alpha_{1},\dotsc,
	      \alpha_{a}+1,\dotsc,\alpha_{k})$. Since $u_{i_{a}}$ is a
	      $\mu$-variable, by a  requirement in
	      Def.~\ref{def:simpleCmuGLEqSys} the corresponding equation
	      is of the form $u_{i_{a}}=_{\mu} u_{i'}$ for some $i'\in
	      [1,m]$.
	      We require that
	      		  there exist ordinals
		  		  $\beta_{1},\dotsc,\beta_{a-1}$ such that
		  \begin{multline*}
		  \approximant_{i_{a}} (\alpha_{1},\dotsc,
		   \alpha_{a}+1,\dotsc,\alpha_{k})
		   \\
		   \sqsubseteq
		  \approximant_{i'} (\beta_{1},\dotsc,\beta_{a-1},
		   \alpha_{a},\dotsc,\alpha_{k})
		  \end{multline*}
		  and
		  $\beta_{1}\le\overline{\alpha_{1}},\dotsc,
		  \beta_{a-1}\le\overline{\alpha_{a-1}}$.
	   % \begin{enumerate}
	   %  \item \textbf{(RHS is a variable)} If the formula
	   % 	  $\varphi_{i_{a}}$ in the
	   % 	  $i_{a}$-th equation $u_{i_{a}}=_{\mu}\varphi_{i_{a}}$
	   % 	  is a variable $u_{i'}$ (for some $i'\in [1,m]$), then
	   % 	  there exist ordinals
	   % 	  		  $\beta_{1},\dotsc,\beta_{a-1}$ such that
	   % 	  \begin{multline*}
	   % 	  \approximant_{i_{a}} (\alpha_{1},\dotsc,
	   % 	   \alpha_{a}+1,\dotsc,\alpha_{k})
	   % 	   \\
	   % 	   \sqsubseteq
	   % 	  \approximant_{i'} (\beta_{1},\dotsc,\beta_{a-1},
	   % 	   \alpha_{a},\dotsc,\alpha_{k})
	   % 	  \end{multline*}
	   % 	  and
	   % 	  $\beta_{1}\le\overline{\alpha_{1}},\dotsc,
	   % 	  \beta_{a-1}\le\overline{\alpha_{a-1}}$.
	   %  \item \textbf{(RHS is a propositional formula) (** no need!!
	   % 	  **)} If the
	   % 	  formula $\varphi_{i_{a}}$ is a propositional formula
	   % 	  $\boxdot_{\gamma}\bigl(u_{j_{1}},\dotsc,u_{j_{n}}\bigr)$,
	   % 	  then
	   % 	  there exist ordinals
	   % 	  		  $\beta_{1},\dotsc,\beta_{a-1}$ such that
	   % 	  \begin{align*}
	   % 	  &\approximant_{i_{a}} (\alpha_{1},\dotsc,
	   % 	   \alpha_{a}+1,\dotsc,\alpha_{k})
	   % 	   \\		
	   % 	   &\sqsubseteq
	   % 	   \sem{\gamma}
	   % 	\left(\,
	   % 	\begin{array}{c}
	   % 	\approximant_{j_{1}} (\beta_{1},\dotsc,\beta_{a-1},
	   % 	 \alpha_{a},\dotsc,\alpha_{k}),
	   % 	 \\
	   % 	 \dotsc,
	   % 	  \\
	   % 	\approximant_{j_{n}} (\beta_{1},\dotsc,\beta_{a-1},
	   % 	\alpha_{a},\dotsc,\alpha_{k})
	   % 	\end{array}
	   % 	\,\right)
	   % 	  \end{align*}
	   % 	  and
	   % 	  $\beta_{1}\le\overline{\alpha_{1}},\dotsc,
	   % 	  \beta_{a-1}\le\overline{\alpha_{a-1}}$.
	   % \end{enumerate}

	 \item\label{item:preprogressMeasDefMuVarLimitCase}
	      \textbf{($\mu$-variables, limit case)}
	       Let $a\in [1,k]$, and
	       let
	       $(\alpha_{1},\dotsc,
	      %\alpha_{a},\dotsc,
	      \alpha_{k})$
	       be a prioritized ordinal
	       such that its $a$-th counter
	       $\alpha_{a}$
	      is a limit ordinal. 
	      % , regarding the approximant
	      %  $\approximant_{i_{a}} (\alpha_{1},\dotsc,
	      % \alpha_{a},\dotsc,\alpha_{k})$,
	      We require 
	       \begin{displaymath}
		\approximant_{i_{a}} (\alpha_{1},\dotsc,
		\alpha_{a},\dotsc,\alpha_{k})
		% \\
		% &\qquad\sqsubseteq\;
		\sqsubseteq\textstyle
		\bigsqcup_{\beta<\alpha_{a}}
				\approximant_{i_{a}} (\alpha_{1},\dotsc,
		\beta,\dotsc,\alpha_{k})\enspace.
	       \end{displaymath}
	      	 \item\label{item:preprogressMeasDefNuVar}
	      \textbf{($\nu$-variables)}
	       Let $i\in [1,m]\setminus\{i_{1},\dotsc, i_{k}\}$ (i.e.\
	       $u_{i}$ is a $\nu$-variable in the
	       system~(\ref{eq:sysOfEq})); let $a\in [1,k]$ such that
	       \begin{displaymath}
		i_{1}<\cdots <i_{a-1}<i<i_{a}<\cdots < i_{k}.
	       \end{displaymath}
	       Let $(\alpha_{1},\dotsc,\alpha_{k})$ be a prioritized ordinal.
	       We require the following on the approximant
		      $\approximant_{i} (\alpha_{1},\dotsc,\alpha_{k})$:
	   \begin{enumerate}
	    \item \textbf{(RHS is a variable)} If the formula
		  $\varphi_{i}$ in the
		  $i$-th equation $u_{i}=_{\nu}\varphi_{i}$
		  is a variable $u_{i'}$ (for some $i'\in [1,m]$), then
		  there exist ordinals
		  		  $\beta_{1},\dotsc,\beta_{a-1}$ such that
		  \begin{multline*}
		  \approximant_{i_{a}} (\alpha_{1},\dotsc,
		   \alpha_{a},\dotsc,\alpha_{k})
		   \\
		   \sqsubseteq
		  \approximant_{i'} (\beta_{1},\dotsc,\beta_{a-1},
		   \alpha_{a},\dotsc,\alpha_{k})
		  \end{multline*}
		  and
		  $\beta_{1}\le\overline{\alpha_{1}},\dotsc,
		  \beta_{a-1}\le\overline{\alpha_{a-1}}$.
	    \item \textbf{(RHS is a propositional formula)} If the
		  formula $\varphi_{i}$ is a propositional formula
		  $\boxdot_{\gamma}\bigl(u_{j_{1}},\dotsc,u_{j_{n}}\bigr)$,
		  then
		  there exist ordinals
		  		  $\beta_{1},\dotsc,\beta_{a-1}$ such that
		  \begin{align*}
		  &\approximant_{i} (\alpha_{1},\dotsc,
		   \alpha_{a},\dotsc,\alpha_{k})
		   \\		
		   &\sqsubseteq
		   \sem{\gamma}
		\left(\,
		\begin{array}{c}
		\approximant_{j_{1}} (\beta_{1},\dotsc,\beta_{a-1},
		 \alpha_{a},\dotsc,\alpha_{k}),
		 \\
		 \dotsc,
		  \\
		\approximant_{j_{n}} (\beta_{1},\dotsc,\beta_{a-1},
		\alpha_{a},\dotsc,\alpha_{k})
		\end{array}
		\,\right)
		  \end{align*}
		  and
		  $\beta_{1}\le\overline{\alpha_{1}},\dotsc,
		  \beta_{a-1}\le\overline{\alpha_{a-1}}$.
	   \end{enumerate}

	\end{enumerate}

  Let $\alpha$ be an ordinal. 
  The collection of all pre-progress measures for a formula $\varphi$,
  whose maximum prioritized ordinal $(
	\overline{\alpha_{1}},\dotsc,
	\overline{\alpha_{k}})
  $ satisfies $\overline{\alpha_{i}}= \alpha$ for each $i\in [1,k]$,
  shall be
  denoted by $\pPM_{\varphi,\alpha}$.
 \end{mydefinition}
 \noindent
Recall that $\Omega$ is the complete lattice of truth values. In the
definition of $\pPM_{\varphi,\alpha}$, the explicit bound by $\alpha$ is
there so that the collection $\pPM_{\varphi,\alpha}$ is a (small) set.

Comparing the previous definition with Def.~\ref{def:progressMeasForEqSys}
of progress measures, what are missing here are the treatment
of modal formulas $\heartsuit_{\lambda}(u_{j_{1}},\dotsc,u_{j_{n}})$
in Cond.%
%~\ref{item:preprogressMeasDefMuVarStepCase}and
~\ref{item:preprogressMeasDefNuVar}---this is precisely the
case where the transition structure of the coalgebra in question
becomes relevant. In the current setting of $\CmuGL$ as a ``branching-time''
logic, this case is taken care of in the following way. MC stands for
``model checking.''
 \begin{mydefinition}[MC progress measure]
  \label{def:branchingTimeProgMeas}
  Assume the setting of Def.~\ref{def:progressMeasForEqSys}, and let
  $c\colon X\to FX$ be a coalgebra in $\Sets$. An \emph{MC progress
  measure}  for $\varphi$ over $c$
  % simply called a \emph{progress
  % measure} when no confusion is likely,
  is given by:
  \begin{itemize}
   \item some ordinal $\alpha$, called the \emph{maximum ordinal}, and
   \item a function $Q\colon X\to \pPM_{\varphi,\alpha}$,
	 %  and
   % \item an $F$-coalgebra structure
   % 	 $q\colon Q\to FQ$
  \end{itemize}
  that are subject to the following condition.
   \begin{enumerate}
  % \item[3(c)]
  % 	   \textbf{($\mu$-variables, step case, RHS is a modal formula)}
  % 	     In the setting of
  % 	     Cond.~\ref{item:preprogressMeasDefMuVarStepCase} of
  % 	     Def.~\ref{def:preProgMeas}, 
  % 	     assume further that the formula $\varphi_{i_{a}}$ is a
  % 	      modal formula:
  % 	      $\varphi_{i_{a}}=
  % 	      \heartsuit_{\lambda}(u_{j_{1}},\dotsc,u_{j_{n}})$. Moreover
  % 	      let
  % 	      $p:=Q(x)$.

  % 	     Now consider
  % 	     the approximant 	     $p_{i_{a}}(\alpha_{1},\dotsc,
  % 	      \alpha_{a}+1,\dotsc,\alpha_{k})\in \Omega$ of $p$.
  % 	      Then there must exist ordinals
  % 	      $\beta_{1},\dotsc,\beta_{a-1}$ such that
  % 		 \begin{align*}
  % 		  &
  % 		 p_{i_{a}}(\alpha_{1},\dotsc,
  % 		  \alpha_{a}+1,\dotsc,\alpha_{k})\sqsubseteq
  % 		  \\&
  % 		\PT_{\heartsuit_{\lambda}(u_{j_{1}},\dotsc,u_{j_{n}})}
  % 		%\PT_{\varphi_{i_{a}}}
  % 		(\beta_{1},\dotsc,\beta_{a-1},
  % 		  \alpha_{a},\dotsc, \alpha_{k})
  % 		  \bigl(\,(FQ\co c)(x)\,\bigr),
  % 		 \end{align*}
  % 	     and $\beta_{1}\le\alpha,\dotsc, \beta_{a-1}\le\alpha$.

    \item[5(c)]
	       \textbf{($\nu$-variables,  RHS is a modal formula)}
	       	      Let $x\in X$ and $p:=Q(x)$ be a pre-progress
	       measure for $\varphi$. Let $i\in [1,m]$ and
			assume the setting of
	     Cond.~\ref{item:preprogressMeasDefNuVar} of
	     Def.~\ref{def:preProgMeas} (i.e.\ $u_{i}$ is a $\nu$-variable), 
	     and further that the formula $\varphi_{i}$ is a modal
	      formula:
		$\varphi_{i}=\heartsuit_{\lambda}(u_{j_{1}},\dotsc,u_{j_{n}})$.

			     Now consider
	     the approximant 	     $p_{i}(\alpha_{1},\dotsc,
	      \alpha_{a},\dotsc,\alpha_{k})\in \Omega$ of $p$.
	      We require there exist ordinals
	      $\beta_{1},\dotsc,\beta_{a-1}$ such that
	       \begin{equation}\label{eq:10260954}
 		 \begin{aligned}
		  &
		 p_{i}(\alpha_{1},\dotsc,
		  \alpha_{a},\dotsc,\alpha_{k})\sqsubseteq
		  \\&
		\PT_{\heartsuit_{\lambda}(u_{j_{1}},\dotsc,u_{j_{n}})}
		%\PT_{\varphi_{i_{a}}}
		(\beta_{1},\dotsc,\beta_{a-1},
	       \alpha_{a},\dotsc, \alpha_{k})\bigl((FQ\co c)(x)\bigr),
		 \end{aligned}
	       \end{equation}	     
and $\beta_{1}\le\alpha,\dotsc,
		\beta_{a-1}\le\alpha$.

	        Note that
	      $(FQ\co c)(x)\in F(\pPM_{\varphi,\alpha})$ since
	      $X\stackrel{c}{\to} FX\stackrel{FQ}{\to} F(\pPMpa)$.
	      %$c(x)\in FX$.
	      % $(F\pi_{2}\co q)(x,p)\in
	       % F(\pPM_{\varphi,\alpha})$.
	      For each
	      $(\alpha'_{1},\dotsc,\alpha'_{k})$, the function
 \begin{align*}
%    &	     \PT_{\varphi_{i_{a}}}(\alpha'_{1},\dotsc,\alpha'_{k})(q)
% \\&=
  \PT_{\heartsuit_{\lambda}(u_{j_{1}},\dotsc,u_{j_{n}})}
  (\alpha'_{1},\dotsc,\alpha'_{k})
  \;\colon\; F(\pPM_{\varphi,\alpha})\to \Omega
 \end{align*}	      
in~(\ref{eq:10260954}) is defined as follows.
	     	     (The name $\PT$ comes from ``predicate
	     transformer.'')
\begin{equation}\label{eq:branchingTimePredTransf}
	      \begin{aligned}
	       &\PT_{\heartsuit_{\lambda}(u_{j_{1}},\dotsc,u_{j_{n}})}(\alpha'_{1},\dotsc,\alpha'_{k})\;:=\;
	       \\
	       &
	       \Bigl[
	       F(\pPMpa)
	       \stackrel{F(\ev(\alpha'_{1},\dotsc,\alpha'_{k}))}{\longrightarrow}
	       F(\Omega^{m})
	      \stackrel{\lambda^{\tuple{j_{1},\dotsc, j_{n}}}}{\longrightarrow}
	       \Omega
	       \Bigr],
	      \end{aligned}
\end{equation}	     where $\lambda^{\tuple{j_{1},\dotsc, j_{n}}}$ is from
	     Lem.~\ref{lem:yonedaLikeCorForPolyadicModality}, and the
	       function
	       \begin{displaymath}
	       \ev(\alpha'_{1},\dotsc,\alpha'_{k})\colon \pPMpa\to
	       \Omega^{m}
	       \end{displaymath}
	       is defined by ``fixing a prioritized
	     ordinal,'' that is,
	     \begin{displaymath}
	      \ev(\vec{\alpha'})(p)
	      :=
	      \bigl(p_{1}(\vec{\alpha'}),\dotsc,
	      p_{m}(\vec{\alpha'})
	      \bigr)\;\in \Omega^{m}.
	     \end{displaymath}

	       % Here $\PT_{\heartsuit_{\lambda}(u_{j_{1}},\dotsc,u_{j_{n}})}(\vec{\alpha'})$ is the same as in~(\ref{eq:branchingTimePredTransf}).

    % \item[6.]
    % 	     	   \textbf{(Compatibility with $c$)} The following
    % 	     diagram commutes. Here $\iota$ is the inclusion map.
    % 	     \begin{displaymath}
    % 	      \vcenter{\xymatrix@R=.6em{
    % 	      {FQ}
    % 	      \ar[r]^-{F\iota}
    % 	      &
    % 	      {F(X\times \pPM_{\varphi,\alpha})}
    % 	      \ar[r]^-{F\pi_{1}}
    % 	      &
    % 	      {FX}
    % 	      \\
    % 	      {Q}
    % 	      \ar[u]^-{q}
    % 	      \ar@{^{(}->}[r]_-{\iota}
    % 	      &
    % 	      {X\times \pPM_{\varphi,\alpha}}
    % 	      \ar[r]_-{\pi_{1}}
    % 	      &
    % 	      {X}
    % 	      \ar[u]_-{c}
    % 	      }}
    % 	     \end{displaymath}

\end{enumerate}
\end{mydefinition}
   The composite in the definition of
   $\PT_{\heartsuit_{\lambda}(u_{j_{1}},\dotsc,u_{j_{n}})}(
   %\alpha'_{1},\dotsc,\alpha'_{k}
   \overrightarrow{\alpha'})$
   in~(\ref{eq:branchingTimePredTransf})
   might seem exotic,
   but the definition here is in fact  a straightforward adaptation of the common 
   interpretation of modal formulas in coalgebraic logics. Recall the interpretation of a modal formula
   $\heartsuit_{\lambda}(\varphi_{1},\dotsc,\varphi_{n})$ in
   Def.~\ref{def:CmuFmlSem}, that is also the standard one in the literature
   (see e.g.~\cite{SchroederP09}). Then it is not hard---by naturality of
   $\lambda$, much like in the proof of
   Thm.~\ref{thm:correctnessOfBranchingTimeProgMeas}---that this
   standard definition of 
   $\sem{\heartsuit_{\lambda}(\vec{\varphi})}_{c}$ is equivalent to the following,
   where
   $\sem{\varphi_{i}}_{c}\colon X\to \Omega$ are the interpretations of the
   constituent
   subformulas (for $i\in[1,n]$), and $\tilde{\lambda}$ is from
   Lem.~\ref{lem:yonedaLikeCorForPolyadicModality}. 
   \begin{displaymath}
    \sem{\heartsuit_{\lambda}(\vec{\varphi})}_{c}
    \;=\;
    \bigl(\,
 		     X\stackrel{c}{\rightarrow} FX
    \stackrel{F\tuple{\sem{\varphi_{i}}_{c}}_{i}}{\longrightarrow}
          F(\Omega^{n})
		     % \stackrel{F(\ev'(\vec{\alpha}))}{\longrightarrow}
		     % F(\Omega^{m})
		     \stackrel{
          %\lambda^{\tuple{{1},\dotsc, {n}}}
                     \tilde{\lambda}}{\longrightarrow}
		     \Omega\,\bigr).
   \end{displaymath}
   This indeed resembles the right-hand side of~(\ref{eq:10260954}),
   namely
   \begin{displaymath}
    \bigl(\,
    X
    \stackrel{c}{\to}
    FX
    \stackrel{F(\ev(\vec{\alpha'})\co Q)}{\longrightarrow}
    F(\Omega^{m})
    \stackrel{\lambda^{\tuple{j_{1},\dotsc,j_{n}}}}{\longrightarrow}
    \Omega
    \,\bigr)(x).
   \end{displaymath}

   \auxproof{
    \begin{displaymath}
        \vcenter{\xymatrix@R=1em@C+2em{
   {\Omega^{n}}
   &
   {(\Omega^{\Omega^{n}})^{n}}
       \ar[r]^-{\lambda_{\Omega^{n}}}
        \ar[d]^{(\Omega^{\tuple{\sem{\varphi_{i}}_{c}}_{i}})^{n}}
   &
   {\Omega^{F(\Omega^{n})}}
       \ar[d]^{\Omega^{F\tuple{\sem{\varphi_{i}}_{c}}_{i}}}
   \\
   X
   \ar[u]_{\tuple{\sem{\varphi_{i}}_{c}}_{i}}
   &
   {(\Omega^{X})^{n}}
   \ar[r]_-{\lambda_{X}}
   &
   {\Omega^{FX}}
%   \mathrlap{\enspace,}
   }}
    \end{displaymath}
 and, since
 $\lambda^{\tuple{1,\dotsc,n}}=\lambda_{\Omega^{n}}(\pi_{1},\dotsc,\pi_{n})$ (Lem.~\ref{lem:yonedaLikeCorForPolyadicModality}),
    \begin{displaymath}
        \vcenter{\xymatrix@R=1em@C+0em{
   % {\Omega^{n}}
   % &
   {(\pi_{1},\dotsc,\pi_{n})}
       \ar@{|->}[r]^-{\lambda_{\Omega^{n}}}
        \ar@{|->}[dd]^{(\Omega^{\tuple{\sem{\varphi_{i}}_{c}}_{i}})^{n}}
   &
   {\lambda^{\tuple{1,\dotsc,n}}}
       \ar@{|->}[d]^{\Omega^{F\tuple{\sem{\varphi_{i}}_{c}}_{i}}}
   \\
   &
   {\bigl(\,FX\xrightarrow{F\tuple{\sem{\varphi_{i}}_{c}}_{i}}
     F(\Omega^{n}) \xrightarrow{\lambda^{\tuple{1,\dotsc,n}}}
     F\Omega\,\bigr)}
     \ar@{=}[d]
   \\
   % X
   % \ar[u]_{\tuple{\sem{\varphi_{i}}_{c}}_{i}}
   % &
   {(\sem{\varphi_{1}}_{c},\dotsc,\sem{\varphi_{n}}_{c})}
   \ar@{|->}[r]_-{\lambda_{X}}
   &
   {\lambda_{X}(\sem{\varphi_{1}}_{c},\dotsc,\sem{\varphi_{n}}_{c})}
   \mathrlap{\enspace.}
   }}
    \end{displaymath}
    }
      
   % end: added

 \begin{mytheorem}[correctness of MC progress measure]
  \label{thm:correctnessOfBranchingTimeProgMeas}
  Assume the setting of Def.~\ref{def:branchingTimeProgMeas}. In
  particular, the formula $\varphi$ is translated to an equational
  system with $m$ variables.
   \begin{enumerate}
  \item\label{item:soundnessBranchingTimeProgressMeas}
       \textbf{(Soundness)}
       Let $Q$ be an MC
       progress measure (with the maximum ordinal $\alpha$), $x\in X$ and $p:=Q(x)$.
       % be an MC progress measure over an ordinal $\alpha$.
       Then
       \begin{displaymath}
	p_{m}(\alpha,\dotsc,\alpha)
	\sqsubseteq\sem{\varphi}_{c}(x),
       \end{displaymath}
 where $\sem{\varphi}_{c}\colon X\to \Omega$
       is from Def.~\ref{def:CmuFmlSem}.
 \item\label{item:completenessBranchingTimeProgressMeas} 
  \textbf{(Completeness)}
      There exists an MC progress measure $Q$
that achieves
      the optimal. That is, an MC progress measure $Q$ such that
      $(Q(x))_{m}(\alpha,\dotsc,\alpha)
      =\sem{\varphi}_{c}(x)$  for each $x\in X$.
      Moreover, $Q$ can be chosen so that its maximum ordinal $\alpha$
      is $\alpha=\ascCL(\Omega^{X})$, where $\ascCL(\Omega^{X})$ is the
      length of the longest strictly ascending chain in $\Omega^{X}$
      (see Thm.~\ref{thm:correctnessOfProgMeasEqSys}.2).
      % \begin{displaymath}
      %  (x,p)\in Q
      %  \;\Longleftrightarrow\;
      % 	p_{m}(\alpha,\dotsc,\alpha)
      % 	=\sem{\varphi}_{c}(x).
	% \end{displaymath}
      \myqed
 \end{enumerate}
 \end{mytheorem}

 \subsection{Algorithms}\label{subsec:branchingTimeAlgo}
 Here we shall further translate the notion of MC progress measure
 (Def.~\ref{def:branchingTimeProgMeas}) to a Jurdzinski-style
 presentation; the latter shall be called a \emph{matrix progress
 measure}.  The correspondence is an extension of the one in
 Appendix~\ref{appendix:parityProgressMeasure};
%===============================================================
\iffalse
%===============================================================
 (in the extended version~\cite{HasuoSC16POPLExtended});
%===============================================================
\fi
%===============================================================
 see also Rem.~\ref{rem:stoneLikeDuality}. We shall then devise a
 model-checking algorithm based on matrix progress measures. Thanks to
 the concrete presentation with matrices, we believe its implementation
 is a fairly straightforward task.

\begin{myassumption}\label{asm:booleanAndFiniteForModelChecking}
  Throughout~\S{}\ref{subsec:branchingTimeAlgo} we focus on the Boolean
 setting (i.e.\ $\Omega=\Bool$), and restrict the state space $X$ of the
 coalgebra $c\colon X\to FX$ (as a system model) to be finite. This is a
 reasonable assumption because we aim at a concrete algorithm.
 In  view of Thm.~\ref{thm:correctnessOfBranchingTimeProgMeas}.2, in
 employing the theoretical machinery developed so far, all the ordinals
 that occur can be restricted to finite (since $\ascCL(\Bool^{X})=|X|$
 is finite).

 Furthermore, we restrict the propositional signature $\Gamma$ to 
 $\Gamma_{n}:=\{\bigwedge_{n}, \bigvee_{n}\}$, where $\bigwedge_{n}$ and
 $\bigvee_{n}$ are the $n$-ary conjunction and disjunction operators with
  obvious interpretations. This signature of $\Gamma$ is functionally
 complete
 in the current monotonic Boolean setting:  any other propositional
 connective $\gamma\colon \Bool^{n}\to \Bool$ can be encoded by
 \begin{align*}\small
  % &\gamma(l_{1},\dotsc,l_{n})
  % =
  % \\
\begin{array}{l}
   \textstyle\bigvee
  \bigl\{
  \textstyle\bigwedge \{l_{i_{1}},\dotsc,l_{i_{k}}\}
  \,\bigl|\bigr.\,
  l_{i_{1}}=\cdots =l_{i_{k}}=\ttrue \,\Rightarrow\,\gamma(l_{1},\dotsc,l_{n})=\ttrue
  \,\bigr\}.
\end{array} 
\end{align*}
\end{myassumption}

  \begin{mydefinition}[prioritized ordinal matrix, POM]
   \label{def:prioritizedOrdMatrix}
   Assume the setting of Def.~\ref{def:preProgMeas}.
   A \emph{prioritized ordinal matrix} is an $m\times k$ matrix
   \begin{displaymath}
    \left[
    \begin{array}{ccc}
     \alpha^{(1)}_{1} & \cdots
      &
      \alpha^{(1)}_{k} 
      \\
     \vdots&\ddots&\vdots
      \\
     \alpha^{(m)}_{1} & \cdots
      &
      \alpha^{(m)}_{k} 
    \end{array}
    \right],
   \end{displaymath}
   where each entry $\alpha^{(i)}_{a}$ is either
   \begin{itemize}
    \item an ordinal, or
    \item the symbol $\NoGood$ for ``failure.''
%	  (cf.\ Def.~\ref{def:preProgMeas}). 
   \end{itemize}
   It is required that, if any entry $\alpha^{(i)}_{a}$ is $\NoGood$
   then all entries on the same row is $\NoGood$, that is,
   $\alpha^{(i)}_{1}=\cdots=\alpha^{(i)}_{k}=\NoGood$.

   The set of  all POMs, such that all the ordinals therein are no
   bigger than $\alpha$, is denoted by $\POM_{\alpha}$.
   % Let $\alpha$ be an ordinal. The set of all POMs such that all the
   % ordinals therein are no bigger than $\alpha$ is denoted by
   % $\POM_{\alpha}$. 
  \end{mydefinition}

    \begin{wrapfigure}[6]{r}{0pt}
     \!\!     \!\!     \!\!     \!\!     \!\!     \!\!     \!\!     \!\!     \!\!
  \begin{math}
   \begin{array}{c}
    u_{1}\\ u_{2}\\u_{3}\\ u_{4} \\ u_{5}
   \end{array}
    \left[
    \begin{array}{ccc}
      3 & 5
      & 
      2
      \\
     *& 1& 0
      \\
      \NoGood &      \NoGood& \NoGood
       \\
     *&*&4
      \\
     *&*&*
    \end{array}
    \right]
  \end{math}
    \end{wrapfigure}
  A POM is therefore an $m$-tuple of prioritized ordinals, where some
  prioritized ordinals can be replaced by $\NoGood$.
  Its $i$-th row will be a prioritized ordinal for
  the $i$-th variable $u_{i}$.
  In view of the monotonicity conditions (in
  Def.~\ref{def:progressMeasForEqSys} and~\ref{def:preProgMeas}) and the
  definition of $\preceq_{i}$ (Def.~\ref{def:prioritizedOrdinal}), we
  can see that some first elements in a row (precisely: those which
  correspond to $\mu$-variables with a smaller priority than $u_{i}$)
  do not make any difference. Such entries can safely be denoted by $*$
  (``arbitrary''). An example is shown in the above: it is a POM
  for an equational system with 5 variables, in which
  $u_{1}, u_{3}, u_{4}$ are $\mu$-variables and $u_{2}, u_{5}$ are
  $\nu$-variables.  We shall however restrict  use of $*$ for providing
  intuitions; it does not appear in the technical
  developments.

  In the current section (\S{}\ref{subsec:branchingTimeAlgo}) where $X$ is
assumed to be a finite set, it is not needed to allow any ordinal as an
entry of a POM (Def.~\ref{def:prioritizedOrdMatrix}). Natural numbers
will just suffice. 
% We will need the generalization later
% in~\S{}\ref{sec:smallModelProperty}. 

    \begin{mydefinition}[matrix progress measure, MPM]
     \label{def:MPM}
    Assume the setting of Def.~\ref{def:branchingTimeProgMeas}. A
    \emph{matrix progress measure (MPM)} for $\varphi$ over $c$, with a
    maximum ordinal $\alpha$, is a function $R\colon X\to \POM_{\alpha}$
    that satisfies the following conditions. Let $x\in X$ be arbitrary,
    and consider $R(x)\in \POM_{\alpha}$.
    \begin{enumerate}
     \item[2.] 	\textbf{($\mu$-variables, base case)}
		Let $a\in [1,k]$ and consider the corresponding
		$\mu$-variable $u_{i_{a}}$.
		Assume  $\alpha^{(i_{a})}_{1}\neq\NoGood.$
		Then we must have
		$(R(x))^{(i_{a})}\succ_{i_{a}}(0,0,\dotsc,0)$. Note
		that the $i_{a}$-th row $(R(x))^{(i_{a})}$ of $R(x)$ is
		a prioritized ordinal, and recall $\succ_{i_{a}}$ from
		Def.~\ref{def:prioritizedOrdinal}. Note also that the required
		inequality is strict.
		% Equivalently:
		% there is $a'\in [1,k]$ such that
		% $a\le a'$ and $(R(x))^{(i)}_{a'}> 0$. 

		(That is,  a row in $R(x)$ that corresponds to a $\mu$-variable must
		not be $(*,\dotsc,*,0,\dotsc,0)$.)
     \item[3.]	\textbf{($\mu$-variables, step case)}
		Let $a\in [1,k]$ and consider the corresponding
		$\mu$-variable $u_{i_{a}}$.
		Assume  $\alpha^{(i_{a})}_{1}\neq\NoGood$.
		Let $u_{i_{a}}=_{\mu}
		u_{i'}$
		(where $i'\in [1,m]$) be the corresponding equation in
		$\varphi$. If $i'\le i_{a}$, then we must have
		$(R(x))^{(i_{a})}\succ_{i}(R(x))^{(i')}$. Note the
		inequality is strict.
		%$(R(x))^{(i_{a})}_{a}\ge (R(x))^{(i')}_{a}+1$.

		% (Note that there is no requirement if $i'> i_{a}$---in
		% which case $(R(x))^{(i')}_{a}$ would be $*$. Note also
		% that $i'=i_{a}$ is prohibited.)

     \item[4.]
	      \textbf{($\mu$-variables, limit case)}
	       Let $a\in [1,k]$, and consider the corresponding
	      $\mu$-variable $u_{i_{a}}$. Then
	      $(R(x))^{(i_{a})}_{a}$ must not be a limit ordinal.
	      (This condition is vacuous when $\alpha$ is finite.)

     \item[5.]
	      \textbf{($\nu$-variables)}
	      Let 
	      $i\in [1,m] \setminus\{i_{1},\dotsc, i_{k}\}$ (i.e.\
	      $u_{i}$ is a $\nu$-variable). Assume that $\alpha^{(i)}_{1}\neq\NoGood$.
	      % 	Let $a\in [1,k]$ be the smallest such that
	      % $i\le i_{a}$.
	      Let $u_{i}=_{\nu}\varphi_{i}$ be the
	      corresponding
	      equation in $\varphi$.
	      \begin{enumerate}
	       \item \textbf{(RHS is a variable)}
		      If the formula
		  $\varphi_{i}$ 
		     is a variable $u_{i'}$ (for some $i'\in [1,m]$).
		     Then $(R(x))^{(i)}\succeq_{i}(R(x))^{(i')}$.
	       \item \textbf{(RHS is a propositional formula)} Recall
		     that we have restricted propositional connectives
		     to $\bigwedge$ and $\bigvee$
		     (Assumption~\ref{asm:booleanAndFiniteForModelChecking}).
		     If $\varphi_{i}=\bigwedge(u_{i_{1}},\dotsc,
		     u_{i_{n}})$
		     then we require all of 
		     \begin{equation}\label{eq:201507081932}
		      (R(x))^{(i)}\succeq_{i}(R(x))^{(i_{1})},
		      \dotsc,
      		      (R(x))^{(i)}\succeq_{i}(R(x))^{(i_{n})}
		     \end{equation}
		     to hold. 
		     If $\varphi_{i}=\bigvee(u_{i_{1}},\dotsc,
		     u_{i_{n}})$
		     then we require at least one
		     of~(\ref{eq:201507081932}) to hold.
	       \item
		    \textbf{(RHS is a modal formula)}
		    Assume that  the formula
		  $\varphi_{i}$ is a modal formula
		    $\varphi_{i}=\heartsuit_{\lambda}(u_{j_{1}},\dotsc,u_{j_{n}})$. Let
		    $\vec{\alpha}=(\alpha_{1},\dotsc,\alpha_{k}):=
		    (R(x))^{(i)}$. 
		    Consider the following composite $h\colon X\to \Bool$:
		    \begin{equation}\label{eq:201507082123}
		      \begin{aligned}
		       h:=
		      \biggl(\,
		      \begin{array}{r}
  		     X\stackrel{c}{\rightarrow} FX\stackrel{FR}{\rightarrow}F(\POM_{\alpha})
		      \stackrel{F(\ev'(\vec{\alpha}))}{\longrightarrow}\quad
		      \\
		     F(\Bool^{m})
		     \stackrel{\lambda^{\tuple{j_{1},\dotsc, j_{n}}}}{\longrightarrow}
		       \Bool\,
		      \end{array}		       
\biggr), 
		      \end{aligned}		    
\end{equation}
		    where $\ev'(\vec{\alpha})\colon \POM_{\alpha}\to
		    \Bool^{m}$
		    is defined by
		    \begin{align*}
		     \Bigl(\ev'(\vec{\alpha})\bigl(\,(\beta^{(i)}_{j})_{i,j}\,\bigr)\Bigr)_{i'}=\ttrue
		     \,\defiff\,
		     \vec{\alpha}\succeq_{i'} (\beta^{(i')}_{1},\dotsc,
		     \beta^{(i')}_{k})
		    \end{align*}
		    for each $i'\in [1,m]$. We require that
		    $h(x)=\ttrue$.
	      \end{enumerate}		   
    \end{enumerate}
    \end{mydefinition}
    % Again, in the current section
    % (\S{}\ref{subsec:branchingTimeAlgo}), Cond.~4 in the previous
    % definition will not be needed since eacy entry will be a natural
		    % number.
   Again, much like for Cond.~5(c) in
   Def.~\ref{def:branchingTimeProgMeas}, the composite in~(\ref{eq:201507082123}) is
   understood as an analogue of the usual interpretation of modal
   formulas in coalgebraic logics (cf.\ Def.~\ref{def:CmuFmlSem}).

    \begin{mytheorem}[correctness of MPM]
     \label{thm:correctnessMPM}
     Assume the setting of Def.~\ref{def:branchingTimeProgMeas}. 
     \begin{enumerate}
      \item        \textbf{(Soundness)}
		   If     there exists an MPM
    $R\colon X\to \POM_{\alpha}$ such that
		   $(R(x))^{(m)}_{k}\neq\NoGood$, then
		   $\sem{\varphi}_{c}(x)=\ttrue$.
      \item        \textbf{(Completeness)} There is an optimal MPM
		   $R_{0}\colon X\to \POM_{|X|}$ such that:
    $\sem{\varphi}_{c}(x)=\ttrue$ if and only if
		   $(R_{0}(x))^{(m)}_{k}\neq\NoGood$.
   \myqed
     \end{enumerate}
    % there exists an MPM
    % $R\colon X\to \POM_{\alpha}$ such that:
    % $\sem{\varphi}_{c}(x)=\ttrue$ if and only if  $(R(x))^{(m)}_{k}\neq\NoGood$.
    % Here $\alpha=|X|$ is the number of
    % states
    % (that is assumed to be finite; see Assumption~\ref{asm:booleanAndFiniteForModelChecking}).
    \end{mytheorem}

   We  follow~\cite{Jurdzinski00} and present an algorithm that looks
   for the optimal MPM. See Algorithm~\ref{algo:branchingTimeMC}.  There
   we use the following functions.

   \begin{algorithm}[tbp]
  \caption{An algorithm for  $\CmuGL$ model checking, in the setting
    of Def.~\ref{def:branchingTimeProgMeas} and
    Assumption~\ref{asm:booleanAndFiniteForModelChecking}.
    % We write
    % $R(x,i,j)$ for $R(x)^{(i)}_{j}$;
    Here $R(x,i)$ denotes the prioritized ordinal
    $\bigl(R(x,i,1),\dotsc,R(x,i,k)\bigr)$. Note that on lines 16, 19
    and 23, $u_{i}$ is necessarily a $\nu$-variable.}
  \label{algo:branchingTimeMC}
  \begin{algorithmic}[1]
   \Require A  $\CmuGL$-formula $\varphi$ presented as an equational system
     \begin{math}
     u_{1}=_{\eta_{1}}\varphi_{1},
  \dotsc,
  u_{m}=_{\eta_{m}} \varphi_{m}
   \end{math} where $u_{i_{1}},\dotsc,u_{i_{k}}$ are $\mu$-variables, and a coalgebra $c\colon X\to FX$
    %----------------------------------------------------
   \Ensure $\sem{\varphi}_{c}\in \Bool^{X}$
       %----------------------------------------------------
   \For{each $x\in X$, $i\in[1,m]$ and $j\in [1,k]$}
   \Comment initialization
   \State $R(x,i,j):=0$
   \EndFor
   \For{each $a\in [1,k]$}
   \Comment  Cond.~2
   \State $R(x,i_{a},a):=1$
   \EndFor
   
   \Repeat
   \Comment the main loop
   % \If{$(R(x))^{i_{a}}_{a}=(R(x))^{i_{a}}_{a+1}=\dotsc=(R(x))^{i_{a}}_{k}=0$}
   % \State $(R(x))^{i_{a}}_{a}:=1$
   % \Comment Force Cond.~2
   % \EndIf

   \For{each $x\in X$ and  $i\in [1,m]$}
%   \Comment  Cond.~5

   \If{$u_{i}$ is a $\mu$-variable, $i=i_{a}$ and $\varphi_{i}=u_{i'}$}
   \Comment  Cond.~3
   \State $R(x,i):=\max_{\preceq_{i}}   \bigl\{\,
      R(x,i),\,\bigr.$
      \Comment cf.\ Def.~\ref{def:maxPrecEqMinPrecEq}
   \State \quad$
      \bigl.
      \bigl(
      R(x,i',1)
      ,\dotsc,
      R(x,i',a)+1,
      \dotsc,
         R(x,i',k)
      \bigr)
      \,\bigr\}
   $
   \EndIf

   \If{$u_{i}$ is a $\nu$-variable and $\varphi_{i}=u_{i'}$}
   \Comment  Cond.~5(a)
   \State $R(x,i):=\max_{\preceq_{i}}   \bigl\{\,
           R(x,i),\,R(x,i')\,\bigr\}$
   \EndIf

   \If{$\varphi_{i}=\bigwedge(u_{j_{1}},\dotsc,u_{j_{n}})$}
   \Comment  Cond.~5(b), the $\bigwedge$-case
   \State $R(x,i):=\max_{\preceq_{i}}   \bigl\{\,
           R(x,i),\,R(x,j_{1}),\dotsc,R(x,j_{n})\,\bigr\}$
   \EndIf

   \If{$\varphi_{i}=\bigvee(u_{j_{1}},\dotsc,u_{j_{n}})$}
   \Comment  Cond.~5(b), the $\bigvee$-case
   \State  $R(x,i):=\max_{\preceq_{i}}   \bigl\{\,
   R(x,i), \,\bigr.\,\bigr.$
   \State \hspace{8em}$\,\bigl.\,\bigl.
   \min_{\preceq_{i}}   \bigl\{\,\,R(x,j_{1}),\dotsc,R(x,j_{n})
   \,\bigr\}\,\bigr\}$
   \EndIf

   \If{$\varphi_{i}=\heartsuit_{\lambda}(u_{j_{1}},\dotsc,u_{j_{n}})$}
   \Comment  Cond.~5(c)
   \State $R(x,i):=\max_{\preceq_{i}}   \bigl\{\,
   R(x,i),\,
   \PTMPM_{i}(x)\,\bigr\}$
   \Comment cf.\ Def.~\ref{def:ptMPM}
   \EndIf

   \State

   \For{each $j\in [1,k]$}
   \If{$R(x,i,j)>|X|$}
   \Comment $u_{i}$ has seen to be false at $x$
   \State $R(x,i):=(\NoGood,\dotsc,\NoGood)$
   \EndIf
   \EndFor

   \EndFor

   \Until{no change is made}
   %---------------------------------------

   \State\Return $\{x\in X\mid
   R(x,m,k)\neq\NoGood\}$
  \end{algorithmic}
\end{algorithm}

  \begin{mydefinition}[$\max_{\preceq_{i}}, \min_{\preceq_{i}}$]
   \label{def:maxPrecEqMinPrecEq}
   In Algorithm~\ref{algo:branchingTimeMC},
  the function
    $\max_{\preceq_{i}}$  takes a set of prioritized
   ordinals (and possibly $(\NoGood,\dotsc,\NoGood)$) and returns a prioritized
   ordinal such that: the first irrelevant entries (due to priorities
   smaller than that of $u_{i}$) are set to $0$; and the
   rest is the maximum (with the lexicographic order, the latter the
   more significant) among the corresponding suffixes of the   prioritized
  ordinals given as input. In case 
   the input set contains
    $(\NoGood,\dotsc,\NoGood)$, then the output is $(\NoGood,\dotsc,\NoGood)$ too.

   For example, in the setting of
   Example~\ref{ex:prioritizedOrd},
   \begin{displaymath}
    \textstyle\max_{\preceq_{3}}\{(1,2,3), (3,4,1)\}=(0,2,3)
   \end{displaymath}
   where the first element of each sequence is irrelevant.

 The function $\min_{\preceq_{i}}$ is defined similarly, by: truncating the
  first irrelevant elements, choosing the smallest one in the
  lexicographic order, and padding the missing elements with $0$. The
   output is $(\NoGood,\dotsc,\NoGood)$ in case the input set
   contains nothing other than  $(\NoGood,\dotsc,\NoGood)$.
  
  The functions $\max_{\preceq_{i}}$  and
   $\min_{\preceq_{i}}$ can be efficiently implemented: if
  the input is the set of $N$ prioritized ordinals then the time
  complexity is $O(Nk)$.
 \end{mydefinition}

  \begin{mydefinition}[$\PTMPM_{i}$]
   \label{def:ptMPM}
  In Algorithm~\ref{algo:branchingTimeMC},
   the function  $\PTMPM_{i}$ takes a state $x\in X$ and
   %an index $i\in [1,m]$, and
   returns
   \begin{equation}\label{eq:201507082125}
\begin{aligned}
 &\PTMPM_{i}(x):=
 \\
 & \textstyle\min_{\preceq_{i}}
    \bigl\{\,
    \vec{\alpha}\in |X|^{k}
    \,\bigl|\bigr.\,
    \bigl(\,\lambda^{\tuple{j_{1},\dotsc,j_{n}}}\co F\bigl(\ev'(\vec{\alpha})\co
    R\bigr)\co c\,\bigr)(x)=\ttrue
    \,\bigr\}
\end{aligned}   
\end{equation}
   where the composite is from~(\ref{eq:201507082123}) and
   $R\colon X\to \POM_{\alpha}$ is given by the current values of
   $\bigl(R(x,i,j)\bigr)_{x,i,j}$ in the algorithm.

   The complexity of $\PTMPM_{i}$ depends greatly on the choice of a functor
   $F$ and a predicate lifting $\lambda$.    A uniform and brute-force
   algorithm for $\PTMPM_{i}$ is possible, however, by enumerating
   all $\vec{\alpha}\in |X|^{k}$ from the smaller ones with respect to
   $\preceq_{i}$, and checking for each $\vec{\alpha}$ whether the
   condition in~(\ref{eq:201507082125}) is satisfied. The worst-case
   complexity
   is $O(km^{2}|X|^{k+1}+C|X|^{k})$ with some constant $C$, on the assumption that the value
\begin{displaymath}
    \bigl(\,\lambda^{\tuple{j_{1},\dotsc,j_{n}}}\co F(\ev'(\vec{\alpha}))\co
 FR\co c\,\bigr)(x)
 \quad\text{that appear    in~(\ref{eq:201507082125}) }
\end{displaymath}
   is computed in time $O(km^{2}|X| +C)$.
   The  last assumption is derived as follows:
the computation of
   $\ev'(\vec{\alpha})$ is in $O(km)$; hence 
   the computation of
   $F(\ev'(\vec{\alpha}))$ is
    in    $O(km|X|)$; that of
   $\lambda^{\tuple{j_{1},\dotsc,j_{n}}}$ is in $O(m)$ (exploiting Lem.~\ref{lem:yonedaLikeCorForPolyadicModality}); and the other
   components like $c$ and application of $F$ have
   only a constant contribution $C$ to the complexity.
  \end{mydefinition}

\begin{myremark}\label{rem:optimizationOfBTMCAlgo}
    Most $F$ and $\Lambda$ allow much better complexity of $\PTMPM_{i}$.
 For example, the choice  $F=\pow(\AP)\times(\place)$  and
 $\lambda=\mathsf{X}$ (the next-time modality) in
 Example~\ref{ex:lambdaAndGamma}.6 (that will yield a logic like LTL in~\S{}\ref{sec:nondetLinearTime}),
the function
   $\PTMPM_{i}$  picks up the prioritized ordinal $R(x',i)$ of the successor
   $x$ and truncates its first irrelevant elements to $0$. This can be
   done in time $O(k)$. More generally, often it is possible to
   ``propagate backwards'' by computing
   \begin{math}
    \{t\in F(\Omega^{m})\mid \lambda^{\tuple{j_{1},\dotsc,j_{n}}}(t)=\ttrue\}
    \end{math}, for which a \emph{one-step complete} set of deduction rules
    can be used (see e.g.~\cite{CirsteaKP09}). Such optimizations by
     deduction rules are left as future work.
\end{myremark}

     \begin{mytheorem}
     %  [correctness of
     % Algorithm~\ref{algo:branchingTimeMC}]
     \label{thm:correctnessOfAlgoBranchingTimeMC}
     Algorithm~\ref{algo:branchingTimeMC} indeed returns
     $\sem{\varphi}_{c}$.
     \myqed
     \end{mytheorem}
 
 \auxproof{
 (*** The following improvement turns out to be not necessary. ***)
     The ``naive'' algorithm in Algorithm~\ref{algo:branchingTimeMC} has an
     advantage that it mirrors the conditions in Def.~\ref{def:ptMPM}
     and hence its correctness is straightforward. Its complexity is not
     optimal, however, because the number of iterations of the main loop
     (lines 7--32) is bounded only by the number of changes made, that
     is
     $|X|^{km|X|}$ (recall that $R(x,i,j)\le |X|$ for each $x\in X,i\in
     [1,m],j\in [1,k]$). This complexity is  worse than the ones
     commonly appear in the model-checking literature (see e.g.~\cite{Wilke01,CleavelandKS92}), where a
     complexity
     is exponential only in the \emph{alternation depth} of
     a formula $\varphi$. The latter is bounded by $k$ in our
     current setting.

By further inspection of the proof of
Thm.~\ref{thm:correctnessOfProgMeasEqSys} we obtain an improved
algorithm. See Algorithm~\ref{algo:branchingTimeMCImproved}. The key is
to specify more explicitly the order of application of the operations in the main loop
(lines 8--31 of Algorithm~\ref{algo:branchingTimeMC}) to different $x,
i$ and $j$. For this purpose we use a counter ($A$ in Algorithm~\ref{algo:branchingTimeMCImproved}).

}

The following complexity result is derived from an analysis of
Algorithm~\ref{algo:branchingTimeMC}. Recall that it assumes a
brute-force algorithm for $\PTMPM_{i}$ (Def.~\ref{def:ptMPM}); fixing
$F$ and $\Lambda$ will allow further
optimization. See Rem.~\ref{rem:optimizationOfBTMCAlgo}.
 It nevertheless achieves a complexity that is exponential only in $k$.
This is much like the most known complexity results for model-checking
 (see e.g.~\cite{Wilke01,CleavelandKS92})---note that $k$ bounds the
 alternation depth of a formula $\varphi$.

\begin{mytheorem}[complexity]
   \label{thm:complexityBranchingTimeMC}
   In the setting
      of Def.~\ref{def:branchingTimeProgMeas} and
      Assumption~\ref{asm:booleanAndFiniteForModelChecking}, the
   model-checking problem can be decided in time
\begin{equation}\label{eq:complexity}
      O\bigl(\,
       m^{2}(km^{2}|X|+C)|X|^{k+2}(|X|+1)^{k}\,\bigr).
       \tag*{\myqed}
\end{equation}
\end{mytheorem}
\noindent
A straightforward optimization is possible: 
%Algorithm~\ref{algo:branchingTimeMC}
each iteration of the inner loop
(lines 8--32) tests all $(x,i)$; this is unnecessary.
Algorithm~\ref{algo:branchingTimeMC} is presented as it is, however,
since the correspondence to Def.~\ref{def:MPM} is clearer.
 It should be possible also to improve the complexity so that it is
 exponential to the alternation depth, instead of to the number $k$ of
 $\mu$-operators, of the given formula $\varphi$.
 % Details are yet to be
 % worked out, though.

\auxproof{
\section{Small-Model Property and Satisfiability, Trial 2 (the most
promising one)}
{\LARGE After all, satisfiability is much harder a problem than
model-checking. For example,~\cite{CirsteaKP09} needed to devise a
tableau system to go from the former to the latter.}
%  \label{sec:smallModelProperty}
What we have done in~\S{}\ref{sec:branchingTimeModelChecking} can be
easily adapted to satisfiability checking. We focus on the Boolean
setting in which $\Omega=\Bool$ and
$\Gamma=\{\bigwedge_{n},\bigvee_{n}\mid n\in\omega\}$, as in
 Assumption~\ref{asm:booleanAndFiniteForModelChecking}. 
\auxproof{
The difficulty in the quantitative case seems to be that the number of
equations can grow
arbitrarily. Cf.~\cite{AlmagorBK14,NakagawaH14arxiv}. 
}

\begin{mydefinition}[$\CmuGL$-satisfiability equation]
Let $\varphi$ be a $\CmuGL$-formula, and
     consider the simple $\CmuGL$-equational system
     \begin{equation}\label{eq:201507092256}
       u_{1}=_{\eta_{1}}\varphi_{1},\;
  \dotsc,\;
  u_{m}=_{\eta_{m}} \varphi_{m}
     \end{equation}
    that arises as the translation of $\varphi$
 (\S{}\ref{subsec:formulasAsEqSys}). Let $i_{1}<\cdots<i_{k}$ be
 the indices of all the $\mu$-variables, as in
 Def.~\ref{def:preProgMeas}.

 We turn the $\CmuGL$-equational system~(\ref{eq:201507092256}) into
 an equational system (in the sense of Def.~\ref{def:eqSys}) over the
 complete lattice $L=\Bool$, in the following way.  The resulting
 equational system is denoted by $\ESatisPhi$.
 \begin{itemize}
  \item An equation of the form $u_{i}=_{\eta_{i}}u_{i'}$ is left as it
	is.
  \item An equation of the form $u_{i}=_{\nu}
	\boxdot_{\gamma}\bigl(u_{j_{1}},\dotsc,u_{j_{n}}\bigr)$
	is turned into the equation 
	$u_{i}=_{\nu}\sem{\gamma}\bigl(u_{j_{1}},\dotsc,u_{j_{n}}\bigr)$,
	where
	$\sem{\gamma}\colon \Omega^{n}\to \Omega$ is a monotone function
	from Def.~\ref{def:modalSigAndPropSig}. 
  \item An equation of the form $u_{i}=_{\nu}
	\heartsuit_{\lambda}(u_{j_{1}},\dotsc,u_{j_{n}})$
	is turned into
	$u_{i}=_{\nu}
	f_{\lambda}(u_{j_{1}},\dotsc,u_{j_{n}})$, where
	the function $f_{\lambda}\colon \Omega^{n}\to \Omega$ is defined by:
\begin{equation}\label{eq:fLambdaDef}
 	\begin{aligned}
	 &f_{\lambda}(l_{1},\dotsc,l_{n}) =\ttrue
	 \\
	 &\defiff
	 \exists t\in F1.\; (\tilde{\lambda}\co F\tuple{\seq{l}{n}})(t)=\ttrue,
	\end{aligned}
	\end{equation}
	where $\tilde{\lambda}\colon F(\Omega^{n})\to \Omega$ is from
	Lem.~\ref{lem:yonedaLikeCorForPolyadicModality}. Note that
			  \begin{math}
			   1\stackrel{\tuple{\seq{l}{n}}}{\to} \Omega^{n}
		  \end{math}, hence
		  \begin{math}
		   F1\stackrel{F\tuple{\seq{l}{n}}}{\to}F(\Omega^{n})
		  \end{math}.
	Monotonicity of $f_{\lambda}$ follows easily from that of
	$\lambda$.

	(*** ... The definition must be modified to take proper
	care of conjunction. They must take the same $t\in F1$ ***)
 \end{itemize}
\end{mydefinition}

\begin{mytheorem}[correctness of satisfiability equations]
 Let $(
l^{\sol}_{1},
\dotsc,
l^{\sol}_{m}
 )\in \Omega^{m}$
 be the solution of $\ESatisPhi$ in the sense of
 Def.~\ref{def:solOfEqSys}. We have: $l^{\sol}_{m}=\ttrue$  if and only
 if there exists a coalgebra $c\colon X\to FX$ and its state $x\in X$
 such that
 $\sem{\varphi}_{c}(x)=\ttrue$.
\end{mytheorem}
\begin{myproof}
 For the `only if' direction, let $X:=\{u_{1},\dotsc,u_{m}\}/\sim$
 where $\sim$ is the equivalence relation generated by:
 \begin{displaymath}
  u_{i}\sim u_{j} \;\text{if there exists ...}
 \end{displaymath}
\end{myproof}
}

\auxproof{
\section{Small-Model Property and Satisfiability, Trial 1 (one that will
not probably work)} 

 *** No, this shouldn't work! consider $\tuple{a}\top\land\tuple{b}\top$
 over a stream automaton. ***

 Ideas:
 \begin{itemize}
  \item probably entries from $\{0,1,\NoGood\}$ are enough
  \item the state space: $\{u_{1},\dotsc,u_{m}\}$, or possibly
	 $\{u_{1},\dotsc,u_{m}\}\times F1$
  \item \emph{Of course} we \emph{cannot} use an arbitrary satisfying coalgebra
	$c$ as a guidance!!
  \item
       \begin{quote}
	   Note that, in the completeness claim, a satisfiability progress
   measure is not uniform over $c$ and $x$. Indeed there are
   (quantitative)
   logics in
   which the truth value of a formula can take an arbitrary number below $1$, but
   not exactly $1$. See e.g.~\cite{AlmagorBK14,NakagawaH14arxiv}.
       \end{quote}
       Therefore in a quantitative case the satisfiability problem is
       \begin{quote}
	Input: $\varphi$ and $l$

	Output: if there is $c$ and $x$ such that $l\sqsubseteq\sem{\varphi}_{c}(x)$
       \end{quote}
       and not
       \begin{quote}
	Input: $\varphi$

	Output: the supremum of $\sem{\varphi}_{c}(x)$ over all $c$ and $x$
       \end{quote}
 \end{itemize}
 
\section{Small-Model Property and Satisfiability}
%  \label{sec:smallModelProperty}
What we have done in~\S{}\ref{sec:branchingTimeModelChecking} can be
easily adapted to satisfiability checking.

The following is a counterpart of Def.~\ref{def:branchingTimeProgMeas}.
 \begin{mydefinition}[satisfiability progress measure]
  \label{def:satisfiabilityProgMeas}
  Assume the setting of Def.~\ref{def:preProgMeas}. 
  A \emph{$\CmuGL$ satisfiability progress measure} for the formula
  $\varphi$
  is a pre-progress measure $p\in \pPM_{\varphi,\alpha}$ (Def.~\ref{def:preProgMeas}) that is
  subject to the following condition.
\begin{itemize}
       \item[5(c)]
	       \textbf{($\nu$-variables, RHS is a modal formula)} Let
		  $i\in [1,m]\setminus\{i_{1},\dotsc,i_{k}\}$ (i.e.\
		  $u_{i}$ is a $\nu$-variable) and that the
		  corresponding equation $u_{i}=_{\nu}\varphi_{i}$ has
		  as $\varphi_{i}$ a modal formula
		  $\varphi_{i}=\heartsuit_{\lambda}(u_{j_{1}},\dotsc,u_{j_{n}})$.

		  We require that
		  there exists an element $t_{i}\in F1$ for each $i\in [1,m]$ such that, 
		  for each approximant $p_{i}(\alpha_{1},\dotsc,
		  \alpha_{a},\dotsc,\alpha_{k})\in \Omega$, there are ordinals
	      $\beta_{1},\dotsc,\beta_{a-1}$ such that
		 \begin{align*}
		  &
		 p_{i}(\alpha_{1},\dotsc,
		  \alpha_{a},\dotsc,\alpha_{k})\sqsubseteq
		  \\&		  
		\PT_{\heartsuit_{\lambda}(u_{j_{1}},\dotsc,u_{j_{n}})}
		%\PT_{\varphi_{i_{a}}}
		(\beta_{1},\dotsc,\beta_{a-1},
	       \alpha_{a},\dotsc, \alpha_{k})\bigl(\,(Fp)(t)\,\bigr),
		 \end{align*}
		  where $\PT_{\heartsuit_{\lambda}(u_{j_{1}},\dotsc,u_{j_{n}})}$
		  is from~(\ref{eq:branchingTimePredTransf}) and
		  $(Fp)(t)\in F(\pPM_{\varphi,\alpha})$ because
		  \begin{math}
		   		    1\stackrel{p}{\to} \pPM_{\varphi,\alpha}
		  \end{math}, hence
		  \begin{math}
		   		    F1\stackrel{Fp}{\to}F(\pPM_{\varphi,\alpha})
		  \end{math}, and
		  \begin{math}
		    t\in F1
		  \end{math}.
		   % \begin{displaymath}
		   %  1\stackrel{p}{\to} \pPM_{\varphi,\alpha},\;
		   %  F1\stackrel{Fp}{\to}F(\pPM_{\varphi,\alpha})
		   %  \;\text{and}\;
		   %  t\in F1.
		   % \end{displaymath}
   \end{itemize}  
  \end{mydefinition}

   \begin{mytheorem}[correctness of  $\CmuGL$ satisfiability progress measure]
    \label{thm:correctnessOfSatisfiabilityProgMeas}
    Assume the setting of Def.~\ref{def:satisfiabilityProgMeas} (hence
    of Def.~\ref{def:preProgMeas}).

       \begin{enumerate}
  \item       \textbf{(Soundness)}
	      If $p\in \pPMpa$ is a satisfiability progress measure,
	      then there exists a coalgebra $c\colon X\to FX$ and its
	      state $x\in X$ such that
	             \begin{math}
	p_{m}(\alpha,\dotsc,\alpha)
	\sqsubseteq\sem{\varphi}_{c}(x)
       \end{math}.
	\item \textbf{(Completeness)}
	      For any coalgebra $c\colon X\to FX$ and its state $x\in
	      X$, there exists an optimal satisfiability progress measure
	      $p\in \pPMpa$ (over some $\alpha$) such that
	      $	p_{m}(\alpha,\dotsc,\alpha)
	=\sem{\varphi}_{c}(x)
	      $.
    \end{enumerate}
   \end{mytheorem}
   \noindent
   Note that, in the completeness claim, a satisfiability progress
   measure is not uniform over $c$ and $x$. Indeed there are
   (quantitative)
   logics in
   which the truth value of a formula can take an arbitrary number below $1$, but
   not exactly $1$. See e.g.~\cite{AlmagorBK14,NakagawaH14arxiv}.
    \begin{myproof}
     For soundness, we take define a coalgebra $c\colon 1\to F1$---whose
     carrier is a singleton $1=\{*\}$---by $c(*):=t$, and let $p\colon
     1\to \pPMpa$ be an MC progress measure
     (Def.~\ref{def:branchingTimeProgMeas}). It is straightforward 
     that $p$ is indeed an MC progress measure. We have
     $p_{m}(\alpha,\dotsc,\alpha)\sqsubseteq\sem{\varphi}_{c}(*)$ by
     Thm.~\ref{thm:correctnessOfBranchingTimeProgMeas}.1.
     
    Consider the simple $\CmuGL$-equational system
     \begin{equation}\label{eq:201507092256}
       u_{1}=_{\eta_{1}}\varphi_{1},\;
  \dotsc,\;
  u_{m}=_{\eta_{m}} \varphi_{m}
     \end{equation}
    that arises as the translation of $\varphi$
    (\S{}\ref{subsec:formulasAsEqSys}). 
    ...
    \end{myproof}
  
  In this section we establish the small-model property of
  $\CmuGL$---that a satisfiable formula $\varphi$ has a model whose size
  is suitably bounded---and present
  an algorithm for satisfiability
  check. Continuing~\S{}\ref{subsec:branchingTimeAlgo} we fix the domain
  of truth values to be $\Omega=\Bool$; and restrict propositional
  connectives to $\bigwedge$ and $\bigvee$. This is much like in
 Assumption~\ref{asm:booleanAndFiniteForModelChecking}. Our developments
  will be based on the notion of matrix progress measure (MPM) from
  Def.~\ref{def:MPM}.

   \begin{mydefinition}[the collapsing function $\collapse$]
   Assume the setting of Def.~\ref{def:preProgMeas}.
   We define a function $\collapse\colon \POM_{\omega}\to \POM_{mk}$---that \emph{collapses} a prioritized ordinal matrix
    (Def.~\ref{def:prioritizedOrdMatrix}) with arbitrary natural numbers
    as its entries into one with entries from $[0,mk]\amalg\{\NoGood\}$---as
    follows. Given a POM $\bigl(\alpha^{(i)}_{j}\bigr)_{i\in [1,m],j\in
    [1,k]}$, the $(i,j)$-entry $\widetilde{\alpha}^{(i)}_{j}$ of
    the POM $\collapse\bigl(\alpha^{(i)}_{j}\bigr)_{i,j}$ is defined as
    follows.
    \begin{itemize}
     \item $\widetilde{\alpha}^{(i)}_{j}:=\NoGood$ if
	   $\alpha^{(i)}_{j}=\NoGood$.	  
     \item $\widetilde{\alpha}^{(i)}_{j}:=0$ if
	   $\alpha^{(i)}_{j}=0$.	  
     \item % For each $j\in [1,k]$, consider the $j$-th column
	   % \begin{math}
	   %  (
	   %  \alpha^{(1)}_{j},\dotsc,
	   %  \alpha^{(k)}_{j}
	   %  )^{\top}
	   % \end{math}
	   % of the input POM.
	   For the other entries,
	   we define
	   $\widetilde{\alpha}^{(i)}_{j}:=N$, for $N\in [1,mk]$, if
	   $\alpha^{(i)}_{j}$ is the $N$-th smallest number among the set
	    \begin{displaymath}
	     \bigl\{\,
	     \alpha^{(i)}_{j}
	    %  ,\dotsc,
	     % \alpha^{(k)}_{j}
	     \,	     \bigl|\bigr.\,
	     i\in [1,m], j\in [1,k]
	     \,\bigr\}\setminus \{0,\NoGood\}.
	    \end{displaymath}
	   Here we do not count duplicates.
    \end{itemize}
   \end{mydefinition} 
 For example:
     \begin{displaymath}
      \collapse
    \left[
    \begin{array}{ccc}
      3 & 6
      & 
      2
      \\
     8& 1& 1
      \\
      \NoGood &      \NoGood& \NoGood
       \\
     8&0&5
    \end{array}
      \right]
      =
          \left[
    \begin{array}{ccc}
      3 & 5
      & 
      2
      \\
     6& 1& 1
      \\
      \NoGood &      \NoGood& \NoGood
       \\
     6&0&4
    \end{array}
      \right].
     \end{displaymath}
     Note also that the size of the set $\POM_{mk}$ is bounded by
     $(mk+2)^{mk}$---each entry is an element of $[0,mk]\amalg\{\NoGood\}$.

 \begin{mytheorem}[small model property of
  $\CmuGL$]
     Assume the setting of Def.~\ref{def:preProgMeas}, and assume
  further that $\Omega=\Bool$ and that
  $\Gamma=\{\bigwedge_{n},\bigvee_{n}\mid n\in\omega\}$. If there exists
  a coalgebra $c\colon X\to FX$ and $x\in X$ such that
  $\sem{\varphi}_{c}(x)=\ttrue$, then there exists one such that $|X|\le
  (mk+2)^{mk}$.
 \end{mytheorem}
   \begin{myproof}
    Let us consider the following (monotone and decreasing) function
    $\Psi\colon \pow(\POM_{mk})\to \pow(\POM_{mk})$. For $Y\subseteq
    \POM_{mk}$, 
   \begin{align*}
    &\Psi(Y)
    :=
    \\
    &
    \left\{
    P\in Y
    \,\left|\,
\begin{minipage}{.38\textwidth}
     $P$ satisfies Cond.~2, 3, 4, 5(a) and 5(b) of Def.~\ref{def:MPM}
    (where we put $P$ in place of $R(x)$), and, considering Cond.~5(c), there exists $t\in FY$
 such that
 \newline
 $\bigl(\lambda^{\tuple{\seq{j}{n}}}\co
 F(\ev'(P^{(i)}))\bigr)(t)=\ttrue$.
\end{minipage}    
\right.\right\}.
   \end{align*}
   Here $P^{(i)}$ denotes the $i$-th row of a POM $P$; note that Cond.~4
    is vacuous. Monotonicity of $\Psi$ is obvious because $Y\supseteq
    Y'\neq \emptyset$ implies $FY\supseteq FY'$ (a well-known property
    of a $\Sets$-functor, see e.g.~\cite{AdamekGT10}).

   Now we repeatedly apply $\Psi$ to the set $\POM_{mk}$; since
    $\POM_{mk}$
    is finite the chain
    \begin{equation}\label{eq:201507091941}
     \POM_{mk}\supseteq \Psi(\POM_{mk})\supseteq
    \Psi^{2}(\POM_{mk})\supseteq\cdots
    \end{equation}    
    eventually stabilizes. Let its limit be denoted by $Y_{0}$.
    By the definition of $\Psi$, for each $P\in Y_{0}$ we can find $t\in
    FY_{0}$ such that
     $\bigl(\lambda^{\tuple{\seq{j}{n}}}\co
 F(\ev'(P^{(i)}))\bigr)(t)=\ttrue$. We define a function  $d\colon
    Y_{0}\to FY_{0}$ by assigning such $t$ to each $P\in Y_{0}$.
    It is straightforward that the inclusion map $\iota\colon
  \widetilde{X}\hookrightarrow\POM_{m}$ constitutes an MPM for $\varphi$
  (Def.~\ref{def:MPM}). Therefore by Thm.~\ref{thm:correctnessMPM}.1
    (soundness of MPMs), we have that
    \begin{equation}\label{eq:201507091934}
     P^{(m)}_{k}\neq \NoGood \;\text{for some $P\in Y_{0}$}
      \;\Longrightarrow\; \varphi \text{ is satisfiable.}
    \end{equation}

    We shall now prove the converse direction
    of~(\ref{eq:201507091934}). Assume $\varphi$ is satisfiable, that
    is, there exists a coalgebra $c\colon X\to FX$ and $x\in X$ such
    that $\sem{\varphi}_{c}(x)=\ttrue$.
    By Thm.~\ref{thm:correctnessMPM}.2 (completeness of MPMs), there exists
  an MPM $R\colon X\to \POM_{|X|}$ such that
  $\bigl(R(x)\bigr)^{(m)}_{k}\neq \NoGood$. Let $P:=\collapse(R(x))\in
    \POM_{mk}$. We shall prove that $P\in Y_{0}$, that is,
    $P$ survives along the chain~(\ref{eq:201507091941}).
    
    It is straightforward that $P$ satisfies Cond.~2, 3, 4, 5(a) and
    5(b)
    of Def.~\ref{def:MPM}
    (where we put $P$ in place of $R(x)$)---the collapsing function
    $\collapse$ preserves the order ($\le$) between different entries of
    $P$,
    hence preserves the order $\preceq_{i}$ between different rows.
    Now it suffices to show the following claim:
    \begin{quote}
     Let
\begin{displaymath}
     Y_{1}:=\bigl\{\,\collapse(R(x'))\,\bigl|\bigr.\,x'\in X\,\bigr\},
\end{displaymath}
     where $X$ is the carrier of the coalgebra that witnesses
     satisfiability of $\varphi$. Then for each $P'\in Y_{1}$, there
     exists $t'\in FY_{1}$ such that
     $\bigl(\lambda^{\tuple{\seq{j}{n}}}\co
 F(\ev'(P'^{(i)}))\bigr)(t')=\ttrue$.
    \end{quote}

================================    
  By Thm.~\ref{thm:correctnessMPM}.2 (completeness of MPM), there exists
  an MPM $R\colon X\to \POM_{|X|}$ such that
  $\bigl(R(x)\bigr)^{m}_{k}\neq \NoGood$. Let us consider the composite
  $\widetilde{R}:= \collapse\co R\colon X\to \POM_{m}$, and define
  \begin{displaymath}
   \widetilde{X}:=\{\widetilde{R}(x')\mid x'\in X\}\;\subseteq \POM_{m}.
  \end{displaymath}
  We shall equip a coalgebraic structure $\widetilde{c}$ to $\widetilde{X}$ and claim that its
  state
  $\widetilde{x}:=\widetilde{R}(x)$ satisfies
  $\sem{\varphi}_{\widetilde{c}}(\widetilde{x})=\ttrue$.

  To define $\widetilde{c}\colon
  \widetilde{X}\to F\widetilde{X}$, let $P\in \widetilde{X}$ and choose
  $x'\in X$ such that $\widetilde{R}(x')=P$. By the definition of
  $\widetilde{X}$, such $x'$ necessarily exists. Then we define
  \begin{displaymath}
   \widetilde{c}(P):=
   \bigl(\,X\stackrel{c}{\to}
   FX\stackrel{F\widetilde{R}}{\longrightarrow}
   F\widetilde{X}\,\bigr)(x')
   \;\in F\widetilde{X};
  \end{displaymath}
  note here that $\widetilde{R}\colon X\to \POM_{m}$ factorizes as $X\to
  \widetilde{X}\hookrightarrow \POM_{m}$ by the definition of
  $\widetilde{X}$ as an image.

  We shall now prove that the inclusion map $\iota\colon
  \widetilde{X}\hookrightarrow\POM_{m}$ constitutes an MPM for $\varphi$
  (Def.~\ref{def:MPM}). The crucial point is that, by the definition,
  the order between different rows---the pointwise extension of $\le$,
  hence $\preceq_{i}$ for any $i$---is preserved by $\collapse$, including whether some entries are
  $0$ or $\NoGood$. This immediately
  derives Cond.~2, 3, 5(a) and~5(b) in Def.~\ref{def:MPM}. Cond.~4 is trivial.

  It remains to show that $\iota$ satisfies Cond.~5(c) in Def.~\ref{def:MPM}.
    ...
   \end{myproof}
}

\section{Coalgebraic $\mu$-Calculus $\Cmu_{\Gamma,\Lambda}$
as a Nondeterministic Linear-Time Logic}
\label{sec:nondetLinearTime}
In this section we adapt the previous results to the setting where we
think of $\CmuGL$ as a (nondeterministic) \emph{linear-time} logic, that is, where a system
in question exhibits nondeterministic branching over transitions of
type $F$. Such a system  is represented
 as a
function $c\colon X\to \pow FX$. 

Our main results here are: 1) categorical characterization of 
the truth value of a linear-time logic formula using progress measures
(Thm.~\ref{thm:correctnessOfNondetExistProgMeasEqSys});
2) a ``smallness'' result that cuts down the search spaces for
linear-time model checking (Thm.~\ref{thm:smallExistentialProgMeas});
and 3) a decision procedure (Thm.~\ref{thm:decidabilityOfLinTimeCmuGLModelChecking}) that depends on the smallness result.
\subsection{Coalgebraic Preliminaries}
\label{subsec:coalgPrelimKleisli}
In what follows we will be dealing  with coalgebras of the type
$c\colon X\to \pow FX$, where $F\colon \Sets\to\Sets$
(that is like in~\S\ref{subsec:coalgPrelim}) is understood as
the type of \emph{linear-time behaviors}, and $\pow$ is the powerset
monad.

This is a common setting taken in the coalgebraic studies of \emph{trace
semantics}. The use of a monad $T$ in a coalgebra $c\colon X\to TFX$
with ``$T$-branching over $F$-linear time behaviors'' originates
in~\cite{PowerT99}, and is subsequently adopted e.g.\
in~\cite{Cirstea11,Jacobs04c,HasuoJS07b,UrabeH15CALCOtoAppear,KerstanK13}.\footnote{Another
common coalgebraic formalization of linear-time semantics is via
\emph{determinization}, and uses Eilenberg-Moore categories (as opposed
to Kleisli) as base categories. See
e.g.~\cite{Bartels04,Jacobs0S15}. 
Adapting the current model-checking
framework to this Eilenberg-Moore approach
seems hard: fixed-point specifications are usually interpreted
over \emph{infinitary} traces such as infinite words; and 
this makes determinization, 
the core of the Eilenberg-Moore approach, much more complicated (like B\"uchi word
automata become Rabin automata, see e.g.~\cite{Vardi95}). }  The
formalization in the current paper most closely follows that
in~\cite{UrabeH15CALCOtoAppear}. We shall again present minimal
preliminaries to this \emph{Kleisli approach} to coalgebraic trace
semantics. See e.g.~\cite{UrabeH15CALCOtoAppear,HasuoJS07b} for further
details; for monads and Kleisli categories see~\cite{MacLane71}.

A \emph{monad} $T$ on $\Sets$ is an endofunctor equipped with natural
transformations $\eta^{T}\colon \id\Rightarrow T$ (\emph{unit}) and $\mu^{T}\colon T\co
T\Rightarrow T$ (\emph{multiplication}) that are subject to certain
``monoid'' commutative diagrams. In our current example of the powerset
monad $\pow$, its unit $\eta^{\pow}$ is the singleton map and its
multiplication $\mu^{\pow}$ is given by union. In the class of examples
of $T$ that are
relevant to us, the unit turns an element into ``a branching with a
unique choice''; and the multiplication ``suppresses'' two
transitions into one (see~\cite{HasuoJS07b}).

The \emph{Kleisli category} $\Kleisli{T}$ has sets as its objects, and
an arrow $X\relto Y$ in $\Kleisli{T}$ is given by a function $X\to TY$. 
It becomes a category using the monad structure of $T$. For example,
given two successive arrows $f\colon X\relto Y$ and $g\colon Y\relto Z$
in $\Kleisli{T}$, its composition $g\Kco f\colon X\relto Z$ is given by
the composite
\begin{math}
 X\stackrel{f}{\to} TY\stackrel{Tg}{\to} T(TZ) \stackrel{\mu_{Z}}{\to } TZ
\end{math}
of functions. It is also easy to see that we have the so-called 
\emph{Kleisli inclusion} functor $J\colon\Sets\to\Kleisli{T}$ by
$JX=X$ and $Jf=\eta^{T}\co f$.

Note that we used the symbols $\relto$ and $\Kco$ (as
opposed to $\to$ and $\co$) for constructs in $\Kleisli{T}$, for
distinction. In what follows we stick to this convention.

Note that for our example of $T=\pow$, the Kleisli category
$\Kleisli{\pow}$
is nothing but the category $\Rel$ of sets and binary relations. We will
however stick to $\Kleisli{\pow}$, hoping that the theory will be
transported to other monads (such as the Giry monad on $\Meas$, for
probabilistic branching).

The following is our current notion of system model.
For technical reasons, we impose certain conditions on $F$. These
conditions  are
common ones and imposed also
in~\cite{Jacobs04c,HasuoJS07b,KerstanK13}.
\begin{mydefinition}[nondeterministic $F$-coalgebra]
\label{def:nondetCoalgebra}
 Let $F\colon \Sets\to\Sets$ be a functor, such that the following hold.
 \begin{enumerate}
  \item\label{item:functorWithFinalCoalg}
%       The functor $F$ is finitary, and hence
       A final
  coalgebra
	$\zeta\colon Z\iso FZ$ exists in $\Sets$.
  \item\label{item:functorWithDistrLaw}
   The functor $F$ comes with a
  distributive law $\xi\colon F\pow\Rightarrow\pow F$ over the powerset
  monad $\pow$ (which, as is well-known~\cite{Jacobs04c}, induces a lifting $\oF\colon
       \Kleisli{\pow}\to\Kleisli{\pow}$ of $F$).
  % \item \label{item:functorPresIntersection}
  % 	The functor $F$ preserves monos.
  % 	, and moreover the
  % intersection of a totally ordered family of subsets. That is, if
  % $\{U_{i}\subset X\}_{i\in I}$ is totally ordered then
  % $F(\bigcap_{i}U_{i})= \bigcap_{i}FU_{i}$.
 \end{enumerate}

 A
 \emph{nondeterministic $F$-coalgebra} is  $c\colon X\to
 \pow FX$ in $\Sets$, that is, an arrow $c\colon
 X\relto \oF X$ in the Kleisli category $\Kleisli{\pow}$.
\end{mydefinition}
Examples of such functors are
polynomial functors inductively generated by
\begin{displaymath}
 F,F_{i}\;::=\;\id \mid A\mid F_{1}\times F_{2}\mid\textstyle
 \coprod_{i\in I}F_{i}
% \quad\text{}
\end{displaymath}
where $A$ is a constant functor that takes any set to $A\in\Sets$.
See
 e.g.~\cite{HasuoJS07b,UrabeH15CALCOtoAppear} for further details on
 Cond.~\ref{item:functorWithFinalCoalg}--\ref{item:functorWithDistrLaw}.
 % ;
 % and~\cite{AdamekGT10} for
 % Cond.~\ref{item:functorPresIntersection}.
 % We will need
 % Cond.~\ref{item:functorWithFinalCoalg}--\ref{item:functorWithDistrLaw}
 % for a coalgebraic characterization of infinite traces; they are 
 % assumptions commonly taken in the coalgebraic literature such
 % as~\cite{Jacobs04c,HasuoJS07b,KerstanK13}.
 % Cond.~\ref{item:functorPresIntersection}
 % will play a crucial role in the proof of our ``smallness'' result
 % (Thm.~\ref{thm:smallExistentialProgMeas}), and it seems that many
 % functors
 % enjoy this property of preservation of (totally-ordered) intersections.

 In view of~\S{}\ref{subsec:coalgPrelim}, in the current setting, we can identify 
 a state $z$
of a final coalgebra $\zeta\colon Z\iso FZ$ with a (possibly infinite,
long-term) \emph{linear-time behavior} of the type $F$. For example, when
$F=\pow(\AP)\times(\place)$ (Example~\ref{ex:lambdaAndGamma}.6), a final coalgebra is carried by the set
$Z=(\pow(\AP))^{\omega}$ of infinite streams of subsets of $\AP$. Such streams
are commonly called \emph{computations} in the context of model
checking.
%, and (linear-time) modal formulas are interpreted over computations.

  %  \begin{wrapfigure}[3]{r}{0pt}
  %   \begin{math}%\label{eq:maximalTrace}       
  % \xymatrix@C+1em@R=.8em{
  %  {\oF X}
  %                \ar@{~>}[r]^{\oF (\tr(c))}|-*\dir{|}
  % &
  %  {\oF Z}
  % \\
  %  {X}
  %                \rar[u]^{c}
  %                \ar@{~>}[r]_{\tr(c)}|-*\dir{|}
  % &
  %  {Z}
  %                \rar[u]_{J\zeta}^{\cong}
  % }
  %   \end{math}
  %  \end{wrapfigure}

  % For characterization of the set of all the (linear-time) $F$-behaviors
  % exhibited by a nondeterministic $F$-coalgebra, we use the result
  % in~\cite{Jacobs04c} that: a final coalgebra $\zeta\colon Z\iso FZ$,
  % when lifted from $\Sets$ to the Kleisli category $\Kleisli{\pow}$,
  % admits the \emph{maximal} homomorphism from an arbitrary
  % nondeterministic $F$-coalgebra.
  
The following result~\cite{Jacobs04c} allows us to characterize, in categorical terms, the set of possible
(linear-time) $F$-behaviors of a nondeterministic
$F$-coalgebra.\footnote{In papers like~\cite{HasuoJS07b} coalgebraic
\emph{finite} trace semantics is studied. Here ``finite'' means
linear-time behaviors that eventually come to halt  within finitely many
steps; and the set of finite $F$-behaviors is identified with the
carrier of an \emph{initial $F$-algebra} in $\Sets$ (as opposed to a final
$F$-coalgebra).} The same holds in a probabilistic setting, too; see e.g.~\cite{UrabeH15CALCOtoAppear}.
\begin{myproposition}[coalgebraic infinitary\footnote{Note that
 ``infinitary''  does not mean that a behavior is necessarily of an infinite length. For
 example, if $F=\{\checkmark\}+A\times (\place)$, a final $F$-coalgebra
 is carried by the set $Z=A^{*}+A^{\omega}$ of all words over $A$ of
 finite or infinite length. All words (finite or infinite) are deemed to be ``infinitary''
 traces. } trace semantics~\cite{Jacobs04c}]
 \label{prop:coalgebraicInfinitaryTrace}
 Let $F\colon \Sets\to\Sets$ be a functor that satisfies the conditions
 in Def.~\ref{def:nondetCoalgebra}; and $c\colon X\to \pow FX$ be a
 nondeterministic $F$-coalgebra. 
 Consider the diagram
  \begin{equation}\label{eq:nondetCoalgTraceSem}
       \vcenter{
  \xymatrix@C+1em@R=.8em{
   {\oF X}
                 \ar@{->}[r]^{\oF f}|-*\dir{|}
  &
   {\oF Z}
  \\
   {X}
                 \rar[u]^{c}
                 \ar@{->}[r]_{f}|-*\dir{|}
  &
   {Z}
                 \rar[u]_{J\zeta}^{\cong}
  }
   }
   \qquad\text{in $\Kleisli{\pow}$;}
  \end{equation}
 then:
% \begin{enumerate}
%  \item there exists at least one function $f\colon X\to \pow Z$ that
%  makes
%  the diagram commute; and
%  \item among such $f$, there exists the greatest
%  one with respect to (the pointwise extension of) the inclusion order in
%  $\pow Z$. 
% \end{enumerate}
 1) there exists at least one function $f\colon X\to \pow Z$ that
 makes
 the diagram commute; and 2) among such $f$, there exists the greatest
 one with respect to (the pointwise extension of) the inclusion order in
 $\pow Z$.
 The greatest one shall be denoted by $\tr(c)\colon X\to \pow
 Z$ and called the \emph{(infinitary) trace semantics} of $c$. Moreover,
 an element $z\in \tr(c)(x)$---identified with a single linear-time
 behavior over time---is referred to as an \emph{infinitary trace} of
 $c$ from $x$.
\end{myproposition}
\noindent
We note that the definition of $\tr(c)$ in
Prop.~\ref{prop:coalgebraicInfinitaryTrace} amounts to the following:
$\tr(c)$ is the greatest fixed point of the monotone function
\begin{equation}\label{eq:monotoneFuncForWhichNondetTraceIsGreatestFixedPoint}
 \Psi\colon \Kleisli{\pow}(X,Z)\to\Kleisli{\pow}(X,Z),
 \quad
 f\mapsto (J\zeta)^{-1}\odot \oF f\odot c
\end{equation}
where $\odot$ denotes composition of arrows in $\Kleisli{\pow}$.

 It has been observed that, for many examples of the functor
  $F$, the greatest homomorphism $\tr(c)$ in Prop.~\ref{prop:coalgebraicInfinitaryTrace} indeed captures the set of all
  possible linear-time behaviors. See e.g.~\cite{Cirstea11}
  and~\cite[Appendix~A.2]{UrabeH15CALCOtoAppear}.

\subsection{$\CmuGL$ as a Linear-Time Logic}
We take a modal language $\CmuGL$ whose modal signature $\Lambda$ is
over $F$. Hence a $\CmuGL$-formula $\varphi$ specifies a property of
$F$-behaviors, where the latter are identified with  elements $z\in Z$
of a final coalgebra $\zeta\colon Z\iso FZ$. See \S{}\ref{subsec:coalgPrelim}.
\begin{mydefinition}[semantics of the logic $\CmuGL$ over nondeterministic
 $F$-coalgebra]
 \label{def:CmuFmlSemNondet}
 Let $\varphi$ be a closed $\CmuGL$-formula, and $c\colon X\to \pow FX$
 be a nondeterministic $F$-coalgebra. The \emph{denotation} of $\varphi$
 over $c$ is given by a function
  \begin{math}
   \sem{\varphi}_{c}\colon X\to \pow(\Omega)
  \end{math}
 defined by
  \begin{displaymath}
   \sem{\varphi}_{c}
   :=\bigl(\,
  X\stackrel{\tr(c)}{\longrelto} Z
   \stackrel{J(\sem{\varphi}_{\zeta})}{\longrelto} \Omega
   \,\bigr)
  \end{displaymath}
  where: $\tr(c)$ is the infinitary trace semantics of $c$
 (Prop.~\ref{prop:coalgebraicInfinitaryTrace});
  $\sem{\varphi}_{\zeta}$ is the denotation of $\varphi$ over the
 (proper) $F$-coalgebra $\zeta\colon Z\to FZ$ defined in
 Def.~\ref{def:CmuFmlSem}; and $J\colon \Sets\to\Kleisli{\pow}$ is the
 Kleisli inclusion functor (\S{}\ref{subsec:coalgPrelimKleisli}).
 \end{mydefinition}

Given a nondeterministic $F$-coalgebra $c$ and its state
$x$,  a typical question is
whether some (or all) of its linear-time behaviors satisfy
a formula $\varphi$. This problem is the \emph{existential} (or
\emph{universal}) model-checking problem, respectively.
In the current paper we focus on existential model checking.
% and this problem is
% what we wish to answer using progress measure-like
% characterizations. The latter are much more amenable to algorithmic
% searches than working directly with Def.~\ref{def:CmuFmlSemNondet}.

  \begin{myexample}
   Take the combination of $F,\Omega,\Gamma$ and $\Lambda$ in
   Example~\ref{ex:lambdaAndGamma}.6. A Kripke structure
 can then be thought of
  as a nondeterministic $F$-coalgebra.\footnote{A Kripke structure is
  most naturally modeled by a function $c'\colon X\to \pow (\AP)\times \pow
  X$. This  gives rise to a function $c\colon X\to \pow\bigl(
  \,\pow(\AP)\times X\,\bigr)$ in an obvious way that turns state-labels
  into transition-labels, namely
   $c(x)=\{((\pi_{1}\co c')(x), x')\mid x'\in (\pi_{2}\co c')(x)\}$.}
   Recall that a final coalgebra is carried by the set
   $(\pow(\AP))^{\omega}$ of computations;
  in this case the infinitary trace semantics $\tr(c)\colon X\to
  \pow\bigl((\pow(\AP))^{\omega}\bigr)$ is precisely the map that
  carries each state $x\in X$ to the set of computations that arise from
   the paths from $x$.

   A $\CmuGL$-formula $\varphi$ is interpreted over elements of a
   final coalgebra, i.e.\ computations. Overall,
   Def.~\ref{def:CmuFmlSemNondet} in this setting yields the set of
   truth values that $\varphi$ can take, ranging over all the possible
   computations $z\in \tr(c)(x)$
   that start from the given state $x\in X$.
  \end{myexample}

\auxproof{
(*** Already introduced***)
It turns out that the two variants (universal vs.\ existential) call for
different semantical frameworks. However they share the following notion
in common. Its definition follows the pattern of that of progress
measure (Def.~\ref{def:progressMeasForEqSys}), but misses some conditions
in Def.~\ref{def:progressMeasForEqSys}.

  \begin{mydefinition}[pre-progress measure, PPM]
   %\label{def:preProgMeas}
  Let $F\colon \Sets\to\Sets$ be a functor as described in
  Def.~\ref{def:nondetCoalgebra}.  Let $\varphi$ be a
  $\CmuGL$-formula---where $\Lambda$ is a modal signature over
  $F$---that is identified with a simple equational system
  \begin{math}
     u_{1}=_{\eta_{1}}\varphi_{1},
  \dotsc,
  u_{m}=_{\eta_{m}} \varphi_{m}
  \end{math}
  as
  in~\S{}\ref{subsec:formulasAsEqSys}.
   Let $i_{1}<\cdots
 <i_{k}$
 enumerate the indices of all the $\mu$-variables.

  A \emph{pre-progress measure (PPM)} $p$ for $\varphi$
  is given by a tuple
 \begin{displaymath}
  p\;=\;
  \bigl(\,
  (
  \overline{\alpha_{1}},\dotsc,
  \overline{\alpha_{k}}),
  \,
  \bigl(\,\approximant_{i}(\alpha_{1},\dotsc,\alpha_{k})\,\bigr)_{i,\seq{\alpha}{k}}
\,\bigr)
 \end{displaymath}
that
consists of:
 \begin{itemize}
  \item the \emph{maximum prioritized ordinal}
 $(\overline{\alpha_{1}},\dotsc, \overline{\alpha_{k}})$; and 
  \item the \emph{approximants} $\approximant_{i}(\alpha_{1},\dotsc,\alpha_{k})\in \Omega$, defined for
	each
	$i\in[1,m]$ and each
	prioritized ordinal
 $(\alpha_{1},\dotsc,\alpha_{k})$
 such that
	$
	\alpha_{1}\le\overline{\alpha_{1}},\dotsc,
	\alpha_{k}\le\overline{\alpha_{k}}
	$. 
% Such a $k$-tuple $(\alpha_{1},\dotsc,\alpha_{k})$
% 	of ordinals is called a \emph{progress measure}.
 \end{itemize}
 The approximants $\approximant_{i}(\alpha_{1},\dotsc,\alpha_{k})$
 are subject to:
 	\begin{enumerate}
	 \item \label{item:preprogressMeasDefMonotonicity}
	      \textbf{(Monotonicity)}
	       Let $i\in[1,m]$ (hence $u_{i}$ is either a $\mu$- or
	       $\nu$-variable). Then 
	       \begin{math}
		(\alpha_{1},\dotsc,\alpha_{k})
		\preceq_{i}
		(\alpha'_{1},\dotsc,\alpha'_{k})
	       \end{math}
	       implies
	       	       \begin{math}
		\approximant_{i}(\alpha_{1},\dotsc,\alpha_{k})
		\sqsubseteq
		\approximant_{i}(\alpha'_{1},\dotsc,\alpha'_{k})
	       \end{math}.
	 \item\label{item:preprogressMeasDefMuVarBaseCase}
	      \textbf{($\mu$-variables, base case)}
	       Let $a\in [1,k]$. Then $\alpha_{a}=0$ implies
%      $\approximant_{i_{a}}(\alpha_{1},\dotsc,
%      \overset{\underparen{a}}{0},\dotsc,\alpha_{k})$
	       $\approximant_{i_{a}}(\alpha_{1},\dotsc,
		  \alpha_{a},\dotsc,\alpha_{k})=\bot$.
	 \item\label{item:preprogressMeasDefMuVarStepCase}
	   \textbf{($\mu$-variables, step case)}
	   	      Let $a\in [1,k]$, and let 
	       $(\alpha_{1},\dotsc,
	       \alpha_{a}+1,\dotsc,\alpha_{k})$ be
	       a prioritized ordinal
	       such that its $a$-th counter
	       $\alpha_{a}+1$ is a successor ordinal. Then, regarding
	       the approximant $\approximant_{i_{a}} (\alpha_{1},\dotsc,
	   \alpha_{a}+1,\dotsc,\alpha_{k})$:
	   \begin{enumerate}
	    \item \textbf{(RHS is a variable)} If the formula
		  $\varphi_{i_{a}}$ on the right-hand side of the
		  $i_{a}$-th equation $u_{i_{a}}=_{\mu}\varphi_{i_{a}}$
		  is a variable $u_{i'}$ (for some $i'\in [1,m]$), then
		  there exist ordinals
		  		  $\beta_{1},\dotsc,\beta_{a-1}$ such that
		  \begin{multline*}
		  \approximant_{i_{a}} (\alpha_{1},\dotsc,
		   \alpha_{a}+1,\dotsc,\alpha_{k})
		   \\
		   \sqsubseteq
		  \approximant_{i'} (\beta_{1},\dotsc,\beta_{a-1},
		   \alpha_{a},\dotsc,\alpha_{k})
		  \end{multline*}
		  and
		  $\beta_{1}\le\overline{\alpha_{1}},\dotsc,
		  \beta_{a-1}\le\overline{\alpha_{a-1}}$.
	    \item \textbf{(RHS is a propositional formula)} If the
		  formula $\varphi_{i_{a}}$ is a propositional formula
		  $\boxdot_{\gamma}\bigl(u_{j_{1}},\dotsc,u_{j_{n}}\bigr)$,
		  then
		  there exist ordinals
		  		  $\beta_{1},\dotsc,\beta_{a-1}$ such that
		  \begin{align*}
		  &\approximant_{i_{a}} (\alpha_{1},\dotsc,
		   \alpha_{a}+1,\dotsc,\alpha_{k})
		   \\		
		   &\sqsubseteq
		   \sem{\gamma}
		\left(\,
		\begin{array}{c}
		\approximant_{j_{1}} (\beta_{1},\dotsc,\beta_{a-1},
		 \alpha_{a},\dotsc,\alpha_{k}),
		 \\
		 \dotsc,
		  \\
		\approximant_{j_{n}} (\beta_{1},\dotsc,\beta_{a-1},
		\alpha_{a},\dotsc,\alpha_{k})
		\end{array}
		\,\right)
		  \end{align*}
		  and
		  $\beta_{1}\le\overline{\alpha_{1}},\dotsc,
		  \beta_{a-1}\le\overline{\alpha_{a-1}}$.
	   \end{enumerate}

	 \item\label{item:preprogressMeasDefMuVarLimitCase}
	      \textbf{($\mu$-variables, limit case)}
	       Let $a\in [1,k]$, and
	       let
	       $(\alpha_{1},\dotsc,
	      %\alpha_{a},\dotsc,
	      \alpha_{k})$
	       be a prioritized ordinals
	       such that its $a$-th counter
	       $\alpha_{a}$
	      is a limit ordinal. Then
	      % , regarding the approximant
	      %  $\approximant_{i_{a}} (\alpha_{1},\dotsc,
	      % \alpha_{a},\dotsc,\alpha_{k})$,
	      we have
	       \begin{displaymath}
		\approximant_{i_{a}} (\alpha_{1},\dotsc,
		\alpha_{a},\dotsc,\alpha_{k})
		% \\
		% &\qquad\sqsubseteq\;
		\sqsubseteq\textstyle
		\bigsqcup_{\beta<\alpha_{a}}
				\approximant_{i_{a}} (\alpha_{1},\dotsc,
		\beta,\dotsc,\alpha_{k})\enspace.
	       \end{displaymath}
	      	 \item\label{item:preprogressMeasDefNuVar}
	      \textbf{($\nu$-variables)}
	       Let $i\in [1,m]\setminus\{i_{1},\dotsc, i_{k}\}$ (i.e.\
	       $u_{i}$ is a $\nu$-variable in the
	       system~(\ref{eq:sysOfEq})); let $a\in [1,k]$ such that
	       \begin{displaymath}
		i_{1}<\cdots <i_{a-1}<i<i_{a}<\cdots < i_{k}.
	       \end{displaymath}
	       Let $(\alpha_{1},\dotsc,\alpha_{k})$ be a prioritized ordinal.
	       Then, regarding the approximant
		      $\approximant_{i} (\alpha_{1},\dotsc,\alpha_{k})$:
	   \begin{enumerate}
	    \item \textbf{(RHS is a variable)} If the formula
		  $\varphi_{i}$ on the right-hand side of the
		  $i$-th equation $u_{i}=_{\nu}\varphi_{i}$
		  is a variable $u_{i'}$ (for some $i'\in [1,m]$), then
		  there exist ordinals
		  		  $\beta_{1},\dotsc,\beta_{a-1}$ such that
		  \begin{multline*}
		  \approximant_{i_{a}} (\alpha_{1},\dotsc,
		   \alpha_{a},\dotsc,\alpha_{k})
		   \\
		   \sqsubseteq
		  \approximant_{i'} (\beta_{1},\dotsc,\beta_{a-1},
		   \alpha_{a},\dotsc,\alpha_{k})
		  \end{multline*}
		  and
		  $\beta_{1}\le\overline{\alpha_{1}},\dotsc,
		  \beta_{a-1}\le\overline{\alpha_{a-1}}$.
	    \item \textbf{(RHS is a propositional formula)} If the
		  formula $\varphi_{i}$ is a propositional formula
		  $\boxdot_{\gamma}\bigl(u_{j_{1}},\dotsc,u_{j_{n}}\bigr)$,
		  then
		  there exist ordinals
		  		  $\beta_{1},\dotsc,\beta_{a-1}$ such that
		  \begin{align*}
		  &\approximant_{i} (\alpha_{1},\dotsc,
		   \alpha_{a},\dotsc,\alpha_{k})
		   \\		
		   &\sqsubseteq
		   \sem{\gamma}
		\left(\,
		\begin{array}{c}
		\approximant_{j_{1}} (\beta_{1},\dotsc,\beta_{a-1},
		 \alpha_{a},\dotsc,\alpha_{k}),
		 \\
		 \dotsc,
		  \\
		\approximant_{j_{n}} (\beta_{1},\dotsc,\beta_{a-1},
		\alpha_{a},\dotsc,\alpha_{k})
		\end{array}
		\,\right)
		  \end{align*}
		  and
		  $\beta_{1}\le\overline{\alpha_{1}},\dotsc,
		  \beta_{a-1}\le\overline{\alpha_{a-1}}$.
	   \end{enumerate}

	\end{enumerate}
  Let $\alpha$ be an ordinal. 
  The collection of all pre-progress measures for a formula $\varphi$,
  whose maximum prioritized ordinal $(
	\overline{\alpha_{1}},\dotsc,
	\overline{\alpha_{k}})
  $ satisfies $\overline{\alpha_{i}}= \alpha$ for each $i\in [1,k]$
  shall be
  denoted by $\pPM_{\varphi,\alpha}$.
  \end{mydefinition}
 \noindent
Recall that $\Omega$ is the complete lattice of truth values. In the
definition of $\pPM_{\varphi,\alpha}$, the explicit bound by $\alpha$ is
there so that the collection $\pPM_{\varphi,\alpha}$ is a (small) set.
Comparing the previous definition with Def.~\ref{def:progressMeasForEqSys}
of progress measures, what are missing here are the treatment
of modal formulas $\heartsuit_{\lambda}(u_{j_{1}},\dotsc,u_{j_{n}})$
in Cond.~\ref{item:preprogressMeasDefMuVarStepCase}
and~\ref{item:preprogressMeasDefNuVar}. These missing cases are
precisely when we need to use the dynamics of a coalgebra; they will be
suitably filled in
later
in~\S{}\ref{subsec:nondetLinearTimeExistential}--\ref{subsec:nondetLinearTimeUniversal}, 
differently for existential and universal model checking.
}

\subsection{(Existential) Linear-Time Model-Checking for $\CmuGL$}
\label{subsec:nondetLinearTimeExistential}
We shall follow essentially the same path as
in~\S{}\ref{subsec:branchingTimeProgressMeasure}. We shall use precisely
the same notion of pre-progress measure
(Def.~\ref{def:preProgMeas}).  The additional compatibility condition
with the dynamic structure of the system in question is 
different reflecting the difference between the 
systems in question ($X\to FX$ in $\Sets$, or $X\relto FX$ in $\Kleisli{\pow}$).

The following is a counterpart of Def.~\ref{def:branchingTimeProgMeas};
LT is for \emph{linear-time}. 
\begin{mydefinition}[LTMC progress measure]\label{def:nondetExistProgMeas}
 Let $\varphi$ be a $\CmuGL$-formula, identified with 
a simple $\CmuGL$-equational system 
  \begin{math}\label{eq:CmuGLEqSysLinearTimeNondet}
     u_{1}=_{\eta_{1}}\varphi_{1},\;
  \dotsc,\;
  u_{m}=_{\eta_{m}} \varphi_{m}.
  \end{math}
 Let $c\colon X\to \pow FX$ be a nondeterministic $F$-coalgebra (with
 some conditions on $F$; see Def.~\ref{def:nondetCoalgebra}).
 An \emph{LTMC progress measure} for $\varphi$ over $c$ is given by a tuple
 $(\alpha,Y\stackrel{q}{\to} FY,r,s)$ of:
 \begin{itemize}
  \item some ordinal $\alpha$,
  % \item a subset $Q\subseteq X\times \pPMpa$ (cf.\ Def.~\ref{def:preProgMeas}), and 
  \item an $F$-coalgebra  $q\colon Y\to FY$, and 
  \item functions $r\colon Y\to \pPMpa$ and $s\colon Y\to X$
 \end{itemize}
 that are subject to the following condition.
 Let $y\in Y$.
 \begin{enumerate}

  \item[5(c)]
	   \textbf{($\nu$-variables,  RHS is a modal formula)}
	     In the setting of
	     Cond.~\ref{item:preprogressMeasDefNuVar} of
	     Def.~\ref{def:preProgMeas}, 
	     % Let $a\in [1,k]$, and let 
	     %   $(\alpha_{1},\dotsc,
	     %   \alpha_{a}+1,\dotsc,\alpha_{k})$ be
	     %   a prioritized ordinal
	     %   such that its $a$-th counter
	     % $\alpha_{a}+1$ is a successor ordinal.
	     assume further that the formula $\varphi_{i}$ is a modal formula:
	     $\varphi_{i}=\heartsuit_{\lambda}(u_{j_{1}},\dotsc,u_{j_{n}})$.

	     Consider
	     the approximant $p_{i}(\alpha_{1},\dotsc,
	      \alpha_{a},\dotsc,\alpha_{k})\in \Omega$ of $p:=r(y)$.  There
	      must exist ordinals $\beta_{1},\dotsc,\beta_{a-1}$ such that
\begin{equation}\label{eq:201507051642}
 	       \begin{aligned}
		&
		p_{i_{a}}(\alpha_{1},\dotsc,
		\alpha_{a},\dotsc,\alpha_{k})
		\sqsubseteq
		\\
		&
		\PT_{\heartsuit_{\lambda}(u_{j_{1}},\dotsc,u_{j_{n}})}
		%\PT_{\varphi_{i_{a}}}
		(\beta_{1},\dotsc,\beta_{a-1},
		\alpha_{a},\dotsc, \alpha_{k})
		\bigl((Fr\co q)(y)\bigr),
	       \end{aligned}
\end{equation}	     
and $\beta_{1}\le\alpha,\dotsc,
	      \beta_{a-1}\le\alpha$.
	      % Note the similarity between~(\ref{eq:201507051642}) and~(\ref{eq:201507051641}).
	      
	           \item[6.] %\label{item:existentialProgMeasCompatibilityWithC}
			    \textbf{(Compatibility with $c$)} For each
			    $y\in Y$ we have $(Fs\co
			    q)(y)\in c(x)$. That is
			     diagrammatically:
	      \begin{equation}\label{eq:existentialProgMeasBwdSim}
	       \vcenter{\xymatrix@R=.6em{
	       {\oF Y}
	       \rar[r]^-{\oF Js}
	       &
	       {\oF X}
	       \\
	       {Y}
	       \rar[u]^{Jq}
	       \rar[r]_-{Js}
	       \ar@{}[ur]|{\subseteq}
	       &
	       {X}
	       \rar[u]_{c}
	       }}
	       \quad\text{in $\Kleisli{\pow}$,}
	      \end{equation}
			     where $Jq\colon Y\to \pow FY$ is given by
			     $(Jq)(y)=\{q(y)\}$
			     (\S{}\ref{subsec:coalgPrelimKleisli}). 

 \end{enumerate}
\end{mydefinition}
% We note that the two additional conditions in the previous definition
% follows the same pattern as in Def.~\ref{def:progressMeasForEqSys}: an
% approximant must be bounded by a suitable approximation of the
% right-hand
% side of the corresponding equation. In the current setting,
% nondeterminism in a coalgebra $c$ is adding complexity.

% Correctness is formulated as in the following way that is suited for
% existential linear-time model-checking.
 \begin{mytheorem}[correctness of LTMC progress measure]
  \label{thm:correctnessOfNondetExistProgMeasEqSys}
  Assume the setting of Def.~\ref{def:nondetExistProgMeas}.
  In
  particular, the formula $\varphi$ is translated to an equational
  system with $m$ variables.
 \begin{enumerate}
  \item\label{item:soundnessNondetExistProgressMeas}
       \textbf{(Soundness)}
       Let
       $(\alpha,Y\stackrel{q}{\to} FY,r,s)$
be an LTMC
       progress measure. Let $y\in Y$ be an arbitrary state, $x:=s(y)$ (a state of the
       coalgebra $c$) and $p:=r(y)$ (a
       pre-progress measure). Then there exists an
       infinitary trace $z\in \tr(c)(x)$ of $x$
       % (cf.\
       % Prop.~\ref{prop:coalgebraicInfinitaryTrace})
       such that
       $p_{m}(\alpha,\dotsc,\alpha)\sqsubseteq
       \sem{\varphi}_{\zeta}(z)$. Here $\sem{\varphi}_{\zeta}\colon Z\to \Omega$
       is from Def.~\ref{def:CmuFmlSem}.
 \item\label{item:completenessNondetExistProgressMeas} 
      \textbf{(Completeness)} Let $x\in X$, and 
       $z\in \tr(c)(x)$    be an
      infinitary trace  from $x$.  
      There is  an LTMC progress measure
$(\alpha,Y\stackrel{q}{\to} FY,r,s)$
      and some $y\in Y$ such that $s(y)=x$, $\beh(q)(y)=z$
      and 
       $p_{m}(\alpha,\dotsc,\alpha)
      =\sem{\varphi}_{\zeta}(z)$ where $p:=r(y)$.
      Here
      $\beh(q)$ is the behavior map induced by
      finality~(\ref{eq:behq}).  \myqed
      % $q_{m}(\alpha,\dotsc,\alpha)\colon X\relto
      % \Omega$ given by
      % \begin{displaymath}
      %  % X\stackrel{q_{m}(\alpha,\dotsc,\alpha)}{\relto} \Omega
      %  % \;\text{given by}\;
      %  q_{m}(\alpha,\dotsc,\alpha)(x)=\bigl\{\,p_{m}(\alpha,\dotsc,\alpha)\,\bigl|\bigr.\,
      %  p\in q(x)\,\bigr\}
      % \end{displaymath}
      % coincides with $\sem{\varphi}_{c}$ in
      % Def.~\ref{def:CmuFmlSemNondet}.
 \end{enumerate}
 \end{mytheorem}

   The completeness result in the last theorem is not totally
   satisfactory, especially from an algorithmic point of view.
   The  question is the size of  an
   LTMC progress measure: in the proof we used $Y\subseteq X\times Z$, but
   this can be very large---$Z$ is an uncountable set for most common
    functors $F$. 
  Fortunately we have the
   following theorem
   that cuts down the set $Y$ from $X\times Z$ to
   $X\times \pPMpa$ (that is potentially much smaller, especially when
   $\Omega=\Bool$).
   \auxproof{The result thus
   opens up a way to a generic coalgebraic
   model-checking algorithm, and  is one of our main technical contributions.}
   
    \begin{mytheorem}[small LTMC progress measure]
     \label{thm:smallExistentialProgMeas}
      Assume the setting of Def.~\ref{def:nondetExistProgMeas}, and let
    $x\in X$. For any infinitary trace $z\in
    \tr(c)(x)$,
     %    such that $l\sqsubseteq \sem{\varphi}_{\zeta}(z)$, then
     there exists
     an LTMC progress measure
 $(\alpha,Y\stackrel{q}{\to} FY,r,s)$ and some $y\in Y$ such that:
     $s(y)=x$, and $p_{m}(\alpha,\dotsc,\alpha)=\sem{\varphi}_{\zeta}(z)$ where
     $p:=r(y)$. Moreover
 $(\alpha,Y\stackrel{q}{\to} FY,r,s)$ can be chosen so that:
 $Y\subseteq X\times \pPMpa$;
     and $r=\pi_{2}$ and $s=\pi_{1}$. 
     \myqed
    \end{mytheorem}
    \noindent Our proof of the last theorem comes in a \emph{pumping}
    flavor. In it, since the relevant set is possibly infinite, we resort to
    Zorn's lemma. 
    % It also shares an intuition with the ``lasso'' proof
    % for the well-known result that: if an $\omega$-regular language $L$
    % is nonempty, then it necessarily contains an ultimately periodic
    % (infinite) word.

\auxproof{
Failure of positionality is already in the simple setting with a Kripke
structure and LTL. Consider a Kripke structure with three states $x_{0},
x_{1}, x_{2}$ with
 $x_{0}\to x_{1}$,
 $x_{0}\to x_{2}$,
 $x_{1}\to x_{0}$,
and $x_{2}\to x_{0}$, and $x_{1}\models P$, $x_{2}\models Q$. Consider
an LTL specification $\mathsf{F}P\land \mathsf{F}Q$. Then a successful
scheduler must not be positional!

}

    \subsection{Decision Procedure}
    \label{subsec:linearTimeAlgo}
    We exploit  the previous results and derive a
    decision procedure for linear-time $\CmuGL$-model checking.
    We make the following assumption; its justification is discussed shortly.
      \begin{myassumption}%[small model property of $\CmuGL$]
       \label{asm:smallModel} In what follows we assume that the
       satisfiability problem of $\CmuGL$ (against $F$-coalgebras) is decidable.

       Moreover we
       assume the \emph{small model property}: for each satisfiable
       $\CmuGL$-formula $\varphi$, we can compute a natural number
       $N_{\varphi}\in \omega$ such that there exists an $F$-coalgebra
       that satisfies $\varphi$ the size of whose state space is no
       greater than $N_{\varphi}$. That is: there exists a coalgebra
       $\varepsilon\colon E\to FE$, its state $e\in E$ and an MC
       progress measure $Q\colon E\to \pPM_{\varphi,\alpha}$
       (Def.~\ref{def:branchingTimeProgMeas}) such that
       $Q(e)(\alpha,\dotsc,\alpha)=\ttrue$ and $|E|\le N_{\varphi}$.  It
       is moreover guaranteed by
       Thm.~\ref{thm:correctnessOfBranchingTimeProgMeas}.2 that we can
       take $\alpha:=N_{\varphi}$.

       Finally, we assume that $F$ preserves finiteness, that is, $FB$
       is finite if $B$ is finite.
      \end{myassumption}

 Assumption~\ref{asm:smallModel} is a mild one. For example,~\cite{CirsteaKP09} shows that
    the assumption holds when the logic $\CmuGL$ comes with a one-step complete, contraction-closed 
    and exponentially-tractable set of deductive
    rules. These conditions hold in well-known modal
    logics, including (the fixed-point extensions of) the normal modal logic
    $\mathsf{K}$,  and monotone modal logic
    (Example~\ref{ex:lambdaAndGamma}).
    \auxproof{Graded and coalition frames do not satisfy preservation of
    finiteness!}
    Of more relevance here is the fact that the assumption holds for
    (coalgebras of) polynomial functors $F$ (with suitable finiteness requirements), which are the ones typically used to specify linear time behavior; modalities and deductive rules for such functors can be modularly derived from their structure, using an approach similar to that of \cite{CirsteaP07}, and proving the tractability of the set of rules is straightforward in this case.

 It also seems that the
    framework in~\S{}\ref{sec:branchingTimeModelChecking} can be adapted
    to satisfiability check and hence to the small-model property. Due
    to lack of space we do not do so in the current paper and just
    assume the small model property.

    %  The last assumption seems to
    % hold in most settings. For example, \cite{CirsteaKP09} shows that
    % the assumption holds when the logic $\CmuGL$ comes with a complete
    % and ``tractable'' (in the complexity sense) set of deductive
    % rules. The latter condition is a mild one and well-known modal
    % logics like the (fixed-point extensions of) the normal modal logic
    % $\mathsf{K}$, the coalition logic, the monotone modal logic, the
    % graded logic and the probabilistic logic satisfy it.
    % In fact it seems that the
    % framework in~\S{}\ref{sec:branchingTimeModelChecking} can be adapted
    % to satisfiability check and also to the small-model property. Due
    % to lack of space we do not do so in the current paper and just
    % impose the assumption.
    
     \begin{mytheorem}[linear-time $\CmuGL$-model checking is decidable]
      \label{thm:decidabilityOfLinTimeCmuGLModelChecking}
      Assume the setting of Def.~\ref{def:nondetExistProgMeas}. Assume
      further  that: $\Omega=\Bool$; and
      $X$ is a finite set. Then it is decidable whether there exists an
      infinitary trace $z\in \tr(c)(x)$ such that
      $\sem{\varphi}_{\zeta}(z)=\ttrue$. \myqed
     \end{mytheorem}

\auxproof{

\subsection{Universal Model-Checking}
\label{subsec:nondetLinearTimeUniversal}

===================

\begin{itemize}
 \item Universal model checking vs.\ existential model checking
 \item Must use different order. 
\end{itemize}

It is crucial that, for universal and existential model-checking,  we
use two different preorders $\SmythLeq$ and $\HoareLeq$, respectively. 
\begin{mydefinition}[the Smyth $\SmythLeq$ and Hoare $\HoareLeq$ preorders] 
 We define preorders
$\SmythLeq$ and
$\HoareLeq$ over $\pow(\Omega^{m})$, as follows. 
% Given  $\alpha,\alpha'\in \pow(\Omega^{m})$.
\begin{equation}\label{eq:smythHoareDefChar}
 \begin{aligned}
  \alpha\SmythLeq\alpha'\;
  &
  \defiff\;
  \forall V'\in\alpha'.\, \exists V\in \alpha.\, V\sqsubseteq V'
  \\
  &\Longleftrightarrow\;
  \upcl\alpha\supseteq\alpha'
  \;\Longleftrightarrow\;
  \upcl\alpha\supseteq\upcl\alpha';
  \\
  \alpha\HoareLeq\alpha'\;
  &
  \defiff\;
  \forall V\in\alpha.\, \exists V'\in \alpha'.\, V\sqsubseteq V'
  \\
  &\Longleftrightarrow\;
  \alpha\subseteq\dwcl\alpha'
  \;\Longleftrightarrow\;
 \dwcl\alpha\subseteq\dwcl\alpha'.
\end{aligned}
\end{equation}
Here $\upcl \alpha:=\{V\mid \exists V'\in \alpha.\, V'\sqsubseteq V\}$ 
and $\dwcl \alpha:=\{V\mid \exists V'\in \alpha.\, V\sqsubseteq V'\}$ 
are the \emph{upward} and \emph{downward} closures, respectively.
\end{mydefinition}
\noindent
 The least and greatest elements are $\{\bot\}$ (that is equivalent to
 $\Omega^{m}$) and
$\emptyset$, for $\SmythLeq$; and $\emptyset$ and $\{\top\}$ 
(that is equivalent to
 $\Omega^{m}$) 
for
$\HoareLeq$. 

The preorders $\SmythLeq$ and $\HoareLeq$ are not partial orders---for example, $\{\ffalse, \ttrue\}$ and $\{\ffalse\}$ are
 equivalent with respect to $\SmythLeq$, for $\Omega=\Bool$. 
(This example suggests use of $\SmythLeq$ for \emph{universal} model
 checking, too.)
The following fact, however, allows us to apply the observations
in~\S{}\ref{sec:progressMeasForCL}. It is derived immediately by 
the characterization in~(\ref{eq:smythHoareDefChar}).
\begin{mylemma}\label{lem:smythHoareCompleteLattice}
 Let $\Omega$ be a complete lattice.
 Then the poset
 $\pow(\Omega^{m})/({\sqsubseteq_{\mathsf{S}}}\cap{\sqsupseteq_{\mathsf{S}}})$
 induced by the preorder $\SmythLeq$ 
 is a complete lattice. Indeed, 
 given a family $(\alpha_{i})_{i\in I}$ such that $\alpha_{i}\subseteq \Omega^{m}$, 
 \begin{align*}\textstyle
  \bigsqcap_{\mathsf{S}} 
  \bigl\{\,[\alpha_{i}]\,\bigl|\bigr.\,i\in I\bigr\}
  &=
  \bigl[\,\textstyle\bigcup_{i} (\upcl\alpha_{i})\,\bigr]
  =
  \bigl[\,\textstyle\bigcup_{i} \alpha_{i}\,\bigr]
 \; \text{and}\;
  \\
  \textstyle
  \bigsqcup_{\mathsf{S}} 
  \bigl\{\,[\alpha_{i}]\,\bigl|\bigr.\,i\in I\bigr\}
  &=
  \bigl[\,\textstyle\bigcap_{i} (\upcl\alpha_{i})\,\bigr]
 \end{align*}
 provide the desired infimum and supremum, respectively. 

 The same holds for the Hoare preorder $\HoareLeq$. The infimums and
 supremums are given by
 \begin{align*}\textstyle
  \bigsqcap_{\mathsf{H}} 
  \bigl\{\,[\alpha_{i}]\,\bigl|\bigr.\,i\in I\bigr\}
  &=
  \bigl[\,\textstyle\bigcap_{i} (\dwcl\alpha_{i})\,\bigr]
 \; \text{and}\;
  \\
  \textstyle
  \bigsqcup_{\mathsf{H}} 
  \bigl\{\,[\alpha_{i}]\,\bigl|\bigr.\,i\in I\bigr\}
  &=
  \bigl[\,\textstyle\bigcup_{i} (\dwcl\alpha_{i})\,\bigr]
  =
  \bigl[\,\textstyle\bigcup_{i} \alpha_{i}\,\bigr].
 \end{align*}
\end{mylemma}
For further details on those preorders
see e.g.~\cite{AbramskyJ94} and~\cite{McIverM01}.

In the current setting where $\CmuGL$ is thought of as a linear-time
logic, unlike in~\S{}\ref{sec:coalgMuCal}, it is not the case that 
a formula $\varphi$ (or its equational presentation) and a coalgebra $c$
directly give rise to an equational system over some complete lattice. 
Our characterization of the (linear-time) semantics of $\varphi$ over
$c$,
instead, relies on the following notion.
\begin{mydefinition}[predicate transformer $\PT^{E,c}_{i}$]
 Let $E=(u_{i}=_{\eta_{i}}\varphi_{i})_{i\in[1,m]}$ be a simple $\CmuGL$-equational system, and $c\colon X\to
 \pow FX$ be a nondeterministic $F$-coalgebra. For each $i\in [1,m]$ we
 define the \emph{$i$-th predicate transformer} induced by $E,c$
 \begin{displaymath}
  \PT^{E,c}_{i}\colon\;
 [X,\pow(\Omega^{m})] 
  \longrightarrow
 [X,\pow(\Omega^{m})] 
 \end{displaymath}
 as follows. Let $p\colon X\to \pow(\Omega^{m})$, that is, an arrow
 $p\colon X\relto \Omega^{m}$ in $\Kleisli{\pow}$.

 When $\varphi_{i}$ is a variable $u_{j}$ , 
 \begin{align*}
  \PT^{E,c}_{i}(p)
  \;:=\;\bigl(\,
  X
  \stackrel{p}{\relto}
  \Omega^{m}
  \stackrel{J\pi_{i,j}}{\relto}
  \Omega^{m}
\,\bigr)
 \end{align*}
 where $J$ is the Kleisli inclusion functor and 
 $\pi_{i,j}\colon \Omega^{m}\to\Omega^{m}$ is
 the function that replaces the $i$-th component of the input
 with its $j$-th, i.e.\ 
 the unique arrow that makes the following diagram commute.
 \begin{displaymath}
  \vcenter{\xymatrix@R=.8em@C+2em{
   &
   {\Omega}
   \\
   {\Omega^{m}}
    \ar[ur]^{\pi_{i'}
    %\text{(for each $i'$ such that $i'\neq i$)}
    }
    \ar@{-->}[r]^{\pi_{i,j}}
    \ar[dr]_{\pi_{j}}
   &
   {\Omega^{m}}
    \ar[u]_{\pi_{i'}}
    \ar[d]^{\pi_{i}}
   \\
   &
   {\Omega}
}}
 \quad\text{where $i'$ ranges $i'\in[1,m]\setminus\{i\}$.}
 \end{displaymath}
 When $\varphi_{i}$ is a propositional formula 
 $\boxdot_{\gamma}(u_{i_{1}},\dotsc, u_{i_{k}})$, 
 \begin{align*}
  \PT^{E,c}_{i}(p)
  \;:=\;\bigl(\,
  X
  \stackrel{p}{\relto}
  \Omega^{m}
  \stackrel{J\gamma_{i,i_{1},\dotsc,i_{k}}}{\relto}
  \Omega^{m}
\,\bigr)
 \end{align*}
 where 
 $\gamma_{i,i_{1},\dotsc,i_{k}}\colon \Omega^{m}\to\Omega^{m}$ is
 the unique arrow that makes the following diagram commute.
 \begin{displaymath}
  \vcenter{\xymatrix@R=.8em@C+2em{
   &
   {\Omega}
   \\
   {\Omega^{m}}
    \ar[ur]^{\pi_{i'}
    %\text{(for each $i'$ such that $i'\neq i$)}
    }
    \ar@{-->}[r]_{\gamma_{i,i_{1},\dotsc,i_{k}}}
    \ar[d]_{\tuple{\pi_{i_{1}},\dotsc,\pi_{i_{k}}}}
   &
   {\Omega^{m}}
    \ar[u]_{\pi_{i'}}
    \ar[d]^{\pi_{i}}
   \\
   {\Omega^{k}}
    \ar[r]_{\gamma}
   &
   {\Omega}
}}
 \quad\text{where $i'$ ranges $i'\in[1,m]\setminus\{i\}$.}
 \end{displaymath}
  When $\varphi_{i}$ is a modal formula 
 $\heartsuit_{\lambda}(u_{i_{1}},\dotsc, u_{i_{k}})$, 
 \begin{align*}
  \PT^{E,c}_{i}(p)
  \;:=\;\bigl(\,
  X
  \stackrel{c}{\relto}
  \overline{F}X
  \stackrel{p}{\relto}
  \overline{F}(\Omega^{m})
  \stackrel{J\sigma}{\relto}
  \Omega^{m}
\,\bigr)
 \end{align*}
where $\sigma$ is...

 The equational system $E_{c}$ on the preorder $\pow(\Omega^{m})$,
 induced by a simple $\CmuGL$-equational system $E$ and a
 nondeterministic
 $F$-coalgebra $c\colon X\to \pow FX$. 
 \begin{displaymath}
  X
  \stackrel{c}{\longrelto}
   \overline{F}X
    \stackrel{q}{\longrelto}
     \overline{F}(\Omega^{m})
 \end{displaymath}
\end{mydefinition}
...
\begin{mytheorem}
 Correctness
\end{mytheorem}

\begin{myexample}
 Existential model-checking of ``an infinite path.'' Sometimes one gets
 a useless witness.
\end{myexample}

\section{Coalgebraic $\mu$-Calculus $\Cmu_{\Gamma,\Lambda}$
as a Probabilistic Linear-Time Logic}
\label{sec:probLinearTime}
The \emph{sub-Giry monad} $\Giry$ is an adaptation of the one
from~\cite{Giry82}---with \emph{subdistributions} that can assign to the
whole space a value that is
strictly less than $1$.

\begin{mydefinition}[the sub-Giry monad $\Giry$]
\label{def:subGiry}
Let $\Meas$ denote the (usual) category of measurable sets and
 measurable functions.
The \emph{sub-Giry monad} is the monad $(\Giry,\eta^{\Giry},\mu^{\Giry})$ on $\Meas$ such that
\begin{itemize}
\item $\Giry(X,\sigalg_X)=(\Giry X, \sigalg_{\Giry X})$, where 
%\begin{itemize}
%\item 
the underling set $\Giry X$ is the set of all \emph{subprobability
      measures} on $(X,\sigalg_X)$. The latter means those measures which assign to the whole
      space $X$ a value in the unit interval $[0,1]$.
 \item 
%\item 
The $\sigma$-algebra $\sigalg_{\Giry X}$ on $\Giry X$ is the smallest $\sigma$-algebra such that, for all $S\in\sigalg_X$, 
the function $\text{ev}_S:\Giry X\to[0,1]$ defined by $\text{ev}_S(P)=P(S)$ is measurable.
%\end{itemize}
\item $\Giry f(\nu)(S)=\nu(f^{-1}(S))$ where $f:(X,\sigalg_X)\to(Y,\sigalg_Y)$ is measurable, $\nu\in\Giry X$, and $S\in\sigalg_Y$.
\item $\eta^{\Giry}_{(X,\sigalg_X)(x)}$ is given by the \emph{Dirac measure}: $\eta^{\Giry}_{(X,\sigalg_X)}(x)(S)$ is $1$ if $x\in S$ and $0$ otherwise.
%$\eta^{\Giry}_{(X,\sigalg_X)}(x)(S)=\begin{cases} 1 & (x\in S) \\ 0 & (x\notin S)\end{cases}$ 
%where $x\in X$ and $S\in\sigalg_X$, and
\item $\mu^{\Giry}_{(X,\sigalg_X)}(\Psi)(S)=\int_{\Giry (X,\sigalg_X)} \text{ev}_S \,d\Psi$ where 
$\Psi\in\Giry^2 X$, $S\in\sigalg_X$ and $\text{ev}_S$ is defined as above.
\end{itemize}
\end{mydefinition}

We use a functor $F\colon \Meas\to\Meas$ for designating the type of
linear-time behaviors; on it we impose the following mild
condition.
The same condition has been exploited
in~\cite{Cirstea10,Schubert09}; and it
allows us to work in the realm of standard Borel spaces---a class of
``well-behaved'' measurable spaces that arise from Polish spaces
(see e.g.~\cite{Doob94}).

\begin{mydefinition}[(standard Borel) polynomial functor]\label{def:polyfuncmeas}
A \emph{standard Borel polynomial functor} $F$ on $\Meas$ is defined by the following BNF notation:
\begin{equation*}\textstyle
F\;::=\; \id\,\mid\, (A,\sigalg_A) \,\mid\, F_1\times F_2 \,\mid\, \coprod_{i\in I} F_i\, ,
\end{equation*} 
%Here, $A\in\Meas$.
%Here, $A$ is a countable set.
where
 $I$ is a countable set and
 $(A,\sigalg_A)\in\Meas$  is a \emph{standard Borel
 space}
 (see e.g.~\cite{Doob94}). 
The set $FX$ has an obvious $\sigma$-algebra $\sigalg_{FX}$ associated
 to it,
 namely: 
%We have to specify $\sigma$-algebras.
% For $(X,\sigalg_X)\in\Meas$, $F(X,\sigalg_X)=(FX,\sigalg_{FX})$ is given as follows:
% the underling set $FX$ is given in the same manner as that for the  polynomial functors on $\Sets$.
% The $\sigma$-algebra $\sigalg_{FX}$ is defined inductively as follows.
 $\sigalg_{F_1 X\times F_2 X}$ is the smallest $\sigma$-algebra that contains $A_1\times A_2$ for 
 each $A_1\in\sigalg_{F_1X}$ and $A_2\in\sigalg_{F_2X}$;
and
$\sigalg_{\coprod_{i\in I} F_i}=\{\coprod_{i\in I} A_i\mid A_i\in
 \sigalg_{F_iX}\}$. The action of $F$ on arrows is obvious.

In what follows, a standard Borel polynomial functor is often referred
 to 
 simply as a \emph{polynomial functor}. 

 A (standard Borel) polynomial functor $F$ comes with a canonical distributive
 law
$\lambda\colon F\Giry\Rightarrow\Giry F$ that is defined inductively on
 the construction of $F$~\cite{Cirstea10}. It follows by a standard
 argument that $F$ has a \emph{lifting} $\overline{F}\colon
 \Kleisli{\Giry}
 \to  \Kleisli{\Giry}$, in the sense that $\overline{F}$ and $F$ are
 compatible via the Kleisli inclusion $J$:
\begin{displaymath}
 \vcenter{\xymatrix@R=.8em@C+3em{
  {\Kleisli{\Giry}}
     \ar[r]^{\overline{F}}
 &
  {\Kleisli{\Giry}}
 \\
  {\Meas}
     \ar[r]_{F}
     \ar[u]^{J}
 &
  {\Meas
   \mathrlap{\; .}}
     \ar[u]_{J}
}}
\end{displaymath}
\end{mydefinition}

\begin{mydefinition}[coalgebra with probabilistic branching]
\label{def:coalgebraWithProbBranching}
 Let $F$ be a (standard Borel) polynomial functor. An
 \emph{$F$-coalgebra with probabilistic branching} is an arrow $c\colon X\to
 \Giry FX$ in $\Meas$, that is equivalently, an arrow $c\colon
 X\relto FX$ in the Kleisli category $\Kleisli{\Giry}$.
\end{mydefinition}
Note that we in fact allow \emph{sub}-probabilistic branching due to 
the choice of $\Giry$ to be the sub-Giry monad.

The following result is from~\cite{UrabeH15CALCOtoAppear} and is closely
related to the observations in~\cite{Jacobs04c,Cirstea10,KerstanK13}.
\begin{myproposition}[probabilistic infinite trace semantics $\tr(c)$]
\label{prop:probInfiniteTrace}
  Let $F$ be a (standard Borel) polynomial functor. Then a final
 $F$-coalgebra $\zeta\colon Z\to FZ$ exists in $\Meas$. Furthermore, 
 the lifting $J\zeta\colon Z\relto FZ$---where $J\colon\Meas\to\Kleisli{\Giry}$ is
 the Kleisli inclusion functor---exhibits
 \emph{weak maximal finality}. The latter means: for each
 $F$-coalgebra  $c\colon X\relto FX$ with  probabilistic branching, there exists 
 a coalgebra morphism $f$ from $c$ to $J\zeta$ as in
\begin{displaymath}
 \vcenter{\xymatrix@R=.8em@C+3em{
  {\overline{F}X}
     \rar[r]^{\overline{F}{f}}
 &
  {\overline{F}Z}
 \\
  {X}
     \rar[r]_{f}
     \rar[u]^{c}
 &
  {Z
%   \mathrlap{\; ;}
 }
     \rar[u]_{\zeta}^{\cong}
}}
 \quad\text{in $\Kleisli{\Giry}$;}
\end{displaymath}
 and furthermore, there
 exists a maximum one among such $f$ with respect to the pointwise
 order between Kleisli arrows $X\relto Z$. Such a maximum morphism from
 $c$ to $J\zeta$ shall be denoted by $\tr(c)$.
\end{myproposition}
\noindent
It turns out that for many examples of the functor $F$ the maximal
homomorphism $\tr(c)$ in Prop.~\ref{prop:probInfiniteTrace} captures the usual
notion of \emph{infinite trace}. This is in particular the case for
\emph{probabilistic tree automata}; see~\cite{UrabeH15CALCOtoAppear}
where the correspondence is established in concrete terms.

\begin{mydefinition}
Let $\varphi$ be a $\CmuGL$-formula, identified with 
a simple $\CmuGL$-equational system 
  \begin{equation}\label{eq:CmuGLEqSysLinearTimeProb}
     u_{1}=_{\eta_{1}}\varphi_{1},\quad
  \dotsc,\quad
  u_{m}=_{\eta_{m}} \varphi_{m}.
  \end{equation}
Let $c\colon X\to \Giry FX$ be a probabilistic $F$-coalgebra.
\end{mydefinition}

In what follows we shall think of the logic $\CmuGL$ 
for the functor
$F\colon \Meas\to\Meas$  as a ``linear-time
logic,'' and interpret  it 
over an $F$-coalgebra $c\colon X\relto
\overline{F}X$ with additional probabilistic branching. Roughly:
a \emph{linear-time behavior} is an element $z$ of the carrier $Z$
of a final $F$-coalgebra in $\Meas$; and,  given a
$\CmuGL$-formula $\varphi$ and a state $x$, 
we aim at defining (and computing)
the probability with which a randomly chosen infinite trace $z\in Z$
from $x$ satisfies the formula $\varphi$. 

We assume $\Omega$ to be a complete lattice with a measurable structure
$\sigalg_{\Omega}$, and that any subset of $\Omega$ of the form 
\begin{math}
 \Omega_{> b_{0}}\{b\in \Omega\mid b > b_{0}\}
\end{math},
 $\Omega_{\ge b_{0}}$,
 $\Omega_{< b_{0}}$ or
 $\Omega_{\le b_{0}}$ (defined similarly) are measurable.
\begin{mylemma}
 The semantics
 $(
l^{\sol}_{1},
\dotsc,
l^{\sol}_{m}
)\in (\Omega^{X})^{m}$ are all measurable.  ...
\end{mylemma}
\begin{myproof}
 Not yet. See~\cite{StinchcombeW92}, or references for probabilistic $\mu$-calculus.
\end{myproof}

Next: use downward closed order
}

% trigger a \newpage just before the given reference number - used to
% balance the columns on the last page adjust value as needed - may
% need to be readjusted if the document is modified later
% \IEEEtriggeratref{8} The "triggered" command can be changed if
% desired: \IEEEtriggercmd{\enlargethispage{-5in}}

% references section

% can use a bibliography generated by BibTeX as a .bbl file BibTeX
% documentation can be easily obtained at:
% http://www.ctan.org/tex-archive/biblio/bibtex/contrib/doc/ The
% IEEEtran BibTeX style support page is at:
% % http://www.michaelshell.org/tex/ieeetran/bibtex/
% \bibliographystyle{./bibtex/IEEEtran} argument is your BibTeX string
% definitions and bibliography database(s) \bibliography{ref.bib}
%
% <OR> manually copy in the resultant .bbl file set second argument
% of \begin to the number of references (used to reserve space for the
%   reference number labels box)
%   \begin{thebibliography}{1}

%   \bibitem{IEEEhowto:kopka} H.~Kopka and P.~W. Daly, \emph{A Guide
%     to \LaTeX}, 3rd~ed.\hskip 1em plus 0.5em minus 0.4em\relax
%     Harlow, England: Addison-Wesley, 1999.

%   \end{thebibliography}

\blindAlt{
}
{
\acks
We thank
Kenta Cho,
Tetsuri Moriya,
Shota Nakagawa,
Jurriaan Rot,
and
Natsuki Urabe
for useful discussions.
I.H.\ and S.S.\ are
 supported by Grants-in-Aid No. 24680001 \& 15KT0012,
 JSPS; 
C.C.\ was supported by a Royal Society International Exchanges Grant (IE131642).
}

% We recommend abbrvnat bibliography style.

 \bibliographystyle{abbrvnat}
 \bibliography{../../../macro/texmf/bibtex/bib/myrefs}
%\bibliography{myrefs}

\newpage
 \appendix

\section{In Case of Parity Games:
Correspondence to Jurdzinski's
  Notion}\label{appendix:parityProgressMeasure}
Here, as a sanity check, we shall show that our notion of progress
  measure (Def.~\ref{def:progressMeasForEqSys}) instantiates to
  Jurdzinski's \emph{parity progress measure}~\cite{Jurdzinski00},
  in the special case where an
  equational system is induced by a parity game.

 The following definitions are all standard. See e.g.~\cite{Jurdzinski00}.
\begin{mydefinition}[parity game]\label{def:parityGame}
A \emph{parity game} is a quadruple 
$G=(X_{\even},X_{\odd},E,\pri)$ of: 
a finite set $X_{\even}$ of \emph{the
 player $\even$'s positions}; 
a finite set $X_{\odd}$ of \emph{the
 player $\odd$'s positions}; 
 a \emph{transition relation} $E\subseteq X\times X$ where
 $X:=X_{\even}\cup X_{\odd}$ is the set of all the positions; and 
 a \emph{priority function} $\pri\colon X\to\{1,2,\dotsc,d\}$
 for some $d\in \omega$. The following are additionally assumed,
 mostly for simplicity:
 the sets $X_{\even}$ and $X_{\odd}$ are disjoint; 
 $X$ is nonempty; each position has at least one $E$-successor; 
 $d$ is an even number. Note, however, that whether $x\in X$ is
 $\even$'s position or $\odd$'s is independent from if $\pri(x)\in [1,d]$ is even
 or odd.

 % ;  $\pri(x)\in\omega$ is an even
 % number if $x\in X_{\even}$; and $\pri(x)$ is an odd number if $x\in
 % X_{\odd}$.
 
 A \emph{play} of $G$ is an infinite sequence $x_{0}x_{1}\dotsc$ of
 positions such that $(x_{i},x_{i+1})\in E$ for each $i\in \omega$.
 A play $x_{0}x_{1}\dotsc$ is \emph{winning} for the player $\even$
 if 
\begin{displaymath}
 \sup\bigl\{k\in\{1,2,\dotsc,d\}\,\bigl|\bigr.\, k=\pri(x_{i}) \text{ for infinitely
 many $i\in\omega$}\bigr\}
\end{displaymath}
is an even number.

 A \emph{strategy} $\sigma$ of  the player $\even$ is a function 
 $\sigma\colon X^{*}\times X_{\even}\to X$,
 with the intuition  that 
 $\even$ chooses his move $\sigma(\vec{x}, x)$ depending on the history 
 $\vec{x}$ of the positions already visited, and the current position
 $x\in X_{\even}$.
 A play $x_{0}x_{1}\dotsc$ \emph{conforms to} a strategy $\sigma$ of
 $\even$ if, for each $i\in\omega$ such that $x_{i}\in X_{\even}$ we
 have
 $x_{i+1}=\sigma(x_{0}x_{1}\dotsc x_{i-1},x_{i})$.
 A \emph{winning strategy} for the player $\even$ is a strategy $\sigma$
 such that every play that conforms to $\sigma$ is winning for $\even$.
 Finally, a position $x\in X$ is \emph{winning}
for  $\even$ if there exists a winning strategy for  $\even$.
\end{mydefinition}

The following notion is precisely the one in~\cite{Jurdzinski00}, modulo
some minor modifications that are made for the fit to the current context.  
\begin{mydefinition}[parity progress measure] 
\label{def:parityProgMeasForParityGame}
%\phantom{hogehoge}\qquad
  Let $G$ be a parity game $G=(X_{\even},X_{\odd},E,\pri)$; let
 $X=X_{\even}\amalg X_{\odd}$. A \emph{parity progress
 measure} for $G$ is a function 
 \begin{math}
  q\colon X\to \omega^{d/2}\amalg\{\NoGood\}
 \end{math}, where $\NoGood$ is a fresh symbol,
 such that: 
 \begin{itemize}
  \item The $a$-th component of the tuple $q(x)$ is never bigger
 than the number $n_{2a-1}$ of the positions of the priority $2a-1$. That is,
$\bigl(q(x)\bigr)_{a}\le n_{2a-1}:= |\pri^{-1}(2a-1)|$ (or $q(x)=\NoGood$) for each $a\in
	[1,d/2]$.
  \item If $x\in X_{\even}$, then there exists a successor $y$ of $x$
	such that:
\begin{align*}
 	&q(x)\succ_{\pri(x)} q(y) &&\text{if $\pri(x)$ is odd;}
        \\
 	&q(x)\succeq_{\pri(x)} q(y) &&\text{if $\pri(x)$ is even.}
\end{align*}	
	Here the order
	$\succ_{\pri(x)}$ is the same as in
	Def.~\ref{def:prioritizedOrdinal}, except that $\NoGood$ is assumed
	to be the greatest element.
  \item If $x\in X_{\odd}$, then for any successor $y$ of $x$ we have
\begin{align*}
 	&q(x)\succ_{\pri(x)} q(y) &&\text{if $\pri(x)$ is odd;}
        \\
 	&q(x)\succeq_{\pri(x)} q(y) &&\text{if $\pri(x)$ is even.}
\end{align*}	
       % The order
       % 	$\succ_{\pri(x)}$ is  the same as in
       % 	Def.~\ref{def:prioritizedOrdinal}, except that $\NoGood$ is the greatest.
 \end{itemize}
\end{mydefinition}

 A parity game gives rise to an equational system. The latter is over 
 (a product of) the
 Boolean lattice $\Bool=\{\ttrue,\ffalse\}$---$\ttrue$ means ``$\even$ is winning.''
\begin{mydefinition}[equaltional system $E_{G}$ from a game $G$]\label{def:fromParityGameToEqSys}
 Let $G=(X_{\even},X_{\odd},E,\pri)$ be a parity game. For each priority
 $i\in [1,d]$, let $n_{i}\in \omega$ be the number of positions with a
 priority $i$, that is, $n_{i}=|\{x\in X\mid \pri(x)=i\}|$. Furthermore,
 let us fix an enumeration
\begin{equation}\label{eq:201506211434}
 x_{1,1},\dotsc,x_{1,n_{1}},
\;
 x_{2,1},\dotsc,x_{2,n_{2}},
\;\dotsc,\;
 x_{d,1},\dotsc,x_{d,n_{d}}
\end{equation}
  of all positions of $G$; it is arranged so that $\pri(x_{i,j})=i$.

 The \emph{equational system  $E_{G}$ induced by $G$} is 
 with variables $u_{1},\dotsc,u_{d}$---we have one variable for each priority
 $i\in [1,d]$---where each variable $u_{i}$ takes its value in the complete
 lattice $\Bool^{n_{i}}$.\footnote{Therefore we shall use
 the extension of the theory developed in the above that allows
 different variables $u_{i}$ to take values in different lattices
 $L_{i}$. As mentioned just after Def.~\ref{def:eqSys}, such  extension is easy. }

 Concretely, the system $E_{G}$ is of the form
\begin{displaymath}
 u_{1}=_{\eta_{1}} f_{1}(\vec{u})\;,\;\dotsc,\;
 u_{d}=_{\eta_{d}} f_{d}(\vec{u})
\end{displaymath}
 where 
$\eta_{i}$ and $f_{i}$ are defined as follows.
\begin{itemize}
 \item The fixed-point symbol $\eta_{i}$ is $\nu$ if $i$ is even; it is $\mu$ if $i$ is odd.
 \item The monotone function
 $f_{i}\colon
% \Bool^{|X|}\cong 
       \Bool^{n_{1}}\times\cdots\times\Bool^{n_{d}}\to \Bool^{n_{i}}$
       is defined, for each $j\in [1,n_{i}]$, by:
  \begin{align*}
  & \bigl(\,
   f_{i}(u_{1},\dotsc,u_{d})
   \,\bigr)_{j}
   \;:=\;
   \\
   &\quad
   \begin{cases}
   \;\bigsqcup \bigl\{\pi_{j'}(u_{i'})\,\bigl|\bigr.\,
    (x_{i,j},x_{i',j'})\in E\bigr\}
    &\text{if $x_{i,j}\in X_{\even}$,}
    \\
   \;\bigsqcap \bigl\{\pi_{j'}(u_{i'})\,\bigl|\bigr.\,
    (x_{i,j},x_{i',j'})\in E\bigr\}
    &\text{if $x_{i,j}\in X_{\odd}$.}
   \end{cases}  
\end{align*}
Here $\bigsqcup$ and $\bigsqcap$ denotes a supremum and an infimum, respectively,
       in the complete lattice $\Bool$.

%       If $i$ is even, we let 
%  $\eta_{i}:=\nu$  and a monotone function
%  $f_{i}\colon
% % \Bool^{|X|}\cong 
%        \Bool^{n_{1}}\times\cdots\times\Bool^{n_{d}}\to \Bool^{n_{i}}$
%  is defined as follows. For each $j\in [1,n_{i}]$,
%   \begin{displaymath}
%    \bigl(\,
%    f_{i}(u_{1},\dotsc,u_{d})
%    \,\bigr)_{j}
%    \;:=\;
%    \begin{cases}
%    \bigsqcup \bigl\{\pi_{j'}(u_{i'})\,\bigl|\bigr.\,
%     (x_{i,j},x_{i',j'})\in E\bigr\}
%     &\text{if $x_{i,j}\in X_{\even}$}
%     \\
%    \bigsqcap \bigl\{\pi_{j'}(u_{i'})\,\bigl|\bigr.\,
%     (x_{i,j},x_{i',j'})\in E\bigr\}
%     &\text{if $x_{i,j}\in X_{\odd}$}
%    \end{cases}  
% \end{displaymath}
 % $f_{i,j}(\vec{x}):=\bigsqcup {\{x\mid
 %       (x_{i,j},x)\in E\}} $. 
% Here $\bigsqcup$ denotes a supremum
%        in the complete lattice $\Bool$.
%  \item Dually, 
% if $i$ is odd, we let
%  $\eta_{i}:=\mu$  and a monotone function
%  $f_{i}\colon
% % \Bool^{|X|}\cong 
%        \Bool^{n_{1}}\times\cdots\times\Bool^{n_{d}}\to \Bool^{n_{i}}$
%  is defined as follows. For each $j\in [1,n_{i}]$,
%   \begin{displaymath}
%    \bigl(\,
%    f_{i}(u_{1},\dotsc,u_{d})
%    \,\bigr)_{j}
%    \;:=\;
%   \bigsqcap \bigl\{\pi_{j'}(u_{i'})\,\bigl|\bigr.\,
%        (x_{i,j},x_{i',j'})\in E\bigr\}.
%   \end{displaymath}
%  % $f_{i,j}(\vec{x}):=\bigsqcup {\{x\mid
%  %       (x_{i,j},x)\in E\}} $. 
% Here $\bigsqcup$ denotes an infimum
%        in $\Bool$.
\end{itemize}
\auxproof{ 
(*** Version 1: this is not convenient for the purpose of complexity
 bound ***)
Let $G=(X_{\even},X_{\odd},E,\pri)$ be a parity game 
and 
\begin{equation}
 x_{1,1},\dotsc,x_{1,m_{1}},
\;
 x_{2,1},\dotsc,x_{2,m_{2}},
\;\dotsc,\;
 x_{d,1},\dotsc,x_{d,m_{d}}
\end{equation}
 be an enumeration of all positions of $G$ such that $\pri(x_{i,j})=i$
 (therefore the positions are organized in the increasing order with
 respect to priorities). 

 The \emph{equational system  $E_{G}$ induced by $G$} is one
 over the complete lattice $\Bool=\{\ttrue,\ffalse\}$, defined as follows.
 Its variables are precisely the positions $x_{1,1},\dotsc,x_{d,m_{d}}$
 of $G$,
 and it is of the form
\begin{multline*}
 x_{1,1}=_{\eta_{1,1}} f_{1,1}(\vec{x})
,\dotsc,
x_{1,m_{1}}=_{\eta_{1,m_{1}}} f_{1,m_{1}}(\vec{x}),
\; \dotsc,
\\
 x_{d,1}=_{\eta_{d,1}} f_{d,1}(\vec{x})
,\dotsc,
x_{d,m_{d}}=_{\eta_{d,m_{d}}} f_{d,m_{d}}(\vec{x}),
\end{multline*}
where $\vec{x}$ stands for $x_{1,1},\dotsc,x_{d,m_{d}}$. We define 
$\eta_{i,j}$ and $f_{i,j}$ by:
\begin{itemize}
 \item If $i$ is even, then 
 $\eta_{i,j}=\nu$  and $f_{i,j}(\vec{x}):=\bigsqcup {\{x\mid
       (x_{i,j},x)\in E\}} $. Here $\bigsqcup$ denotes a supremum
       in the complete lattice $\Bool$.
 \item If $i$ is odd, then 
 $\eta_{i,j}=\mu$  and $f_{i,j}(\vec{x}):=\bigsqcap {\{x\mid
       (x_{i,j},x)\in E\}} $. Here, similarly, $\bigsqcap$ is an infimum
       in  $\Bool$.
\end{itemize}
}
 \end{mydefinition}

We are ready to show that our notion of progress measure generalizes 
Jurdzinski's~\cite{Jurdzinski00}. We use the extension in Def.~\ref{def:extendedProgressMeasForEqSys}.
\begin{myproposition}\label{prop:progMeasInParityGameVSProgMeasForEqSys}
   Let $G$ be a parity game; let $E_{G}$ be the equational system that
 arises from $G$ (Def.~\ref{def:fromParityGameToEqSys}). 
 Let $x$ be an arbitrary position of $G$, and assume that 
 $x=x_{i,j}$ in the enumeration~(\ref{eq:201506211434}) (in particular $\pri(x)=i$).
 
Let 
 $q\colon X\to \omega^{d/2}$ be a parity
 progress measure for $G$
 (Def.~\ref{def:parityProgMeasForParityGame}) such that $q(x)\neq
 \NoGood$. Then this $q$
gives rise to an extended progress
 measure $p$
 (in the sense of Def.~\ref{def:extendedProgressMeasForEqSys}) such that:
 regarding the maximum prioritized ordinal we have
 $\overline{\alpha_{a}}\le n_{2a-1}$ for each $a\in [1,d/2]$; and 
 $\bigl(\,p_{i}(\overline{\alpha_{1}},\dotsc,\overline{\alpha_{d/2}})
\,\bigr)_{j}=\ttrue$.
 
Conversely, let $p$ be 
a progress measure
 for $E_{G}$  (in the sense of Def.~\ref{def:progressMeasForEqSys}) 
that satisfies
\begin{math}
  \bigl(\,p_{i}(\overline{\alpha_{1}},\dotsc,\overline{\alpha_{d/2}})
 \,\bigr)_{j}=\ttrue
\end{math}. It
gives rise to 
a parity progress measure $q$ for the parity game $G$ such that 
$q(x)\neq
 \NoGood$.
\end{myproposition}
\begin{myproof}
 In what follows we shall identify an element $p_{i}\in \Bool^{n_{i}}$
 with a subset $p_{i}\subseteq \{x_{i,1},\dotsc,x_{i,n_{i}}\}$, and
 furthermore identify a tuple $(p_{1},\dotsc,p_{d})\in
 \Bool^{n_{1}}\times\cdots\times
 \Bool^{n_{d}}$ with a subset $p\subseteq X$.

 For the former direction, assume that $q\colon X\to \omega^{d/2}$ is 
 such a parity progress measure. We define a progress measure $p$ by:
$\overline{\alpha_{a}}:= n_{2a-1}$ for each $a\in [1,d/2]$, and
 \begin{align*}
  x_{i,j}\in p_{i}(\alpha_{1},\dotsc,\alpha_{d/2})
  \;\defiff\;
  q(x_{i,j})\preceq_{i} 
  (\alpha_{1},\dotsc,\alpha_{d/2}),
 \end{align*}

Checking that thus defined $p$ is indeed an (extended) progress measure is not
 hard. Monotonicity (Cond.~\ref{item:progressMeasDefMonotonicity} of
 Def.~\ref{def:progressMeasForEqSys}) follows from the transitivity of
 $\preceq_{i}$. 
Cond.~\ref{item:progressMeasDefMuVarBaseCase}' (see
 Def.~\ref{def:extendedProgressMeasForEqSys}) is easy, too.
To check
 Cond.~\ref{item:progressMeasDefMuVarStepCase}, let us first consider
 the case when
 $x_{2a-1,j}\in X_{\odd}$.
\begin{equation}\label{eq:201506211744}
  \begin{aligned}
&  x_{2a-1,j}\in 
  p_{2a-1}(\alpha_{1},\dotsc,\alpha_{a}+1,\dotsc,\alpha_{d/2})
\\
&\Longleftrightarrow\;
 q(x_{2a-1,j})\preceq_{2a-1}
  (\alpha_{1},\dotsc,\alpha_{a}+1,\dotsc,\alpha_{d/2})
\\
&\Longrightarrow\;
 \text{for each successor $y$ of $x_{2a-1,j}$,}\;
\\
&\qquad\quad
  q(y)\prec_{2a-1}
  (\alpha_{1},\dotsc,\alpha_{a}+1,\dotsc,\alpha_{d/2})
\\
&\qquad\qquad\qquad\qquad\qquad\qquad\qquad
\text{(by $q(x_{2a-1,j}) \succ_{2a-1}q(y)$)}
\\
&\stackrel{(\dagger)}{\Longrightarrow}\;
 \text{for each successor $y$ of $x_{2a-1,j}$,}\;
\\
&\qquad\quad
  q(y)\preceq_{2a-1}
  (\alpha_{1},\dotsc,\alpha_{a},\dotsc,\alpha_{d/2})
\\
&\Longleftrightarrow\;
 \text{for each successor $y$ of $x_{2a-1,j}$,}\;
\\
&\qquad\quad
  y\in p  (\alpha_{1},\dotsc,\alpha_{a},\dotsc,\alpha_{d/2})
 \quad\text{(by def.\ of $p$)}
\\
&\Longleftrightarrow\;
 x_{2a-1,j}\in f_{2a-1}
\left(
\begin{array}{c}
  p_{1}  (\alpha_{1},\dotsc,\alpha_{a},\dotsc,\alpha_{d/2}),
\\
 \dotsc,
\\
 p_{d}  (\alpha_{1},\dotsc,\alpha_{a},\dotsc,\alpha_{d/2})
\end{array}
\right),
 \end{aligned}
\end{equation}
as required. Here $(\dagger)$ holds since 
$(\alpha_{1},\dotsc,\alpha_{a},\dotsc,\alpha_{d/2})$ is the greatest
 among those which are strictly $\prec_{2a-1}$-smaller than the
 prioritized ordinal
 $(\alpha_{1},\dotsc,\alpha_{a}+1,\dotsc,\alpha_{d/2})$.
 The case when  $x_{2a-1,j}\in X_{\even}$ is similar, replacing the
 three occurrences of ``each'' by ``some'' in~(\ref{eq:201506211744}).
There is no need of showing
 Cond.~\ref{item:progressMeasDefMuVarLimitCase} since every $\alpha_{a}$
 in question is finite.
 Cond.~\ref{item:progressMeasDefNuVar} is shown much like
 in~(\ref{eq:201506211744}). Finally, that $\bigl(\,p_{i}(\overline{\alpha_{1}},\dotsc,\overline{\alpha_{d/2}})
\,\bigr)_{j}=\ttrue$, that is, 
 $x=x_{i,j}\in
 p_{i}(\overline{\alpha_{1}},\dotsc,\overline{\alpha_{d/2}})$, follows 
 immediately by $q(x)\neq\NoGood$.

 For the opposite direction, we first use
 Thm.~\ref{thm:correctnessOfProgMeasEqSys}---soundness, and then
 completeness---to obtain a ``small'' progress measure, that is, one
 such that $\overline{\alpha_{a}}\le \ascCL(L_{2a-1})$ for each
 $a\in[1,d/2]$. (Recall that we are using an immediate generalization of
 Thm.~\ref{thm:correctnessOfProgMeasEqSys} where different variables
 $u_{i}$ are allowed to take values in different lattices $L_{i}$). Now
 $L_{2a-1}=\Bool^{n_{2a-1}}$ yields $\ascCL(L_{2a-1})=n_{2a-1}$: in a
 strictly ascending chain, each subset is at least one element bigger than
 its previous step. This gives us a small progress measure $p$ such that
$\overline{\alpha_{a}}\le n_{2a-1}$. 
 
 This $p$ is used to define a parity progress measure 
 $q\colon X\to \omega^{d/2}$ by:
 \begin{align*}
  q(x)
  :=
   \min\bigl\{\,
    (\seq{\alpha}{d})
   \,\bigl|\bigr.\,
    x\in p(\seq{\alpha}{d})
\,\bigr\},
 \end{align*}
 where the minimum is taken with respect to the lexicographic order 
 $\preceq$ (the latter an element is, the more significant it is), and the minimum of the
 empty set is defined to be $\NoGood$. 
 \auxproof{We still have to check this!}
It is easy to check that the $q$ is indeed what is desired.
 \myqed
\end{myproof}

\begin{mycorollary}\label{cor:parityGameAndEqSys}
 A position $x=x_{i,j}$ of a parity game $G$ is winning for the player $\even$
 if and only if the solution (Def.~\ref{def:solOfEqSys}) of $E_{G}$
 has it that  $l^{\sol}_{i}(j)=\ttrue$.
\end{mycorollary}
\begin{myproof}
 Immediately from
 Prop.~\ref{prop:progMeasInParityGameVSProgMeasForEqSys},
 Thm.~\ref{thm:correctnessOfProgMeasEqSys} and
 the correctness of parity progress measures~\cite[Cor.~7--8]{Jurdzinski00}. 
 \myqed
\end{myproof}

%-----------------------------------------------
% begin auxproof
%-----------------------------------------------
 \auxproof{
 \section{An Improved Algorithm for $\CmuGL$ Model Checking}
    \begin{algorithm}[tbp]
  \caption{Improved  algorithm for  $\CmuGL$ model checking, in the setting
    of Def.~\ref{def:branchingTimeProgMeas} and
    Assumption~\ref{asm:booleanAndFiniteForModelChecking}.
    We use a counter $A=\bigl(A(1),\dotsc,A(k)\bigr)\in [0,|X|]^{k}$.
    % We write
    % $R(x,i,j)$ for $R(x)^{(i)}_{j}$;
    }
  \label{algo:branchingTimeMCImproved}
  \begin{algorithmic}[1]
   \Require the same as in Algorithm~\ref{algo:branchingTimeMC}
   % A  $\CmuGL$-formula $\varphi$ presented as an equational system
  %    \begin{math}
  %    u_{1}=_{\eta_{1}}\varphi_{1},
  % \dotsc,
  % u_{m}=_{\eta_{m}} \varphi_{m}
  %  \end{math} where $u_{i_{1}},\dotsc,u_{i_{k}}$ are $\mu$-variables, and a coalgebra $c\colon X\to FX$
    %----------------------------------------------------
   \Ensure the same as in Algorithm~\ref{algo:branchingTimeMC}
   %$\sem{\varphi}_{c}\in \Bool^{X}$
   %----------------------------------------------------
   \State lines 1--6 of Algorithm~\ref{algo:branchingTimeMC}
      \Comment initialization and Cond.~2
   % \For{each $x\in X$, $i\in[1,m]$ and $j\in [1,k]$}
   % \Comment initialization
   % \State $R(x,i,j):=0$
   % \EndFor
   % \For{each $a\in [1,k]$}
   % \Comment  Cond.~2
   % \State $R(x,i_{a},a):=1$
   % \EndFor

   \State $A(1):=0,\dotsc,A(k):=0$
   \State
   \Comment initialize a counter $A=\bigl(A(1),\dotsc,A(k)\bigr)\in [0,|X|]^{k}$

   \Repeat
   \Comment the main loop, executed for each $A$
   % \If{$(R(x))^{i_{a}}_{a}=(R(x))^{i_{a}}_{a+1}=\dotsc=(R(x))^{i_{a}}_{k}=0$}
   % \State $(R(x))^{i_{a}}_{a}:=1$
   % \Comment Force Cond.~2
   % \EndIf

   \Repeat
   \For{each $x\in X$ and $i\in [1,m]$ }
   \If{$ R(x,i)=A$}
   \Comment use the counter $A$ %, cf.\ Def.~\ref{def:nextPO}
   \If{$u_{i}$ is a $\mu$-variable and $\varphi_{i}=u_{i'}$}
   \Comment Cond.~3
   \State lines 9--10 of Algorithm~\ref{algo:branchingTimeMC}
   \EndIf
   \If{$u_{i}$ is a $\nu$-variable}
      \Comment Cond.~5
   \State lines 13--25 of Algorithm~\ref{algo:branchingTimeMC}
   \EndIf
   \State lines 28--30 of Algorithm~\ref{algo:branchingTimeMC}
   % \Else
   % \State skip
   \EndIf
      \EndFor
   \Until{no change is made}
   
   \State $A:=\nextPO(A)$
   \Comment update $A$; see Def.~\ref{def:nextPO}
   \Until{$A=(|X|,\dotsc,|X|)$}

   \State\Return $\{x\in X\mid
   R(x,m,k)\neq\NoGood\}$
  \end{algorithmic}
   \end{algorithm}
   
   Algorithm~\ref{algo:branchingTimeMCImproved} is one that improves
   Algorithm~\ref{algo:branchingTimeMC}. In it we additionally use the following
function.
 \begin{mydefinition}[$\nextPO$]\label{def:nextPO}
    In Algorithm~\ref{algo:branchingTimeMCImproved},
 the function $\nextPO$ takes a prioritized ordinal
 $(\alpha_{1},\dotsc,\alpha_{k})\in [0,|X|]^{k}$ and returns the
  smallest prioritized ordinal
  $(\alpha'_{1},\dotsc,\alpha'_{k})$
  (with respect to $\preceq$, the
  lexicographic order in Def.~\ref{def:prioritizedOrdinal}) such that
  \begin{displaymath}
   (\alpha_{1},\dotsc,\alpha_{k})     \prec(\alpha'_{1},\dotsc,\alpha'_{k}).
  \end{displaymath}
  We define $\nextPO(|X|,\dotsc,|X|)=(|X|,\dotsc,|X|)$ for convenience.
  For example, when $k=3$,  $\nextPO(2,|X|,0)=(0,0,1)$.
\end{mydefinition}

    \begin{mytheorem}[correctness of
     Algorithm~\ref{algo:branchingTimeMCImproved}]
     \label{thm:correctnessOfAlgoBranchingTimeMCImproved}
     Algorithm~\ref{algo:branchingTimeMCImproved} indeed returns
     $\sem{\varphi}_{c}$.
    \end{mytheorem}
      \begin{myproof}
       By establishing that Algorithms~\ref{algo:branchingTimeMC}
       and~\ref{algo:branchingTimeMCImproved} yield the same output. Let
       $R_{1}(x,i,j)$ denote the value of $R(x,i,j)$ after line 34 of
       Algorithm~\ref{algo:branchingTimeMC}. It is obvious that,
       throughout the execution of
       Algorithm~\ref{algo:branchingTimeMCImproved}, we have
       $R(x,i) \preceq_{i}R_{1}(x,i)$.
       
      The following is not hard to see by induction on $A$. In each iteration of the
      main loop, after
      line 17,
       we have
       \begin{equation}\label{eq:201507091337}
	R(x,i)\preceq A
	\;\Longleftrightarrow\;
	R(x,i)=R_{1}(x,i).
       \end{equation}
       To see it, we proceed in the following way.
       \begin{itemize}
	\item In Algorithm~\ref{algo:branchingTimeMCImproved}, every
	      time we come to line 4,
	      the set $\{R(x,i)\mid x\in X,i\in [1,m]\}$ is classified
	      into:
	      \begin{itemize}
	       \item those which are strictly smaller than $A$ with respect to
		     $\preceq$, in which case $R(x,i)=R_{1}(x,i)$ by
		     the induction hypothesis;
	       \item those which are equal to $A$, and
	       \item those which are strictly greater than $A$.
	      \end{itemize}
	\item 
	      It suffices to show the following. Assume that $x,i$ are
	      so that $R(x,i)$ is
	      in the second
	      category in the above (i.e.\ $R(x,i)=A$). If it is
	      unchanged after execution of lines 5--17, then
	      $R(x,i)=R_{1}(x,i)$.
	\item The last claim is not hard to see, by examining how it is
	      determined whether $R(x,i)$ should be updated or not, in
	      Algorithms~\ref{algo:branchingTimeMC}
	      and~\ref{algo:branchingTimeMCImproved}. Namely, it is
	      determined based on whether the relevant prioritized
	      ordinals are strictly greater than  $R(x,i)$, and it does
	      not matter how greater they are.
       \end{itemize}
       Since the main loop ranges all possible $A$'s (from
       $(0,\dotsc,0)$ to $(|X|,\dotsc,|X|)$),
       by~(\ref{eq:201507091337}),  after line 19
       of Algorithm~\ref{algo:branchingTimeMCImproved} we have
       $R(x,i)=R_{1}(x,i)$ for all $x,i$. This proves the claim. \myqed
      \end{myproof}

 \begin{mytheorem}[complexity]
  \label{thm:complexityBranchingTimeMCImproved}
 In the setting
    of Def.~\ref{def:branchingTimeProgMeas} and
    Assumption~\ref{asm:booleanAndFiniteForModelChecking}, the
 model-checking problem can be decided in time
\begin{equation}
 O\left(\,|X|^{k+2}m
  \left( 
\begin{array}{r}
   k^{3}
  +
 km
 +
 km^{2}(m-k)|X|^{k+1}
 \\
 +C(m-k)|X|^{k}
\end{array}
\right) \,\right),
\end{equation}
 where $C$ is the constant from Def.~\ref{def:ptMPM}.
 \end{mytheorem}
\begin{myproof}
  In Algorithm~\ref{algo:branchingTimeMC} the main loop (lines 4-19)
  executes $|X|^{k}$ times. Now consider the for-loop in lines 6--16.
  In one execution of the for-loop:
 \begin{itemize}
  \item  Lines 8--10 are executed at most $|X|\cdot k$ times and each
	 takes $O(k(k+1))$ times (see
	 Def.~\ref{def:maxPrecEqMinPrecEq}), therefore taking overall
	 time
	 $O(
	 |X|k\cdot k (k+1))=
	 O(k^{3}|X|)$;

  \item Lines 11--13 are executed at most $|X|\cdot (m-k)$ times and
	each takes time $O(km^{2}|X|^{k+1}+C|X|^{k})$
	(see Def.~\ref{def:ptMPM}). Therefore
	lines 11--13 incur time consumption	
	$O(km^{2}(m-k)|X|^{k+2}+C(m-k)|X|^{k+1})$.
	  \item Line 14 takes time $O(km|X|)$.
 \end{itemize}
  Overall, one execution of the for-loop (lines 6--16) takes
  $O(k^{3}|X|+km^{2}(m-k)|X|^{k+2}+C(m-k)|X|^{k+1}+
  km|X|)$ time. This is repeated at most $|X|\cdot m$ times: the
  for-loop in lines 6--16 is repeated until no change is made (lines
  5--17); and as long as it does not terminate, each execution of the
  for-loop makes at least one $R(x,i)$ from $A$ to something strictly
  greater (with respect to $\preceq$). Line 18 takes $O(k)$ time and
  this is overwhelmed by the other parts.
  Combining all these yields the claimed complexity.
  \myqed
 \end{myproof}
}
%-----------------------------------------------
% end auxproof
%-----------------------------------------------
 
\section{Omitted Proofs}
\label{appendix:proofs}

\subsection{Proof of Thm.~\ref{thm:correctnessOfProgMeasEqSys}}
\begin{myproof}
\fbox{\bf  The item~\ref{item:soundnessProgressMeas} (soundness).} We shall prove the
 following claim $(*)$
 by induction on $i$. This obviously suffices, by the definition of $(
l^{\sol}_{1},
\dotsc,
l^{\sol}_{m}
)$ (Def.~\ref{def:solOfEqSys}).
\begin{quote}
 $(*)$  
For each $i\in[1,m]$ the following holds: 
 for each $j\in [1,i]$, and
for each prioritized ordinal $(\alpha_{1},\dotsc,\alpha_{k})$
 such that
 $\alpha_{1}\le\overline{\alpha_{1}},\dotsc,\alpha_{k}\le\overline{\alpha_{k}}$,
 \begin{equation}\label{eq:201506161832}
  \approximant_{j}(
  %\overline{\alpha_{1}}
   \alpha_{1}
  ,\dotsc,
%  \overline{\alpha_{k}}
  \alpha_{k}
)
  \sqsubseteq 
  l^{(i)}_{j}
  		\left(\,
		\begin{array}{c}
		\approximant_{i+1} (  
		 %\overline{\alpha_{1}}
		 \alpha_{1}
		 ,
		 \dotsc,
		 %\overline{\alpha_{k}}
		 \alpha_{k}
		 ),
		 \\
		 \dotsc,
		  \\
		\approximant_{m} (  
		 %\overline{\alpha_{1}}
		 \alpha_{1}
		 ,\dotsc,
		 %\overline{\alpha_{k}}
		 \alpha_{k}
		 )
		\end{array}
		\,\right),
 \end{equation}
 where $l^{(i)}_{j}\colon L^{m-i}\to L$ is the $i$-th interim solution
 in Def.~\ref{def:solOfEqSys}.
\end{quote}
In the following proof by induction on $i$,
we distinguish cases, according to whether $u_{i}$ is a $\mu$-variable
 or a $\nu$-variable. There is no need of distinguishing the 
base case ($i=1$) from the step case: it is easy to take proper care of
the occurrences of $i-1$ in the proof below.

\underline{\bf Case: $u_{i}$ is a $\mu$-variable.} 
% We shall first show the special case of the claim $(*)$ in the
%  above where $j=i$; this specialized claim is referred to as
%  $(*)_{j=i}$.
Let us choose $a\in [1,k]$  so
 that $i=i_{a}$, that is, the variable $u_{i}$ currently in question 
 has the $a$-th smallest priority among all the $\mu$-variables.

Our proof of the claim $(*)$ shall follow the following path. We will first
consider the special case $(*)_{j=i}$ of the claim $(*)$ where $j$ is
 fixed to be $j=i$.
Then we will show the inequality~(\ref{eq:201506161832}) by
 (transfinite) induction.  The general claim $(*)$ for the other  $j\in
 [0,i-1]$ will be
 derived from this special case $j=i$.

Let us fix $j=i$ in the claim $(*)$; thus we set out to prove
the following claim $(*)_{j=i}$.
\begin{quote}
 $(*)_{j=i}$  \quad
For each prioritized ordinal $(\alpha_{1},\dotsc,\alpha_{k})$
 such that
 $\alpha_{1}\le\overline{\alpha_{1}},\dotsc,\alpha_{k}\le\overline{\alpha_{k}}$,
 \begin{equation}\label{eq:201506162111}
  \approximant_{i}(
  %\overline{\alpha_{1}}
   \alpha_{1}
  ,\dotsc,
%  \overline{\alpha_{k}}
  \alpha_{k}
)
  \sqsubseteq 
  l^{(i)}_{i}
  		\left(\,
		\begin{array}{c}
		\approximant_{i+1} (  
		 %\overline{\alpha_{1}}
		 \alpha_{1}
		 ,
		 \dotsc,
		 %\overline{\alpha_{k}}
		 \alpha_{k}
		 ),
		 \\
		 \dotsc,
		  \\
		\approximant_{m} (  
		 %\overline{\alpha_{1}}
		 \alpha_{1}
		 ,\dotsc,
		 %\overline{\alpha_{k}}
		 \alpha_{k}
		 )
		\end{array}
		\,\right).
 \end{equation}
\end{quote}

 We first note a
fact
that follows from the definition of $l^{(i)}_{i}$
 (Def.~\ref{def:solOfEqSys}) and
 Lem.~\ref{lem:FPLowerApprox}.\ref{item:lemFPLowerApproxLFP}:
 the right-hand side $l^{(i)}_{i}(\dotsc)$ of the claimed inequality~(\ref{eq:201506162111})
 is given a lower bound by
 \begin{math}
 \Phi^{\alpha}(\bot)
 \end{math}, 
 where $\alpha$ is an arbitrary ordinal and $\Phi\colon L\to L$ is
 defined by
\begin{equation}\label{eq:201506162145}
  \begin{aligned}
  \Phi(l)
  &:=
  f^{\ddagger}_{i}
  		\left(\,
		\begin{array}{c}
		 l,
		  \\
		\approximant_{i+1} (  
		 %\overline{\alpha_{1}}
		 \alpha_{1}
		 ,\dotsc,
		 %\overline{\alpha_{k}}
		 \alpha_{k}
		 ),
		 \\
		 \dotsc,
		  \\
		\approximant_{m} (  
		 %\overline{\alpha_{1}}
		 \alpha_{1}
		 ,\dotsc,
		 %\overline{\alpha_{k}}
		 \alpha_{k}
		 )
		\end{array}
		\,\right)
   \\
  &\stackrel{(\dagger)}{=}
  f^{\ddagger}_{i}
  		\left(\,
		\begin{array}{c}
		 l,
		  \\
		\approximant_{i+1} (  
		 0
		 ,\dotsc,
		 0,
		 \alpha_{a+1},
		 \dotsc,
		 \alpha_{k}
		 ),
		 \\
		 \dotsc,
		  \\
		\approximant_{m} (  
		 0
		 ,\dotsc,
		 0,
		 \alpha_{a+1},
		 \dotsc,
		 \alpha_{k}
		 )
		\end{array}
		\,\right)
   \\
   &=
  f_{i}
  		\left(
		\begin{array}{c}
		 %----------------------------------
		 l^{(i-1)}_{1}
                 \left(
		  \begin{array}{c}
		  		 l,
		  \\
		\approximant_{i+1} (
		 0
		 ,\dotsc,
		 0,
		 \alpha_{a+1},
		 \dotsc,
		 \alpha_{k}
		 ),
		 \\
		 \dotsc,
		  \\
		\approximant_{m} (  
		 0
		 ,\dotsc,
		 0,
		 \alpha_{a+1},
		 \dotsc,
		 \alpha_{k}
		 )
		  \end{array}
		 \right),
%		 \mathrlap{\qquad {\footnotesize \text{1st}}}
		 %----------------------------------
		 \\
		 \dotsc,
		 \\
		 %----------------------------------
		 l^{(i-1)}_{i-1}
                 \left(
		  \begin{array}{c}
		  		 l,
		  \\
		\approximant_{i+1} (  
		 0
		 ,\dotsc,
		 0,
		 \alpha_{a+1},
		 \dotsc,
		 \alpha_{k}
		 ),
		 \\
		 \dotsc,
		  \\
		\approximant_{m} (  
		 0
		 ,\dotsc,
		 0,
		 \alpha_{a+1},
		 \dotsc,
		 \alpha_{k}
		 )
		  \end{array}
		 \right),
%		 \mathrlap{\qquad {\footnotesize \text{$(i-1)$-th}}}
		 %----------------------------------
		 \\
		 l,
		  \\
		\approximant_{i+1} (  
		 0
		 ,\dotsc,
		 0,
		 \alpha_{a+1},
		 \dotsc,
		 \alpha_{k}
		 ),
		 \\
		 \dotsc,
		  \\
		\approximant_{m} ( 
		 0
		 ,\dotsc,
		 0,
		 \alpha_{a+1},
		 \dotsc,
		 \alpha_{k}
		 )
		\end{array}
		\right),
  \end{aligned}
\end{equation}
where the second equality $(\dagger)$ holds due to 
Cond.~\ref{item:progressMeasDefMonotonicity} (monotonicity) of
  Def.~\ref{def:progressMeasForEqSys}, and that
\begin{math}
 (\alpha_{1},\dotsc,\alpha_{k})
   =_{b}
  (  
		 0
		 ,\dotsc,
		 0,
		 \alpha_{a+1},
		 \dotsc,
		 \alpha_{k}
		 )
\end{math}
for each $b\in [i+1,m]$ (cf.\ Def.~\ref{def:prioritizedOrdinal}).
In particular, the definition of $\Phi$ relies on
$\alpha_{a+1},\dotsc,\alpha_{k}$
but not on $\alpha_{a}$.

 We shall show, by 
 (transfinite) induction on the ordinal $\alpha_{a}$, that
\begin{equation}\label{eq:201506161816}
\approximant_{i}(
%\underset{\text{1st}}{\mathstrut\overline{\alpha_{1}}}
\alpha_{1}
,\dotsc,
%\underset{\text{$a$-th}}{\mathstrut\alpha_{a}}
\alpha_{a}
,\dotsc,
%\underset{\text{$k$-th}}{\mathstrut\overline{\alpha_{k}}}
\alpha_{k}
)
\;\sqsubseteq \;
\Phi^{\alpha_{a}}(\bot)
\end{equation}
for each ordinal $\alpha_{a}$ such that $\alpha_{a}\le \overline{\alpha_{a}}$. 
 If $\alpha_{a}=0$, then the left-hand side
       of~(\ref{eq:201506161816}) is $\bot$---this is 
       Cond.~\ref{item:progressMeasDefMuVarBaseCase} of
       Def.~\ref{def:progressMeasForEqSys}. 

 If $\alpha_{a}$ is a successor ordinal
       $\alpha'_{a}+1$, then
       \begin{align*}
&\approximant_{i}(
\alpha_{1},\dotsc,
\alpha'_{a}+1,
\alpha_{a+1}
\dotsc,
\alpha_{k}
)
\\
&
\sqsubseteq
f_{i}
		\left(\,
		\begin{array}{c}
		 \approximant_{1} (\beta_{1},\dotsc,\beta_{a-1},
		  \alpha'_{a},\alpha_{a+1},\dotsc,\alpha_{k}),
		 \\
		 \dotsc,
		  \\
		 \approximant_{i-1} (\beta_{1},\dotsc,\beta_{a-1},
		 \alpha'_{a},\alpha_{a+1},\dotsc,
		      \alpha_{k}),
		  \\
		 \approximant_{i} (\beta_{1},\dotsc,\beta_{a-1},
		 \alpha'_{a},\alpha_{a+1},\dotsc,
		      \alpha_{k}),
		  \\
		 \approximant_{i+1} (\beta_{1},\dotsc,\beta_{a-1},
		 \alpha'_{a},\alpha_{a+1},\dotsc,
		      \alpha_{k}),
		  \\
		 \dotsc,
		  \\
		 \approximant_{m} (\beta_{1},\dotsc,\beta_{a-1},
		  \alpha'_{a},\alpha_{a+1},\dotsc,\alpha_{k})
		\end{array}
		\,\right)
\\
&\qquad\text{for some ordinals $\beta_{1},\dotsc, \beta_{a-1}$, by 
	Cond.~\ref{item:progressMeasDefMuVarStepCase} of
	Def.~\ref{def:progressMeasForEqSys}
	} 
\\
&
\sqsubseteq
f_{i}
		\left(\,
		\begin{array}{c}
		 %----------------------------------
		 l^{(i-1)}_{1}
		  \left(
		   \begin{array}{c}
		    \approximant_{i} (\beta_{1},\dotsc,\beta_{a-1},
		     \alpha'_{a},\alpha_{a+1},\dotsc,
		      \alpha_{k}),
		    \\
		    \approximant_{i+1} (\beta_{1},\dotsc,\beta_{a-1},
		     \alpha'_{a},\alpha_{a+1},\dotsc,
		      \alpha_{k}),
		    \\
		    \dotsc,
		    \\
		    \approximant_{m} (\beta_{1},\dotsc,\beta_{a-1},
		     \alpha'_{a},\alpha_{a+1},\dotsc,
		      \alpha_{k})
		   \end{array}
		  \right),
		 %----------------------------------
		 \\
		 \dotsc,
		  \\
		 %----------------------------------
		 l^{(i-1)}_{i-1}
		  \left(
		   \begin{array}{c}
		    \approximant_{i} (\beta_{1},\dotsc,\beta_{a-1},
		     \alpha'_{a},\alpha_{a+1},\dotsc,
		      \alpha_{k}),
		    \\
		    \approximant_{i+1} (\beta_{1},\dotsc,\beta_{a-1},
		     \alpha'_{a},\alpha_{a+1},\dotsc,
		      \alpha_{k}),
		    \\
		    \dotsc,
		    \\
		    \approximant_{m} (\beta_{1},\dotsc,\beta_{a-1},
		     \alpha'_{a},\alpha_{a+1},\dotsc,
		      \alpha_{k})
		   \end{array}
		  \right),
		 %----------------------------------
		  \\
		 \approximant_{i} (\beta_{1},\dotsc,\beta_{a-1},
		 \alpha'_{a},\alpha_{a+1},\dotsc,
		      \alpha_{k}),
		  \\
		 \approximant_{i+1} (\beta_{1},\dotsc,\beta_{a-1},
		 \alpha'_{a},\alpha_{a+1},\dotsc,
		      \alpha_{k}),
		  \\
		 \dotsc,
		  \\
		 \approximant_{m} (\beta_{1},\dotsc,\beta_{a-1},
		  \alpha'_{a},\alpha_{a+1},\dotsc,\alpha_{k})
		\end{array}
		\,\right)
\\
&\qquad\text{by the induction hypothesis (the claim $(*)$ for  $i-1$)}
\\
&
=
f_{i}
		\left(\,
		\begin{array}{c}
		 %----------------------------------
		 l^{(i-1)}_{1}
		  \left(
		   \begin{array}{c}
		    \approximant_{i} (\beta_{1},\dotsc,\beta_{a-1},
		     \alpha'_{a},\alpha_{a+1},\dotsc,
		      \alpha_{k}),
		    \\
		    \approximant_{i+1} (0,\dotsc,0,
		     \alpha_{a+1},\dotsc,
		      \alpha_{k}),
		    \\
		    \dotsc,
		    \\
		    \approximant_{m} (0,\dotsc,0,\alpha_{a+1},\dotsc,
		      \alpha_{k})
		   \end{array}
		  \right),
		 %----------------------------------
		 \\
		 \dotsc,
		  \\
		 %----------------------------------
		 l^{(i-1)}_{i-1}
		  \left(
		   \begin{array}{c}
		    \approximant_{i} (\beta_{1},\dotsc,\beta_{a-1},
		     \alpha'_{a},\alpha_{a+1},\dotsc,
		      \alpha_{k}),
		    \\
		    \approximant_{i+1} (0,\dotsc,0,
		     \alpha_{a+1},\dotsc,
		      \alpha_{k}),
		    \\
		    \dotsc,
		    \\
		    \approximant_{m} (0,\dotsc,0,\alpha_{a+1},\dotsc,
		      \alpha_{k})
		   \end{array}
		  \right),
		 %----------------------------------
		  \\
		 \approximant_{i} (\beta_{1},\dotsc,\beta_{a-1},
		 \alpha'_{a},\alpha_{a+1},\dotsc,
		      \alpha_{k}),
		    \\
		    \approximant_{i+1} (0,\dotsc,0,
		     \alpha_{a+1},\dotsc,
		      \alpha_{k}),
		    \\
		    \dotsc,
		    \\
		    \approximant_{m} (0,\dotsc,0,\alpha_{a+1},\dotsc,
		      \alpha_{k})
		\end{array}
		\,\right)
\\
&\qquad\text{by monotonicity
 (Cond.~\ref{item:progressMeasDefMonotonicity} of
 Def.~\ref{def:progressMeasForEqSys}), much like in $(\dagger)$ 
 in~(\ref{eq:201506162145})}
\\
&=
\Phi\bigl(\,
		 \approximant_{i} (\beta_{1},\dotsc,\beta_{a-1},
		 \alpha'_{a},\alpha_{a+1},\dotsc,
		      \alpha_{k})
\,\bigr)
\\
&\qquad\text{by def.\ of $\Phi$, see~(\ref{eq:201506162145})}
\\
&=
\Phi\bigl(\,
		 \approximant_{i} (\alpha_{1},\dotsc,\alpha_{a-1},
		 \alpha'_{a},\alpha_{a+1},\dotsc,
		      \alpha_{k})
\,\bigr)
\\
&\qquad\text{by monotonicity, much like the second last equality}
\\
&\sqsubseteq
\Phi^{\alpha'_{a}+1}(\bot)
\qquad\text{by the induction hypothesis~(\ref{eq:201506161816}).}
       \end{align*}

 Finally, if $\alpha_{a}$ in~(\ref{eq:201506161816}) is a limit 
ordinal, then
\begin{align*}
& \approximant_{i}(
\alpha_{1},\dotsc,
\alpha_{a},
\alpha_{a+1}
\dotsc,
\alpha_{k}
)
\\
&
\sqsubseteq
\bigsqcup_{\beta < \alpha_{a}}
\approximant_{i}(
\alpha_{1},\dotsc,
\beta,
\alpha_{a+1}
\dotsc,
\alpha_{k}
)
\quad
\text{by Cond.~\ref{item:progressMeasDefMuVarLimitCase} of 
 Def.~\ref{def:progressMeasForEqSys}}
\\
&
\sqsubseteq
\bigsqcup_{\beta < \alpha_{a}}
\Phi^{\beta}(\bot)
\quad
\text{by the induction hypothesis~(\ref{eq:201506161816})}
\\
&
=
\Phi^{\alpha_{a}}(\bot)
\quad
\text{by def.\ of $\Phi^{\alpha_{a}}$. }
\end{align*}
This concludes the proof that the inequality (\ref{eq:201506161816})
holds for any ordinal $\alpha_{a}$. By the fact 
that $\Phi^{\alpha_{a}}(\bot)$ is a lower bound for the right-hand 
side of~(\ref{eq:201506162111}) (that is
 argued after the inequality~(\ref{eq:201506162111})), we have now shown 
that the claim $(*)_{j=i}$ holds.

Finally,  the general claim $(*)$ for any $j\in [1,i-1]$ (other than $j=i$) is shown as follows.
\begin{align*}
 & \approximant_{j}(
\alpha_{1},\dotsc,
\alpha_{k}
)
\\
&
\sqsubseteq
l^{(i-1)}_{j}
\left(
\begin{array}{c}
\approximant_{i}(
\alpha_{1},\dotsc,
\alpha_{k}
),
 \\
\approximant_{i+1}(
\alpha_{1},\dotsc,
\alpha_{k}
)
,
\\
\dotsc,
\\
\approximant_{m}(
\alpha_{1},\dotsc,
\alpha_{k}
)
\end{array}
\right)
\\
&\qquad\text{by the induction hypothesis (the claim $(*)$ for  $i-1$)}
\\
&
\sqsubseteq
l^{(i-1)}_{j}
\left(
\begin{array}{c}
l^{(i)}_{i}\left(
\begin{array}{c}
 \approximant_{i+1}(\alpha_{1},\dotsc,
\alpha_{k}
),
\\
\dotsc,
\\
 \approximant_{m}(\alpha_{1},\dotsc,
\alpha_{k}
)
\end{array}
\right),
 \\
\approximant_{i+1}(
\alpha_{1},\dotsc,
\alpha_{k}
)
,
\\
\dotsc,
\\
\approximant_{m}(
\alpha_{1},\dotsc,
\alpha_{k}
)
\end{array}
\right)
\\
&\quad\text{by the claim $(*)_{j=i}$ (that we have 
 already shown for the current $i$)}
\\
&=
l^{(i)}_{j}
\left(
\begin{array}{c}
\approximant_{i+1}(
\alpha_{1},\dotsc,
\alpha_{k}
)
,
\\
\dotsc,
\\
\approximant_{m}(
\alpha_{1},\dotsc,
\alpha_{k}
)
\end{array}
\right)
\\
&\quad\text{by the 
definition~(\ref{eq:defInterimSolutionForSmallerJ}) 
of $l^{(i)}_{j}$; see Def.~\ref{def:solOfEqSys}.}
\end{align*}
This concludes one case of our proof of the claim $(*)$ 
(by induction on $i$), where $u_{i}$ is a $\mu$-variable.

\underline{\bf Case: $u_{i}$ is a $\nu$-variable.} 
Let us choose $a\in [1,k]$  so
 that
\begin{displaymath}
 		i_{1}<\cdots <i_{a-1}<i<i_{a}<\cdots < i_{k}.
\end{displaymath} 

We shall prove the special case of $(*)$ where $j=i$. The claim $(*)$ 
for general $j\in [1,i]$ follows from this special case, much like in
the above for the case where $u_{i}$ is a $\mu$-variable.

By the definition of $l^{(i)}_{i}$
 (Def.~\ref{def:solOfEqSys}) and
 Lem.~\ref{lem:FPLowerApprox}.\ref{item:lemFPLowerApproxGFP}, 
showing the following (i.e.\ 
$    \approximant_{i}(
   \alpha_{1}
  ,\dotsc,
  \alpha_{k}
)$
is a suitable postfixed point) suffices for establishing
 the desired inequality~(\ref{eq:201506161832}):
\begin{equation}\label{eq:201506162250}
    \approximant_{i}(
   \alpha_{1}
  ,\dotsc,
  \alpha_{k}
)
 \sqsubseteq
 f^{\ddagger}_{i}
 \left(
 \begin{array}{c}
    \approximant_{i}(
   \alpha_{1}
  ,\dotsc,
  \alpha_{k}
  ),
  \\
    \approximant_{i+1}(
   \alpha_{1}
  ,\dotsc,
  \alpha_{k}
  ),
  \\
  \dotsc,
\\
    \approximant_{m}(
   \alpha_{1}
  ,\dotsc,
  \alpha_{k}
)
 \end{array} 
\right).
\end{equation}
We proceed as follows.
\begin{align*}
&
    \approximant_{i}(
   \alpha_{1}
  ,\dotsc,
  \alpha_{k}
)
\\
&\sqsubseteq
		f_{i}
		\left(\,
		\begin{array}{c}
		\approximant_{1} (\beta_{1},\dotsc,\beta_{a-1},
		 \alpha_{a},\dotsc,\alpha_{k}),
		 \\
		 \dotsc,
		  \\
		\approximant_{m} (\beta_{1},\dotsc,\beta_{a-1},
		\alpha_{a},\dotsc,\alpha_{k})
		\end{array}
		\,\right)
\\
&\qquad\text{for some ordinals $\beta_{1},\dotsc, \beta_{a-1}$, by 
	Cond.~\ref{item:progressMeasDefNuVar} of
	Def.~\ref{def:progressMeasForEqSys}
	} 
\\
&
\sqsubseteq
f_{i}
		\left(
		\begin{array}{c}
		 %----------------------------------
		 l^{(i-1)}_{1}
		  \left(
		   \begin{array}{c}
		    \approximant_{i} (\beta_{1},\dotsc,\beta_{a-1},
		     \alpha_{a}\dotsc,
		      \alpha_{k}),
		    \\
		    \approximant_{i+1} (\beta_{1},\dotsc,\beta_{a-1},
		     \alpha_{a}\dotsc,
		      \alpha_{k}),
		    \\
		    \dotsc,
		    \\
		    \approximant_{m} (\beta_{1},\dotsc,\beta_{a-1},
		     \alpha_{a}\dotsc,
		      \alpha_{k})
		   \end{array}
		  \right),
		 %----------------------------------
		 \\
		 \dotsc,
		  \\
		 %----------------------------------
		 l^{(i-1)}_{i-1}
		  \left(
		   \begin{array}{c}
		    \approximant_{i} (\beta_{1},\dotsc,\beta_{a-1},
		     \alpha_{a}\dotsc,
		      \alpha_{k}),
		    \\
		    \approximant_{i+1} (\beta_{1},\dotsc,\beta_{a-1},
		     \alpha_{a}\dotsc,
		      \alpha_{k}),
		    \\
		    \dotsc,
		    \\
		    \approximant_{m} (\beta_{1},\dotsc,\beta_{a-1},
		     \alpha_{a}\dotsc,
		      \alpha_{k})
		   \end{array}
		  \right),
		 %----------------------------------
		  \\
		 \approximant_{i} (\beta_{1},\dotsc,\beta_{a-1},
		 \alpha_{a}\dotsc,
		      \alpha_{k}),
		  \\
		 \dotsc,
		  \\
		 \approximant_{m} (\beta_{1},\dotsc,\beta_{a-1},
		  \alpha_{a}\dotsc,\alpha_{k})
		\end{array}
		\right)
\\
&\qquad\text{by the induction hypothesis (the claim $(*)$ for  $i-1$)}
\\
&
=
f^{\ddagger}_{i}
\left(
\begin{array}{c}
		 \approximant_{i} (\beta_{1},\dotsc,\beta_{a-1},
		 \alpha_{a}\dotsc,
		      \alpha_{k}),
		  \\
		 \dotsc,
		  \\
		 \approximant_{m} (\beta_{1},\dotsc,\beta_{a-1},
		  \alpha_{a}\dotsc,\alpha_{k})
		\end{array}
		\right)
\\
&\qquad\text{by def.\ of $f^{\ddagger}_{i}$ (Def.~\ref{def:solOfEqSys})}
\\
&=
 f^{\ddagger}_{i}
 \left(
 \begin{array}{c}
    \approximant_{i}(
   \alpha_{1}
  ,\dotsc,
  \alpha_{k}
  ),
  \\
  \dotsc,
\\
    \approximant_{m}(
   \alpha_{1}
  ,\dotsc,
  \alpha_{k}
)
 \end{array} 
\right),
\end{align*}
where the last equality is, once again, a consequence of 
monotonicity
 (Cond.~\ref{item:progressMeasDefMonotonicity} of
 Def.~\ref{def:progressMeasForEqSys}) and that 
\begin{math}
 		 (\beta_{1},\dotsc,\beta_{a-1},
		 \alpha_{a}\dotsc,
		      \alpha_{k})
=_{i}
(
   \alpha_{1}
  ,\dotsc,
  \alpha_{k}
)
\end{math}. This concludes the proof of 
Thm.~\ref{thm:correctnessOfProgMeasEqSys}.\ref{item:soundnessProgressMeas}.

\fbox{\bf  The item~\ref{item:completenessProgressMeas} (completeness).} 
The proof is by induction on the number $m$ of equations in the equational
 system $E$. 
For each $l\in L$, let $E^{(l)}$ be 
 the equational system obtained from
 $E$ in~(\ref{eq:sysOfEq}) by removing the last equation and
 substituting $l$ for the last variable $u_{m}$. That is,
\begin{equation}\label{eq:truncatedEquation}
\begin{aligned}
&  E^{(l)}
 \;:=\;
 \left[
\begin{array}{c}
  u_{1} =_{\eta_{1}} f_{1}(u_{1},\dotsc, u_{m-1}, l),
\\
 \dotsc,
\\
  u_{m-1} =_{\eta_{m-1}} f_{m-1}(u_{1},\dotsc, u_{m-1}, l)
\end{array} 
\right].
\end{aligned}
\end{equation}

For the rest of the proof we distinguish cases, depending on whether the last variable $u_{m}$ 
is a $\mu$-variable or a $\nu$-variable.

\underline{\bf Case: $u_{m}$ is a $\mu$-variable.} 
By the induction hypothesis there exists a progress measure that
 achieves the exact solution of the system $E^{(l)}$, for each $l\in L$. 
Such a progress measure shall be denoted by
 \begin{align*}
 & p^{(l)}\;=\;
 \\
 &\quad
  \bigl(\,
  (
  \overline{\alpha^{(l)}_{1}},\dotsc,
  \overline{\alpha^{(l)}_{k-1}}),
  \,
  \bigl(\,\approximant^{(l)}_{i}(\alpha_{1},\dotsc,\alpha_{k-1})\,\bigr)_{i\in
  [1,m-1],\seq{\alpha}{k-1}}
\,\bigr).
 \end{align*}
(Note here that $E^{(l)}$ has $k-1$ $\mu$-variables, since $E$ has $k$ of
 such and we assumed that $u_{m}$ is a $\mu$-variable.)
It is easily seen to  satisfy, for each $i\in [1,m-1]$:
\begin{equation}\label{eq:201506181551}
 p^{(l)}_{i} 
   \bigl(\,
  \overline{\alpha^{(l)}_{1}},\dotsc,
  \overline{\alpha^{(l)}_{k-1}}\,\bigr)
 \;= \;
 l^{(m-1)}_{i}(l),
\end{equation}
where $l^{(m-1)}_{i}(l)$ is (a component of) the $(m-1)$-th
 interim solution of the original
 equational system $E$ (Def.~\ref{def:solOfEqSys}). 
Furthermore, by induction, we assume that 
\begin{math}
   \overline{\alpha^{(l)}_{a}} \le \ascCL(L)
\end{math}
for each $a\in[1,k-1]$.

Using the above data, we shall construct a desired progress measure $p$
 for the system $E$. We define its approximants
 $p_{i}(\alpha_{1},\dotsc,\alpha_{k})$ by induction on the ordinal
 $\alpha_{k}$. For the base case:
\begin{align*}
 p_{m}
 (\alpha_{1},\dotsc,\alpha_{k-1},0)
 &:=\bot,
 \\
p_{i}
 (\alpha_{1},\dotsc,\alpha_{k-1},0)
 &:=
p^{( p_{m}
 (\alpha_{1},\dotsc,\alpha_{k-1},0)
)}_{i}
 (\alpha_{1},\dotsc,\alpha_{k-1})
 \\
&
\!\!\!\!\!\!\!\!\!\!\!\!\!\!\!\!\!\!\!\!\!\!\!\!\!\!\!\!
=
p^{(\bot
)}_{i}
 (\alpha_{1},\dotsc,\alpha_{k-1})
\qquad\text{for each $i\in[1,m-1]$.}
\end{align*}
For the step case we 
use
the function $f^{\ddagger}_{m-1}$ from Def.~\ref{def:solOfEqSys} 
and define
\begin{align*}
 p_{m}
 (\alpha_{1},\dotsc,\alpha_{k-1},\alpha_{k}+1)
 &:=
 f^{\ddagger}_{m}\bigl(\,
 p_{m}
 (\alpha_{1},\dotsc,\alpha_{k-1},\alpha_{k})
 \,\bigr)
,
 \\
p_{i}
 (\alpha_{1},\dotsc,\alpha_{k-1},\alpha_{k}+1)
 &:=
% p^{(\,p_{m}
%  (\alpha_{1},\dotsc,\alpha_{k-1},\alpha_{k}+1)
% \,)}_{i}
%  (\alpha_{1},\dotsc,\alpha_{k-1})
 \\
&
\!\!\!\!\!\!\!\!\!\!\!\!\!\!\!\!\!\!\!\!\!\!\!\!\!\!\!\!
\!\!\!\!\!\!\!\!\!\!\!\!\!\!\!\!\!\!\!\!\!\!\!\!\!\!\!\!
\!\!\!\!\!\!
p^{(\,p_{m}
 (\alpha_{1},\dotsc,\alpha_{k-1},\alpha_{k}+1)
\,)}_{i}
 (\alpha_{1},\dotsc,\alpha_{k-1})
\quad\text{for each $i\in[1,m-1]$.}
\end{align*}
For the limit case, similarly, we 
 define
\begin{equation}\label{eq:201506181814}
 \begin{aligned}
 p_{m}
 (\alpha_{1},\dotsc,\alpha_{k-1},\alpha_{k})
 &:=
 \bigsqcup_{\beta<\alpha_{k}}
 p_{m}
 (\alpha_{1},\dotsc,\alpha_{k-1},\beta)
,
 \\
p_{i}
 (\alpha_{1},\dotsc,\alpha_{k-1},\alpha_{k})
 &:=
p^{(\,p_{m}
 (\alpha_{1},\dotsc,\alpha_{k-1},\alpha_{k})
\,)}_{i}
 (\alpha_{1},\dotsc,\alpha_{k-1})
 \\
&\qquad\text{for each $i\in[1,m-1]$.}
\end{aligned}
\end{equation}
In the above definitions it may happen that
 $p^{(l)}(\alpha_{1},\dotsc,\alpha_{k-1})$ is not defined because
the ordinals $\alpha_{1},\dotsc,\alpha_{k-1}$ exceed the domain of 
$p^{(l)}$, that is,
$\overline{\alpha^{(l)}_{i}}< \alpha_{i}$ for some $i\in[1,k-1]$.
In such a case we use, in place of $(\alpha_{1},\dotsc,\alpha_{k-1})$, 
the greatest prioritized ordinal that is smaller than it (with respect
 to
the lexicographic order $\preceq$ in Def.~\ref{def:prioritizedOrdinal}).

Regarding the maximum prioritized ordinal
 $(\overline{\alpha_{1}},\dotsc, \overline{\alpha_{k}})$ of $p$:
\begin{itemize}
 \item On the ordinal $\overline{\alpha_{k}}$ for the last 
       $\mu$-variable $u_{m}$, the approximants for $u_{m}$ form a (transfinite) ascending chain
 \begin{equation}\label{eq:201506181609}
 p_{m}
 (\alpha_{1},\dotsc,\alpha_{k-1},0)
 \;\sqsubseteq\;
 p_{m}
 (\alpha_{1},\dotsc,\alpha_{k-1},1)
 \;\sqsubseteq\;\cdots\enspace;
 \end{equation}
 because the chain is nothing but $\bot\sqsubseteq f^{\ddagger}_{m}(\bot) 
 \sqsubseteq f^{\ddagger}_{m}(f^{\ddagger}_{m}(\bot))
       \sqsubseteq\cdots$.
 This chain in the complete lattice $L$ eventually stabilizes (bounded
       by the ordinal $\ascCL(L)$, by the definition of the latter); we let $\overline{\alpha_{k}}$ to be
       an ordinal, such that $\alpha_{k}\le \ascCL(L)$, at which the chain~(\ref{eq:201506181609}) has
       stabilized. 
 \item On the other ordinals $\overline{\alpha_{a}}$ for $a\in [1,k-1]$, 
       we define
       \begin{equation}\label{eq:201506181759}
	\overline{\alpha_{a}}
	:=
	\bigvee_{\beta\le \overline{\alpha_{k}}}
	\overline{\alpha^{( p_{m}
 (\alpha_{1},\dotsc,\alpha_{k-1},\beta))
	}_{a}};
       \end{equation}
it is obvious, from the induction hypothesis that 
\begin{math}
   \overline{\alpha^{(l)}_{a}} \le \ascCL(L)
\end{math}
for each $l\in L$ and $a\in[1,k-1]$, that $\overline{\alpha_{a}}$
in~(\ref{eq:201506181759}) is no bigger than $\ascCL(L)$.
\end{itemize}

We have to check that the data $p$ thus defined is indeed a progress
 measure (Def.~\ref{def:progressMeasForEqSys}). On 
Cond.~\ref{item:progressMeasDefMonotonicity} (monotonicity): it holds if
 $i=m$ because~(\ref{eq:201506181609}) is an ascending chain; 
 for $i\in[1,m-1]$ the claim follows from the induction hypothesis
 that $p^{(l)}$ is a progress measure for $E^{(l)}$.
 Cond.~\ref{item:progressMeasDefMuVarBaseCase} is easy, distinguishing
 cases for $a=k$ (obvious by definition) and $a\in[1,k-1]$ (by the
 induction hypothesis).

 On  Cond.~\ref{item:progressMeasDefMuVarStepCase}, let $a=k$ (hence $i_{a}=m$). 
 Then
\begin{align*}
& p_{m}
 (\alpha_{1},\dotsc,\alpha_{k-1},\alpha_{k}+1)
\\
 &=
 f^{\ddagger}_{m}\bigl(\,
 p_{m}
 (\alpha_{1},\dotsc,\alpha_{k-1},\alpha_{k})
 \,\bigr)
 \quad\text{by definition}
 \\
&=
f_{m}
		\left(\,
		\begin{array}{c}
		 %----------------------------------
		 l^{(m-1)}_{1}
		  \left(\,
		    \approximant_{m} 
		     (\alpha_{1},\dotsc,\alpha_{k-1},\alpha_{k})
		  \,\right),
		 %----------------------------------
		 \\
		 \dotsc,
		  \\
		 %----------------------------------
		 l^{(m-1)}_{m-1}
		  \left(\,
		    \approximant_{m} 
		     (\alpha_{1},\dotsc,\alpha_{k-1},\alpha_{k})
		  \,\right),
		 %----------------------------------
		  \\
		    \approximant_{m} 
 (\alpha_{1},\dotsc,\alpha_{k-1},\alpha_{k})
		\end{array}
		\,\right)
 \\
&=
f_{m}
		\left(\,
		\begin{array}{c}
		 %----------------------------------
 p^{(		    \approximant_{m} 
		     (\alpha_{1},\dotsc,\alpha_{k-1},\alpha_{k})
)}_{1} 
  (\beta_{1},\dotsc,\beta_{k-1}),
		 %----------------------------------
		 \\
		 \dotsc,
		  \\
		 %----------------------------------
 p^{(		    \approximant_{m} 
		     (\alpha_{1},\dotsc,\alpha_{k-1},\alpha_{k})
)}_{m-1} 
  (\beta_{1},\dotsc,\beta_{k-1}),
		 %----------------------------------
		  \\
		    \approximant_{m} 
 (\alpha_{1},\dotsc,\alpha_{k-1},\alpha_{k})
		\end{array}
		\,\right)
\\
&\qquad
\text{for some ordinals $\beta_{1},\dotsc,\beta_{k-1}$, by~(\ref{eq:201506181551})}
 \\
&\stackrel{(\dagger)}{=}
f_{m}
		\left(\,
		\begin{array}{c}
		 %----------------------------------
 p^{(		    \approximant_{m} 
		     (\beta_{1},\dotsc,\beta_{k-1},\alpha_{k})
)}_{1} 
  (\beta_{1},\dotsc,\beta_{k-1}),
		 %----------------------------------
		 \\
		 \dotsc,
		  \\
		 %----------------------------------
 p^{(		    \approximant_{m} 
		     (\beta_{1},\dotsc,\beta_{k-1},\alpha_{k})
)}_{m-1} 
  (\beta_{1},\dotsc,\beta_{k-1}),
		 %----------------------------------
		  \\
		    \approximant_{m} 
 (\beta_{1},\dotsc,\beta_{k-1},\alpha_{k})
		\end{array}
		\,\right)
\\
&=
f_{m}
		\left(\,
		\begin{array}{c}
		 %----------------------------------
 p_{1} 
 (\beta_{1},\dotsc,\beta_{k-1},\alpha_{k}),
		 %----------------------------------
		 \\
		 \dotsc,
		  \\
		 %----------------------------------
 p_{m-1} 
 (\beta_{1},\dotsc,\beta_{k-1},\alpha_{k}),
		 %----------------------------------
		  \\
		    \approximant_{m} 
 (\beta_{1},\dotsc,\beta_{k-1},\alpha_{k})
		\end{array}
		\,\right)
\\
&\qquad\text{by def.\ of $p_{1},\dotsc,p_{m-1}$,}
\end{align*}
as required. Here the equality $(\dagger)$ holds since the first $k-1$ arguments do
 not matter in the definition of $\approximant_{m}$, that is, 
\begin{math}
 \approximant_{m} 
		     (\alpha_{1},\dotsc,\alpha_{k-1},\alpha_{k})
=
\approximant_{m} 
		     (\beta_{1},\dotsc,\beta_{k-1},\alpha_{k})
\end{math} (this is easily seen by induction on $\alpha_{k}$).
On Cond.~\ref{item:progressMeasDefMuVarStepCase}, if $a\in[1,k-1]$, the
 claim follows immediately by the induction hypothesis.
Cond.~\ref{item:progressMeasDefMuVarLimitCase} is easy, by definition
 for $a=k$ and by the induction hypothesis for $a\in
 [1,k-1]$. Cond.~\ref{item:progressMeasDefNuVar} is easy, too, by the
 induction hypothesis (we have assumed that $u_{m}$ is a
 $\mu$-variable).

Finally, we have to show that the progress measure $p$ defined in the
 above
indeed achieves the exact solution, that is, 
 \begin{math}
  \approximant_{i}(
  \overline{\alpha_{1}},\dotsc,
  \overline{\alpha_{k}})
  \;=\;
  l^{\sol}_{i}
 \end{math}
 for each $i\in[1,m]$. For $i=m$ this is easy: $  l^{\sol}_{m}$ 
 is characterized as the supremum of the chain
 $\bot\sqsubseteq f^{\ddagger}_{m}(\bot) 
 \sqsubseteq f^{\ddagger}_{m}(f^{\ddagger}_{m}(\bot))
       \sqsubseteq\cdots$ (Def.~\ref{def:solOfEqSys} and
 Lem.~\ref{lem:FPLowerApprox}.\ref{item:lemFPLowerApproxLFP}); and the
 last chain coincides, by definition, with the one
 in~(\ref{eq:201506181609}).
 For the other $i$ (i.e.\ $i\in [1,m-1]$):
 \begin{align*}
  l^{\sol}_{i}
  &=
  l^{m-1}_{i}(l^{\sol}_{m})
  \quad\text{by Def.~\ref{def:solOfEqSys}}
  \\
  &=
 p^{(l^{\sol}_{m})}_{i} 
   \bigl(\,
  \overline{\alpha^{(l^{\sol}_{m})}_{1}},\dotsc,
  \overline{\alpha^{(l^{\sol}_{m})}_{k-1}}\,\bigr)
  \quad\text{by~(\ref{eq:201506181551})}
  \\
  &=
 p^{(  \approximant_{m}(
  \overline{\alpha_{1}},\dotsc,
  \overline{\alpha_{k}})
)}_{i} 
   \bigl(\,
  \overline{\alpha^{(\approximant_{m}(
  \overline{\alpha_{1}},\dotsc,
  \overline{\alpha_{k}}))}_{1}},\dotsc,
  \overline{\alpha^{(\approximant_{m}(
  \overline{\alpha_{1}},\dotsc,
  \overline{\alpha_{k}}))}_{k-1}}\,\bigr)
 \\
 &=
 p^{(  \approximant_{m}(
  \overline{\alpha_{1}},\dotsc,
  \overline{\alpha_{k}})
)}_{i} 
   \bigl(\,
  \overline{\alpha_{1}},\dotsc,
  \overline{\alpha_{k-1}}\,\bigr)
 \\
 &\qquad\text{by the convention we adopted just after (\ref{eq:201506181814})}
 \\
 &=
 p_{i} 
   (
  \overline{\alpha_{1}},\dotsc,
  \overline{\alpha_{k}}) 
 \quad\text{by def.\ of $p_{i}$.}
 \end{align*}
 This concludes the case where
$u_{m}$ is a $\mu$-variable.

\underline{\bf Case: $u_{m}$ is a $\nu$-variable.} 
The proof in this case is similar and simpler. We shall therefore
 describe
only its main points; further details can be easily filled out.

By the induction hypothesis there exists a progress measure $p^{(l)}$ that
 achieves the exact solution of the system $E^{(l)}$, for each $l\in L$:
 \begin{align*}
 & p^{(l)}\;=\;
 \\
 &\quad
  \bigl(\,
  (
  \overline{\alpha^{(l)}_{1}},\dotsc,
  \overline{\alpha^{(l)}_{k}}),
  \,
  \bigl(\,\approximant^{(l)}_{i}(\alpha_{1},\dotsc,\alpha_{k})\,\bigr)_{i\in
  [1,m-1],\seq{\alpha}{k}}
\,\bigr).
 \end{align*}
(Note that $E^{(l)}$ has $k$ $\mu$-variables.)
We use such $p^{(l)}$ to define a desired progress measure $p$ of $E$. 
Its approximants are defined by:
\begin{align*}
 p_{m}
 (\alpha_{1},\dotsc,\alpha_{k})
 &:=
 l^{\sol}_{m},
\\
 p_{i}
 (\alpha_{1},\dotsc,\alpha_{k})
 &:=
p^{(\,l^{\sol}_{m}
\,)}_{i}
 (\alpha_{1},\dotsc,\alpha_{k})
 \quad\text{for each $i\in[1,m-1]$.}
\end{align*}
Regarding
 the maximum prioritized ordinal
 $(\overline{\alpha_{1}},\dotsc, \overline{\alpha_{k}})$, we define
 $\overline{\alpha_{a}}:=
\overline{\alpha^{(l^{\sol}_{m})}_{a}} 
$ for each $a\in [1,k]$. Obviously, from the induction hypothesis,
 $\overline{\alpha_{a}}$ can be chosen so that 
$\overline{\alpha_{a}}\le \ascCL(L)$.

In seeing that $p$ is indeed a progress measure, checking the conditions
 of Def.~\ref{def:progressMeasForEqSys} is mostly straightforward by the
 induction hypothesis. 
 The only nontrivial point is  Cond.~\ref{item:progressMeasDefNuVar}, in
 case $i=m$. We have to show that
 \begin{equation}\label{eq:201506181840}
   l^{\sol}_{m}
  \;\sqsubseteq\;
  f_{m}
  		\left(\,
		\begin{array}{c}
		\approximant_{1} (\beta_{1},\dotsc,\beta_{k}),
		 \\
		 \dotsc,
		  \\
		\approximant_{m-1} (\beta_{1},\dotsc,\beta_{k}),
		 \\
		   l^{\sol}_{m} 
		\end{array}
		\,\right)
 \end{equation}
 for some ordinals $\seq{\beta}{k}$ (note that all the $\mu$-variables
 have priorities smaller than that of $u_{m}$).
 We set $\beta_{a}:=\overline{\alpha^{(l^{\sol}_{m})}_{a}} $ for each
 $a\in [1,k]$. Then the right-hand side of~(\ref{eq:201506181840})
 becomes:
  \begin{align*}
  &
   f_{m}
   		\left(\,
 		\begin{array}{c}
 		\approximant^{(l^{\sol}_{m})}_{1} (
 		 \overline{\alpha^{(l^{\sol}_{m})}_{1}}
 		 ,\dotsc,
 		 \overline{\alpha^{(l^{\sol}_{m})}_{k}}
 ),
 		 \\
 		 \dotsc,
 		  \\
 		\approximant^{(l^{\sol}_{m})}_{m-1} (
 		 \overline{\alpha^{(l^{\sol}_{m})}_{1}}
 		 ,\dotsc,
 		 \overline{\alpha^{(l^{\sol}_{m})}_{k}}
 ),
 		 \\
 		   l^{\sol}_{m} 
 		\end{array}
 		\,\right)
  \\
  &=
   f_{m}
   \bigl(\,
l^{(m-1)}_{1}(l^{\sol}_{m}),\,
\dotsc,\,
l^{(m-1)}_{m-1}(l^{\sol}_{m}),\,
 		   l^{\sol}_{m} 
\,\bigr)
  \\
  &=
  f^{\ddagger}_{m}(l^{\sol}_{m} )
  \quad\text{by Def.~\ref{def:solOfEqSys}}
  \\
  &=l^{\sol}_{m}, 
  \end{align*}
 where the last equality is because of the definition of 
 $l^{\sol}_{m} = l^{(m)}_{m}$ as a greatest fixed point
 (Def.~\ref{def:solOfEqSys}). This proves~(\ref{eq:201506181840}).

It is straightforward to show that the progress measure $p$ indeed
 achieves the exact solution. This completes the proof. \myqed
\end{myproof}

\subsection{Proof of Prop.~\ref{prop:soundnessOfExtProgMeas}}
\begin{myproof}
 The structure of the proof remains the same; we shall focus on what is
 changed. 

 Let us denote the function $\Phi$ in~(\ref{eq:201506162145}) by
  $\Phi_{\alpha_{a+1},\dotsc,\alpha_{k}}$, so that its dependence on 
  $\alpha_{a+1},\dotsc,\alpha_{k}$ becomes explicit.
 We shall prove, instead of~(\ref{eq:201506161816}), the following claim:
 \begin{equation}\label{eq:201506211659}
\approximant_{i}(
\alpha_{1}
,\dotsc,
\alpha_{a}
,\dotsc,
\alpha_{k}
)
\;\sqsubseteq \;
(\Phi_{\alpha_{a+1},\dotsc,\alpha_{k}})^{\alpha}(\bot)
\quad\text{for some $\alpha$}
\end{equation}
by  transfinite induction on a tuple
$\alpha_{a},\alpha_{a+1},\dotsc,\alpha_{k}$ (ordered lexicographically
 with $\preceq$,
 with the latter being the more significant). 

For one base case, assume $\alpha_{a}=0$ and
$p_{i_{a}}(\alpha_{1},\dotsc,\alpha_{k})=\bot$. For example this must be
 the case, by  Def.~\ref{def:extendedProgressMeasForEqSys},  when
$\alpha_{a}=\alpha_{a+1}=\cdots=\alpha_{k}=0$---this is because 
 there is no 
$(\alpha'_{1},\dotsc,\alpha'_{k})$ such that
 $(\alpha'_{1},\dotsc,\alpha'_{k})\prec_{i_{a}}
 (\alpha_{1},\dotsc,\alpha_{k})$.
In this case~(\ref{eq:201506211659})
 holds with $\alpha=0$.

For the other base case, assume that $\alpha_{a}=0$
and there exists
$(\alpha'_{1},\dotsc,\alpha'_{k})$ such that
 $(\alpha'_{1},\dotsc,\alpha'_{k})\prec_{i_{a}}
 (\alpha_{1},\dotsc,\alpha_{k})$ and
 $p_{i_{a}}(\alpha_{1},\dotsc,\alpha_{k}) \sqsubseteq
  p_{i_{a}}(\alpha'_{1},\dotsc,\alpha'_{k})$.
 The former condition means 
 $(\alpha'_{a},\dotsc,\alpha'_{k})\prec
 (\alpha_{a},\dotsc,\alpha_{k})$ with the lexicographic order $\prec$; 
 by the induction hypothesis, therefore, we have
 $\approximant_{i}(
\alpha'_{1}
,\dotsc,
\alpha'_{a}
,\dotsc,
\alpha'_{k}
)
\;\sqsubseteq \;
(\Phi_{\alpha'_{a+1},\dotsc,\alpha'_{k}})^{\alpha'}(\bot)
$ for some $\alpha'$. It follows easily from Cond.~\ref{item:progressMeasDefMonotonicity}  that 
$\Phi_{\alpha_{a+1},\dotsc,\alpha_{k}}$ is monotonic with respect to 
the ordinals $\alpha_{a+1},\dotsc,\alpha_{k}$. Summarizing, we have 
\begin{multline*}
 p_{i_{a}}(\alpha_{1},\dotsc,\alpha_{k}) \sqsubseteq
  p_{i_{a}}(\alpha'_{1},\dotsc,\alpha'_{k})
\\
  \sqsubseteq
(\Phi_{\alpha'_{a+1},\dotsc,\alpha'_{k}})^{\alpha'}(\bot)
 \sqsubseteq
(\Phi_{\alpha_{a+1},\dotsc,\alpha_{k}})^{\alpha'}(\bot),
\end{multline*}
showing the claim.

The other cases (when $\alpha_{a}$ is a successor or limit ordinal) are
 the same as in the proof of Thm.~\ref{thm:correctnessOfProgMeasEqSys}.
This concludes the proof.
\myqed
\end{myproof}

\subsection{Proof of Lem.~\ref{lem:invariantUnderCoalgMor}}
 \begin{myproof}
  We prove the following (more general) statement by induction on $\varphi$:
 for any formula $\varphi$ and for each $V'_{1},\dotsc,V_{m}\colon Y\to \Omega$, we
 have
 \begin{equation}\label{eq:201507061247}
 \sem{\varphi}_{c}(f^{*}(V'_{1}),\dotsc,f^{*}(V'_{m}))
 = f^{*}\bigl(\,
 \sem{\varphi}_{d}(V'_{1},\dotsc,V'_{m})
 \,\bigr) 
 \end{equation}
  where $f^{*}(V')$ is defined by $X\stackrel{f}{\to}
  Y\stackrel{V'}{\to} \Omega$.

  For the cases other than  fixed-point operators the proof is as usual
  in coalgebraic modal logic. For the modal formula case we exploit the
  naturality of $\lambda$.

  For the case of a $\mu$-formula $\mu u.\,\varphi$, we shall rely on the Cousot-Cousot
  characterization of least fixed points and prove the following
  by transfinite induction on the ordinal $\alpha$:
\begin{equation}\label{eq:201507061251}
    \begin{aligned}
   &\bigl[\,\sem{\varphi}_{c}(\overrightarrow{f^{*}(V')},
   \place)\,\bigr]^{\alpha}(\bot_{X\to \Omega})
     =\\
     &\qquad
     f^{*}\bigl(\,
   \bigl[\,\sem{\varphi}_{d}(\vec{V'},
   \place)\,\bigr]^{\alpha}(\bot_{Y\to \Omega})
   \,\bigr)
   \quad
   \text{for each ordinal $\alpha$.}
    \end{aligned}
\end{equation}  
The base case follows from the induction hypothesis (for the proof
  of~(\ref{eq:201507061247})), since $f^{*}(\bot_{Y\to
  \Omega})=\bot_{X\to \Omega}$. For the step case,
   \begin{align*}
    &   \bigl[\,\sem{\varphi}_{c}(\overrightarrow{f^{*}(V')},
   \place)\,\bigr]^{\alpha+1}(\bot_{X\to \Omega})
   \\
    &=
    \bigl(\,\sem{\varphi}_{c}(\overrightarrow{f^{*}(V')},
    \place)\,\bigr)
\Bigl(\,    f^{*}\bigl(\,
   \bigl[\,\sem{\varphi}_{d}(\vec{V'},
   \place)\,\bigr]^{\alpha}(\bot_{Y\to \Omega})
   \,\bigr)
    \,\Bigr)
    \\&
    \quad\text{by ind.\ hyp. (for~(\ref{eq:201507061251}))}
    \\
    &=
    \sem{\varphi}_{d}
    \bigl(\,\vec{V'},\,
    \bigl[\,\sem{\varphi}_{d}(\vec{V'},
   \place)\,\bigr]^{\alpha}(\bot_{Y\to \Omega})
    \,\bigr)
    \;\text{by ind.\ hyp. (for~(\ref{eq:201507061247}))}
    \\
    &=
         f^{*}\bigl(\,
   \bigl[\,\sem{\varphi}_{d}(\vec{V'},
   \place)\,\bigr]^{\alpha+1}(\bot_{Y\to \Omega})
   \,\bigr),
   \end{align*}
  as required. For the limit case (when $\alpha$ is a limit ordinal)
  the claim follows from the fact that $f^{*}$ preserves supremums.

  The case of a $\nu$-formula is symmetric to the last case. This
  concludes the proof. \myqed
 \end{myproof}

 \subsection{Proof of Lem.~\ref{lem:yonedaLikeCorForPolyadicModality}}
  \begin{myproof}
  Straightforward from the naturality of $\lambda$ along the arrow
  $\tuple{\pi_{j_{1}},\dotsc,\pi_{j_{n}}}\colon
  \Omega^{m}\to \Omega^{n}$. Specifically, consider the diagram
  \begin{displaymath}
      \vcenter{\xymatrix@R=1em@C+2em{
   {\Omega^{n}}
   &
   {(\Omega^{\Omega^{n}})^{n}}
       \ar[r]^-{\lambda_{\Omega^{n}}}
        \ar[d]_{(\Omega^{f})^{n}}
   &
   {\Omega^{F(\Omega^{n})}}
       \ar[d]^{\Omega^{Ff}}
   \\
   {\Omega^{m}}
   \ar[u]^{f}
   &
   {(\Omega^{\Omega^{m}})^{n}}
   \ar[r]_-{\lambda_{\Omega^{m}}}
   &
   {\Omega^{F(\Omega^{m})}}
%   \mathrlap{\enspace,}
   }}
  \end{displaymath}
  where $f$ stands for $\tuple{\pi_{j_{1}},\dotsc,\pi_{j_{n}}}$.
  Starting from the element $\tuple{\pi_{1},\dotsc,\pi_{n}}\in
  (\Omega^{\Omega^{n}})^{n}$ on the top-left corner proves the claim.
  \myqed
 \end{myproof}

\subsection{Proof of Thm.~\ref{thm:correctnessOfBranchingTimeProgMeas}}
 \begin{myproof}
  In view of Prop.~\ref{prop:translationIsCorrect} and
  Thm.~\ref{thm:correctnessOfProgMeasEqSys}, it suffices to show that:
\begin{itemize}
 \item 
   an MC progress measure
  (Def.~\ref{def:branchingTimeProgMeas}) gives rise to 
 \item a progress
  measure (in the sense of Def.~\ref{def:progressMeasForEqSys}) for the
  equational system $E_{\varphi,c}$ over $\Omega^{X}$ that arises from
  $\varphi$ and $c$,
\end{itemize}
  and vice versa. 
  
  The correspondence between the two notions is straightforward by
  (un)Currying. In particular,
  Thm.~\ref{thm:correctnessOfProgMeasEqSys}.2 gives the bound for a
  maximal ordinal $\alpha$ by $\ascCL(\Omega^{X})$.
  % the maximum ordinals
  % %  $\alpha,\alpha_{1},\dotsc,\alpha_{k}$
  % can be suitably chosen since
  % we are dealing only with small sets.
  We must check that Cond.~1--5
  (in each notion) are suitably transferred to each other;
  we shall focus on the cases handled in
  Cond.~5(c) of Def.~\ref{def:branchingTimeProgMeas}. The other cases are straightforward.

  It is not hard to see that, for this case, what needs to be shown is
  the following claim that informally reads
  ``$\PT_{\heartsuit_{\lambda}(u_{j_{1}},\dotsc,u_{j_{n}})}$ properly
  imitates the semantics of
  $\heartsuit_{\lambda}(u_{j_{1}},\dotsc,u_{j_{n}})$'': 
\begin{equation}\label{eq:201507051408}
   \begin{aligned}
&\bigl(\,\PT_{\heartsuit_{\lambda}(u_{j_{1}},\dotsc,u_{j_{n}})}(\vec{\alpha'})\co
   FQ\co c\,\bigr)(x)
   %--------------------
   =
   \\&\qquad
   \lambda_{X}\bigl(
   \pi_{j_{1}}\co\ev(\vec{\alpha'})\co Q,\dotsc, \pi_{j_{n}}\co\ev(\vec{\alpha'})\co Q
   \bigr) (c(x))
%   \colon X\to \Omega
  \end{aligned}
\end{equation}  
for each $x\in X$. Here recall that $\pi_{j}\co\ev(\vec{\alpha'})\co
  Q$ is of type  $X \to \Omega$, and $\lambda_{X}\colon (\Omega^{X})^{n}\to
  \Omega^{FX}$; the right-hand side of~(\ref{eq:201507051408}) therefore coincides with
  $\sem{\heartsuit_{\lambda}(u_{j_{1}},\dotsc,u_{j_{n}})}_{c}(\ev(\vec{\alpha'})\co
  Q)$ (see Def.~\ref{def:CmuFmlSem}).

  Now let us prove the equality~(\ref{eq:201507051408}). First notice
  the naturality of $\lambda$, where we write $Q'$ for $\ev(\vec{\alpha'})\co
  Q$:
  \begin{equation}\label{eq:201507051429}
   \vcenter{\xymatrix@R=1em@C+2em{
   {\Omega^{m}}
   &
   {(\Omega^{\Omega^{m}})^{n}}
       \ar[r]^-{\lambda_{\Omega^{m}}}
        \ar[d]_{(\Omega^{Q'})^{n}}
   &
   {\Omega^{F(\Omega^{m})}}
       \ar[d]^{\Omega^{FQ'}}
   \\
   X
   \ar[u]_{Q'}
   &
   {(\Omega^{X})^{n}}
   \ar[r]_-{\lambda_{X}}
   &
   {\Omega^{FX}}
%   \mathrlap{\enspace,}
   }}
  \end{equation}
  that is used in:
  \begin{align*}
   &\bigl(\,\PT_{\heartsuit_{\lambda}(u_{j_{1}},\dotsc,u_{j_{n}})}(\vec{\alpha'})\co
   FQ\co c\,\bigr)(x)
   \\
   &=
   \bigl(\,\lambda^{\tuple{j_{1},\dotsc, j_{n}}}\co F(\ev(\vec{\alpha'}))\co
   FQ\co c\,\bigr)(x)   
   \\
   &\qquad\quad\text{by def.\ of $\PT_{\heartsuit_{\lambda}(u_{j_{1}},\dotsc,u_{j_{n}})}(\vec{\alpha'})$}
   \\
   &=
   \bigl(\,\lambda^{\tuple{j_{1},\dotsc, j_{n}}}\co FQ'\co c\,\bigr)(x)
   \quad\text{by $Q'=\ev(\vec{\alpha})\co Q$}
   \\
   &=
   \bigl(\,\lambda_{\Omega^{m}}(\pi_{j_{1}},\dotsc, \pi_{j_{n}})\co FQ'\co c\,\bigr)(x)
   \quad\text{by $Q'=\ev(\vec{\alpha})\co Q$}
      \\
   &=
   \bigl(\,\lambda_{X}(\pi_{j_{1}}\co Q',\dotsc, \pi_{j_{n}}\co Q')\co c\,\bigr)(x)
   \quad\text{by~(\ref{eq:201507051429}),}   
  \end{align*}
  as required. This proves~(\ref{eq:201507051408}) and hence the
  theorem.
  \myqed
 \end{myproof}

\subsection{Proof of Thm.~\ref{thm:correctnessMPM}}
   \begin{myproof}
    The proof, much like in Appendix~\ref{appendix:parityProgressMeasure}, is by showing that MPMs and MC
 progress measures induce each other. We then appeal to
    Thm.~\ref{thm:correctnessOfBranchingTimeProgMeas} to obtain the
    claim.

    The mutual construction between an MPM $R$ and an MC
 progress measure $Q$ is much like in
    Prop.~\ref{prop:progMeasInParityGameVSProgMeasForEqSys}. The essence
    is: for any prioritized ordinal $(\alpha_{1},\dotsc,\alpha_{k})$,
    \begin{displaymath}
     (R(x))^{(i)}\preceq_{i}(\alpha_{1},\dotsc,\alpha_{k})
     \;\Longleftrightarrow\;
     (Q(x))_{i}(\alpha_{1},\dotsc,\alpha_{k})=\ttrue,
    \end{displaymath}
    and a row $(\NoGood,\dotsc,\NoGood)$ in an MPM handles an
    exceptional case that
    $(Q(x))_{i}(\alpha_{1},\dotsc,\alpha_{k})=\ffalse$ for every
    $(\alpha_{1},\dotsc,\alpha_{k})$.
    Then Cond.~1--5 of both notions are easily seen to be mutually
    transferred. Note that, for MPMs, Cond.~1 is not needed.
    \myqed
   \end{myproof}

   \subsection{Proof of Thm.~\ref{thm:correctnessOfAlgoBranchingTimeMC}}
       \begin{myproof}
      Let $R_{0}\colon X\to \POM_{|X|}$ be the optimal MPM guaranteed in
      Thm.~\ref{thm:correctnessMPM}.2. It is easy to see that, at any time
      during the execution of the algorithm, we have $R(x,i,j)\le
      (R_{0}(x))^{(i)}_{j}$ for any $x,i,j$. Here $\le$ is the usual
      inequality between natural numbers, where $\NoGood$ is deemed to be
      the greatest.
      Therefore we have
      \begin{equation}\label{eq:201507082245}
       \{x\in X\mid
	R(x,m,k)\neq\NoGood\}
	\supseteq
	       \{x\in X\mid
	R_{0}(x,m,k)\neq\NoGood\}
	=\sem{\varphi}_{c}.
      \end{equation}
      It is also easy to see that, once the algorithm
      terminates,
      the data $\bigl(R(x,i,j)\bigr)_{x,i,j}$ defines an MPM. By
      Thm.~\ref{thm:correctnessMPM}.1 (soundness) we have
      the opposite inclusion $\subseteq$
      in~(\ref{eq:201507082245}). This proves the claim. \myqed
     \end{myproof}

\subsection{Proof of
  Thm.~\ref{thm:complexityBranchingTimeMC}}

\begin{myproof}
  It can be easily seen that each iteration of the main loop (lines
 7--33) strictly increases $R(x,i)$ for at least one $(x,i)$ with respect to the
  preorder $\preceq_i$ (except for the last iteration). Since each
 $R(x,i)$ belongs to $\bigl[0,|X|\,\bigr]^{k}\amalg\{(\NoGood,\dotsc,\NoGood)\}$, each $R(x,i)$ increases at most $(|X|+1)^k$
 times. There are $m|X|$ of $(x,i)$'s; therefore the main loop iterates at most
 $m|X|(|X|+1)^k$ times. It is obvious that inner loop (lines 8--32) iterates $m|X|$
 times.

  The complexities of lines 9--12 and lines 13--15 are $O(k)$, and those
 of lines 16--18 and lines 19--22 are $O(km)$, by bounding $n$ by
 $m$. The complexity of lines 23--25 is $O(km^{2}|X|^{k+1}+C|X|^{k})$,
 as noted in Definition~\ref{def:ptMPM}; it dominates the overall
 complexity of the inner loop. From these
 we derive the claimed complexity. \myqed
\end{myproof}

\subsection{Proof of
  Thm.~\ref{thm:correctnessOfNondetExistProgMeasEqSys}}
    \begin{myproof}
   For the item~\ref{item:soundnessNondetExistProgressMeas} (soundness),
    the desired infinitary trace  $z$ is obtained by $z:=\beh(q)(y)\in
    Z$, where $\beh(q)$ is from~(\ref{eq:behq}). 
    We shall first establish that $z$ is indeed an
    infinitary trace of $c$ from $x$, that is, $z\in \tr(c)(x)$.
    \begin{itemize}
     \item
	      Cond.~6 of
	  Def.~\ref{def:nondetExistProgMeas} asserts that $Js$ is
	  a \emph{backward Kleisli simulation}, a notion
	  from~\cite{Hasuo06a,UrabeH15CALCOtoAppear}. Its soundness
	  against infinitary trace semantics---the latter being coalgebraically formalized
	  in Prop.~\ref{prop:coalgebraicInfinitaryTrace}---has been
	  established in~\cite{UrabeH15CALCOtoAppear}, under the
	  conditions of \emph{nonemptiness} and
	  \emph{image-finiteness}. These conditions are obviously
	  satisfied
	  by the arrow $Js\colon Y\relto X$ in $\Kleisli{\pow}$, since
	  it is the graph relation of a function $s\colon Y\to X$.
	  From this we conclude that the inequality
	  \begin{displaymath}
	   \vcenter{\xymatrix@R=.8em@C+2em{
	   {X}
	   \rar[r]^-{\tr(c)}
	   &
	   {Z}
	   \\
	   {Y}
	   \rar[u]^{Js}
	   \rar@/_/[ur]_-{\tr(Jq)}^{\supseteq}
	   }}
	  \end{displaymath}
	  holds; see~\cite[Thm.~4.6]{UrabeH15CALCOtoAppear}.
     \item Next we compare two arrows $\tr(Jq)$ and $J(\beh(q))$ of the type
	   $Y\relto Z$. We aim at $\tr(Jq)\supseteq J(\beh(q))$.
	   
    By the characterization of $\tr(Jq)$ as a greatest fixed point
    (Prop.~\ref{prop:coalgebraicInfinitaryTrace}), it suffices to show
    that $J(\beh(q))$ is a fixed point
    of the function $\Psi$
    in~(\ref{eq:monotoneFuncForWhichNondetTraceIsGreatestFixedPoint}). This
    is shown as follows.
    \begin{align*}
     &\Psi(J\beh(q))
     \\
     &=
     (J\zeta)^{-1}\odot \oF J\beh(q)\odot Jq
     \\
     &=
     (J\zeta)^{-1}\odot J(F\beh(q) \co q)
     \quad\text{by $\oF J=JF$, see
     e.g.~\cite{UrabeH15CALCOtoAppear}}
     \\
     &=
     (J\zeta)^{-1}\odot J(\zeta\co \beh(q))
     \quad\text{by def.~(\ref{eq:behq}) of $\beh(q)$}
     \\
     &=
     J\beh(q).
    \end{align*}
    \end{itemize}
    Combining the two items in the above, we conclude
    \begin{displaymath}
     \tr(c)\odot Js\;\supseteq\; J\beh(q)
     \quad\colon\; Y\relto Z,
    \end{displaymath}
    and equivalently $\beh(q)(y)\in \tr(c)(s(y))=\tr(c)(x)$ because $(J\beh(q))(y)=\bigl\{\beh(q)(y)\bigr\}$.

    It remains to be shown that
    $p_{m}(\alpha,\dotsc,\alpha)\sqsubseteq\sem{\varphi}_{\zeta}(z)$.
 It is crucial here that $r\colon Y\to \pPMpa$ forms an
	   MC progress measure
	   (Def.~\ref{def:branchingTimeProgMeas}) for $\varphi$ over
	   $q\colon Y\to FY$. This fact is obvious when one
	   compares Cond~5(c) in
	   Def.~\ref{def:branchingTimeProgMeas}
	   and~\ref{def:nondetExistProgMeas}.  It follows
	   from Thm.~\ref{thm:correctnessOfBranchingTimeProgMeas} that
	   $p_{m}(\alpha,\dotsc,\alpha)\sqsubseteq\sem{\varphi}_{q}(y)$.
    Then our goal follows from the fact that
    $\sem{\varphi}_{q}(y)=\sem{\varphi}_{\zeta}(z)$; the latter is a
    consequence of
    Lem.~\ref{lem:invariantUnderCoalgMor}. This concludes the proof of
    the item 1 (soundness).

    For the item 2 (completeness),
        let us first fix an optimal
    MC
    progress measure $Q^\zeta\colon Z\to \pPMpa$, i.e.\ one such that
    $\bigl(Q^{\zeta}(z')\bigr)_{m}(\alpha,\dotsc,\alpha)=\sem{\varphi}_{\zeta}(z')$
    for each $z'\in Z$. By
    Thm.~\ref{thm:correctnessOfBranchingTimeProgMeas}.\ref{item:completenessBranchingTimeProgressMeas}
    such $Q^{\zeta}$ exists.
    We define an LTMC progress
    measure
    as follows. 

 For each infinitary
	   trace $z\in Z$, there exists an MC progress
	   measure $p^{(z)}$ such that
	   $p^{(z)}_{m}(\alpha^{(z)},\dotsc,\alpha^{(z)})=\sem{\varphi}_{\zeta}(z)$
	   (by completeness,
	   Thm.~\ref{thm:correctnessOfBranchingTimeProgMeas}).
	   An ordinal $\alpha$ is chosen so that $\alpha^{(z)}\le
	   \alpha$ for each $z\in Z$; this is possible since $Z$ is a
	   (small) set.
 We define $Y$ by
	   \begin{displaymath}
	    Y:=\bigl\{\,(x',z')\in X\times Z\,\bigl|\bigr.\,z'\in
	    \tr(c)(x')\, \bigr\}.
	   \end{displaymath}
	   The functions $r$ and $s$ are defined by $r(x',z'):=
	   Q^{\zeta}(z')$ and $s(x',z'):=x'$.

    The construction of the coalgebra structure $q\colon Y\to FY$ is as follows.
    Let $y'=(x',z')\in Y$ be an arbitrary element of $Y$, so that $z'\in \tr(c)(x')$. We can pick $t\in FX$
    such that: $t\in c(x')$ and
\begin{equation}\label{eq:201507061746}
 \zeta(z')\in\bigl(\xi_{Z}\co F(\tr(c))\bigr)(t)
  \quad\text{where }
  FX\stackrel{F(\tr(c))}{\to} F\pow Z \stackrel{\xi_{Z}}{\to} \pow FZ.
\end{equation}
    Recall that $\xi\colon F\pow\Rightarrow \pow F$ is a distributive
    law (Def.~\ref{def:nondetCoalgebra}).
    Indeed, $z'\in \tr(c)(x')$ implies
\begin{equation}\label{eq:201507061748}
     \begin{aligned}
      \zeta(z')&\in 
      (\mu^{\pow}_{FZ}\co \pow \xi_{Z}\co \pow F\tr(c)\co
      c)(x')
      \\
      &\quad
      \text{by def.~(\ref{eq:nondetCoalgTraceSem}) of $\tr(c)$, expanded
      in $\Sets$}
      \\
      &=       \bigl(\,\mu^{\pow}_{FZ}\co \pow (\xi_{Z}\co
      F\tr(c))\,\bigr)(c(x'))
            \\
      &=    \textstyle\bigcup_{t\in c(x')}     (\xi_{Z}\co  F\tr(c))(t)
      \\
      &\quad
      \text{by def.\ of $\mu^{\pow}$ (union) and $\pow$'s
      action on arrows (direct image);}
    \end{aligned}
\end{equation}
    hence there must be some $t\in c(x')$ such that~(\ref{eq:201507061746})
    holds. 

    Now consider the following diagram:
     \begin{equation}\label{eq:201507062152}
      \vcenter{\xymatrix@R=.8em{
      {FX}
      \ar[r]^-{F\tuple{X,\tr(c)}}
      \ar@/_/@{-->}[rrd]
      &
      {F(X\times \pow Z)}
      \ar[r]^-{F\str}
      &
      {F\pow (X\times Z)}
      \ar[r]^-{\xi_{X\times Z}}
      &
      {\pow F(X\times Z)}
      \\
      &&
      {F\pow Y}
      \ar[u]_{F\pow \iota}
      \ar[r]_-{\xi_{Y}}
      &
      {\pow F Y}
            \ar[u]^{\pow F\iota}
      }}
% \\      
%      FX\xrightarrow{F\tuple{X,\tr(c)}}
%      F(X\times \pow Z)
%      \xrightarrow{F\str}
%      F\pow(X\times Z)
%      \xrightarrow{\xi_{X\times Z}}
%      \pow F(X\times Z),
    \end{equation}
    where $\str(x',U):=\{(x',z')\mid z'\in U\}$ equips the monad $\pow$ with
    a \emph{strength}~\cite{Kock72}, and $\iota\colon Y\hookrightarrow
    X\times Z$ is the 
    inclusion function. Factorization via the dashed arrow can be seen
    by
    the restriction
    \begin{displaymath}
           \vcenter{\xymatrix@R=.8em{
      {X}
      \ar[r]^-{\tuple{X,\tr(c)}}
      \ar@/_/@{-->}[rrd]
      &
      {X\times \pow Z}
      \ar[r]^-{\str}
      &
      {\pow (X\times Z)}
      \\
      &&
      {\pow Y}
      \ar[u]_{\pow \iota}
      }}
    \end{displaymath}
    which obviously follows from
    the definition of $Y$ and $\str$.  The arrow $FX\to
    \pow FY$ that arises in~(\ref{eq:201507062152}) shall be denoted by
    $h$.
    
    Let us now note that the following diagrams commute---we use naturality
    of $\xi$ and compatibility of $\xi$ and $\str$ with the
    monad structure of $\pow$.
    % The above composite~(\ref{eq:201507062152})
    % appears in the second row.
    \begin{equation}\label{eq:201507062141}
\begin{aligned}
 \vcenter{\xymatrix@R=.8em{
       {}
      &
      {FX}
      \ar[r]^-{F\eta^{\pow}_{X}}
      \ar`u[rr] `/9.99pt[rr]^-{\eta^{\pow}_{FX}} [rr]
      &
      {F\pow X}
      \ar[r]^{\xi_{X}}
      &
      {\pow FX}
\\
      {FX}
      \ar[r]^-{F\tuple{X,\tr(c)}}
 \ar@/_/@{-->}[rrd]
 \ar@/^/@{=} [ru]
      \ar `d[rrrd]+/d1.8em/ `[rrrd]_{h} [rrrd]
      &
      {F(X\times \pow Z)}
 \ar[r]^-{F\str}
      \ar[u]^{F\pi_{1}}
      &
      {F\pow (X\times Z)}
 \ar[r]^-{\xi_{X\times Z}}
      \ar[u]^{F\pow \pi_{1}}
      &
 {\pow F(X\times Z)}
      \ar[u]^{\pow F \pi_{1}}
      \\
      &&
      {F\pow Y}
      \ar[u]_{F\pow \iota}
      \ar[r]_-{\xi_{Y}}
      &
      {\pow F Y}
            \ar[u]^{\pow F\iota}
      }}
\\
 \vcenter{\xymatrix@R=.8em{
      &&
      {F\pow Z}
      \ar[r]^-{\xi_{Z}}
      &
      {\pow FZ}
\\
      {FX}
      \ar[r]^-{F\tuple{X,\tr(c)}}
 \ar@/_/@{-->}[rrd]
      \ar@/^.7pc/[rru]^-{F(\tr(c))}
      \ar `d[rrrd]+/d1.8em/ `[rrrd]_{h} [rrrd]
      &
      {F(X\times \pow Z)}
 \ar[r]^-{F\str}
      \ar[ru]^{F\pi_{2}}
      &
      {F\pow (X\times Z)}
 \ar[r]^-{\xi_{X\times Z}}
      \ar[u]^{F\pow \pi_{2}}
      &
 {\pow F(X\times Z)}
      \ar[u]^{\pow F \pi_{2}}
      \\
      &&
      {F\pow Y}
      \ar[u]_{F\pow \iota}
      \ar[r]_-{\xi_{Y}}
      &
      {\pow F Y}
            \ar[u]^{\pow F\iota}
 }}
 \\
 %     \vcenter{\xymatrix@R=.8em{
 %      {}
 %      &
 %      {FX}
 %      \ar[r]^-{F\eta^{\pow}_{X}}
 %      \ar`u[rr] `/9.99pt[rr]^-{\eta^{\pow}_{FX}} [rr]
 %      &
 %      {F\pow X}
 %      \ar[r]^{\xi_{X}}
 %      &
 %      {\pow FX}
 %      \\
 %     {FX}
 %      \ar[r]^-{F\tuple{X,\tr(c)}}
 %      \ar@/^/@{=} [ru]
 %      \ar@/_.7pc/[rrd]_-{F(\tr(c))}
 %      &
 %     {F(X\times \pow Z)}
 %     \ar[r]^-{F\str}
 %     \ar[u]^{F\pi_{1}}
 %           \ar[rd]_-{F\pi_{2}}
 %     &
 %     {F\pow(X\times Z)}
 %     \ar[r]^-{\xi_{X\times Z}}
 %     \ar[u]^{F\pow \pi_{1}}
 %                \ar[d]^-{F\pow\pi_{2}}
 %     &
 %     {\pow F(X\times Z)}
 %     \ar[u]^{\pow F \pi_{1}}
 %     \ar[d]_-{\pow F\pi_{2}}
 %     \\
 %     &&
 %      {F\pow Z}
 %      \ar[r]^-{\xi_{Z}}
 %      &
 %      {\pow FZ}
 % }}
\end{aligned}    
\end{equation}
    We claim that the set $h(t)\subseteq FY$,
    % \begin{equation}\label{eq:201507062200}
    %  \bigl(\,\xi_{X\times Z}\co F\str\co F\tuple{X,\tr(c)}\,\bigr)(t)
    %  \;\subseteq F(X\times Z),
    % \end{equation}
    where
    %the function is the one in~(\ref{eq:201507062152}) and
    $t\in FX$ is
    the one we chose so that~(\ref{eq:201507061746}) holds, contains an
    element $t'$ such that $(F\pi_{1})(t')=t$ and
    $(F\pi_{2})(t')=\zeta(z')$. Indeed, we have
     \begin{align*}
      &(\pow F\pi_{2}\co h)(t)
     % &     \bigl(\,\pow F\pi_{2}\co\xi_{X\times Z}\co F\str\co
     % F\tuple{X,\tr(c)}\,\bigr)(t)
     \\
     &=
     \bigl(\xi_{Z}\co F(\tr(c))\bigr)(t)
     \quad\text{by the second diagram in~(\ref{eq:201507062141})}
     \\
     &\ni \zeta(z')
          \quad\text{by~(\ref{eq:201507061746});}
     \end{align*}
    therefore there exists an element $t'$ of the set
    $h(t)$ such that $(F\pi_{2})(t')=\zeta(z')$.
    We also have $(F\pi_{1})(t')=t$---in fact $(F\pi_{1})(t'')=t$ holds
    for any element $t''$ of the set $h(t)$, since $(\pow F\pi_{1}\co
    h)(t)=\{t\}$ by the first diagram in~(\ref{eq:201507062141}).
    % \begin{align*}
    %  &     \bigl(\,\pow F\pi_{1}\co\xi_{X\times Z}\co F\str\co
    %  F\tuple{X,\tr(c)}\,\bigr)(t)
    %  \\
    %  &= \{t\} \quad\text{by~(\ref{eq:201507062141}).}
    % \end{align*}

    Finally, we define $q(y')\in FY$ by $q(y')=q(x',z'):=t'\in h(t)\subseteq
    FY$. It is not hard to see that the data
    $(\alpha,Y\stackrel{q}{\to} FY,r,s)$ thus
    defined
    satisfies the conditions in
    Def.~\ref{def:nondetExistProgMeas}. Specifically, Cond.~5(c) follows immediately from the fact that $Q^\zeta\colon Z\to
    \pPMpa$ is an MC
    progress measure (Def.~\ref{def:branchingTimeProgMeas}) over the
    final coalgebra $\zeta$. In its course we use the fact that, for
    each $y'=(x',z')\in Y$, 
    \begin{align*}
     (Fr\co q)(y') &= \bigl(F(Q^{\zeta}\co \pi_{2})\bigr)(q(y'))
     \quad\text{by def.\ of $r$}
     \\
     &= (FQ^{\zeta}\co \zeta)(z')
     \quad\text{by def.\ of $q$.}     
    \end{align*}
    Cond.~6 follows from $(F\pi_{1})(q(y'))=t\in c(x')$ (see~(\ref{eq:201507061746})).

    Thus we have obtained an LTMC progress measure $q\colon
    Y\to FY$. Let $y\in Y$ required in the statement to be $y:=(x,z)$.

    Now we shall check that the data thus obtained indeed satisfies the
    conditions in the statement. That $s(y)=x$ is by definition of $s$
    and $y$. For the condition that $\beh(q)(y)=z$, we observe that
    \begin{displaymath}
     \vcenter{\xymatrix@R=.8em{
     {FY}
     \ar[r]^-{F\pi_{2}}
     &
     {FZ}
     \\
     {Y}
     \ar[r]_-{\pi_{2}}
     \ar[u]^{q}
     &
     {Z}
          \ar[u]^{\text{final}}_{\zeta}
     }}
    \end{displaymath}
    commutes---recall that $t'=q(x',z')$ is chosen so that
    $(F\pi_{2})(t')=\zeta(z')$ (see above). Therefore $\pi_{2}$ in the
    diagram is a
    coalgebra homomorphism and we have
    \begin{align*}
     \beh(q)(y)&=\beh(\zeta)(\pi_{2}(y))\quad\text{by finality}
     \\&=\beh(\zeta)(z)=z,
    \end{align*}
    as required. To see that
           $p_{m}(\alpha,\dotsc,\alpha)
    =\sem{\varphi}_{\zeta}(z)$ where $p:=r(y)$, we have
    \begin{align*}
     &p_{m}(\alpha,\dotsc,\alpha)
     \\
     &=
     \bigl(Q^{\zeta}(z)\bigr)_{m}(\alpha,\dotsc,\alpha)
     \quad\text{by def.\ of $r$}
     \\
     &=
      \sem{\varphi}_{\zeta}(z)
     \quad\text{by the choice of $Q^{\zeta}$.}
    \end{align*}
    This concludes the proof.
    \myqed
   \end{myproof}

   \subsection{Proof of Thm.~\ref{thm:smallExistentialProgMeas}}
    \begin{myproof}
 Let  $(\alpha,Y_{0}\stackrel{q_{0}}{\to} FY_{0},r_{0},s_{0})$ denote, for the sake of
     distinction,  the LTMC
     progress measure that we constructed in the proof of
     Thm.~\ref{thm:correctnessOfNondetExistProgMeasEqSys}.2. 
     \textbf{So note that $Y,q,\dotsc$ in the proof of
     Thm.~\ref{thm:correctnessOfNondetExistProgMeasEqSys}.2, and those
     which are
     here, are different.} Recall that
     $Y_{0}\subseteq X\times Z$. We shall define $Y\subseteq X\times
     \pPMpa$
     as (an image of) a subset of $Y_{0}$.

     Let $\mathcal{Y}$ be the family of subsets
     $U\subseteq Y_{0}$ (hence $U\subseteq X\times Z$) that satisfy the following conditions.
     \begin{enumerate}
 \item 	   \textbf{(Initial state)}
      $(x,z)\in U$
     (where $x\in X$ and  $z\in\tr(c)(x)$ are both from  the statement);
      \item \textbf{(No redundancy)}
	    In case $(x',z'_{1})\in U$, $(x',z'_{2})\in U$ and 
	    $Q^{\zeta}(z'_{1})=Q^{\zeta}(z'_{2})$ hold, then
	    $z'_{1}=z'_{2}$. 
     \end{enumerate}
     On the family $\mathcal{Y}$ we define an order $\unlhd$ by:
     $U\unlhd U'$ if
\begin{itemize}
 \item 
      $U= U'$, or 
 \item $U\subsetneq U'$, and
	 \begin{equation}\label{eq:201507071341}
  \begin{aligned}
   \text{ for any $(x',z')\in U$, there exists $t''\in FU'$ such that}
   \!\!\!\!\!\!\!\!\!\!\!\!\!\!\!\!\!\!\!\!\!\!\!\!\!\!\!\!\!\!\!\!
   \!\!\!\!\!\!\!\!\!\!\!\!\!\!\!\!\!
   \!\!\!\!\!\!\!\!\!\!\!\!\!\!\!\!\!
            \!\!\!\!\!\!\!\!\!\!\!\!\!\!
   &
   \\
 	  (F\pi_{1})(t'')&=(F\pi_{1}\co q_{0})(x',z'),
	   \quad\text{and}\;
	   \\
	(FQ^{\zeta}\co F\pi_{2})(t'')&=
	(FQ^{\zeta}\co\zeta)(z').
  \end{aligned}	 
 \end{equation}
\end{itemize}
     Reflexivity, antisymmetry and transitivity of $\unlhd$ are straightforward.
    We aim at applying Zorn's lemma to obtain a maximal element of
     $\mathcal{Y}$ with respect to $\unlhd$. 

     Note first that $\mathcal{Y}$ is nonempty; indeed
     $\{(x,z)\}\in\mathcal{Y}$. Let $\{U_{i}\}_{i\in I}$ be a totally
     ordered subset of $\mathcal{Y}$. Then $\bigcup_{i}U_{i}$---where
     $\bigcup$ is the set-theoretic union---is
     an upper bound of $\{U_{i}\}_{i}$ in $\mathcal{Y}$. Indeed,
     $\bigcup_{i}U_{i}$ belongs to $\mathcal{Y}$: Cond.~1 is obvious;
     and Cond.~2 is easy too, because we can find $i_{1},i_{2}\in I$
     such that
     $(x',z'_{1})\in U_{i_{1}}$
     and
     $(x',z'_{2})\in U_{i_{2}}$, and $U_{i_{1}}\subseteq U_{i_{2}}$
     (without loss of generality) because $\{U_{i}\}_{i\in I}$ is
     totally ordered. It remains to be shown that
     $U_{i}\unlhd\bigcup_{i}U_{i}$
     for each $i\in I$. In case $U_{i}$ is the maximum (with respect to
     $\unlhd$) in $\{U_{i}\}_{i\in I}$, then it is so with respect to
     the inclusion order $\subseteq$; therefore
     $U_{i}=\bigcup_{i}U_{i}$.
     Assume otherwise, in which case there exists $j\in I$ such that
     $U_{i}\unlhd U_{j}$ and $U_{i}\neq U_{j}$. Now $U_{j}\subseteq
     \bigcup_{i}U_{i}$ and hence $FU_{j}\subseteq F(\bigcup_{i}U_{i})$
     (as subsets of $FY_{0}$,
     since any $\Sets$-functor preserves monos with a nonempty domain).
It is now easy to check the condition~(\ref{eq:201507071341}), and to see that
     $U_{i}\unlhd\bigcup_{i}U_{i}$.

      What we have shown so far allows us to appeal to Zorn's lemma and
     to conclude that $\mathcal{Y}$ has a maximal element with respect
     to $\unlhd$. We pick
     one
     and that is denoted by $Y'$.

     On such a maximal element $Y'$ we shall show that the
     condition~(\ref{eq:201507071341}) holds with $U=U'=Y'$.
     Assume otherwise. By the proof of 
     Thm.~\ref{thm:correctnessOfNondetExistProgMeasEqSys}.2, there
     does exist an element $t'':=q_{0}(x',z')\in FY_{0}$ such that
     the two equalities in~(\ref{eq:201507071341}) hold. Hence the
     condition~(\ref{eq:201507071341}) holds for $U=Y',U'=Y_{0}$.
     Now let us define
     \begin{align*}
      Y'_{\rm r}&:=\bigl\{\,(x',z')\in Y_{0}\,\bigl|\bigr.\, \exists (x',z'')\in Y'
	     \text{ s.t.\ } Q^{\zeta}(z')=Q^{\zeta}(z'')\,\bigr\}\setminus
      Y',
      \\
      Y'_{\rm n}&:=
      Y_{0}\setminus (Y'\cup Y'_{\rm r}).
     \end{align*}
     We shall show that $Y'_{\rm n}\neq\emptyset$, and that the
     condition~(\ref{eq:201507071341}) holds with $U=Y'$ and
     $U'=Y'\cup Y'_{\rm n}$. This will yield contradiction with
     the maximality of $Y'$ with respect to $\unlhd$. Note here that
     $Y_{0}=Y'\amalg Y'_{\rm r}\amalg Y'_{\rm n}$.

     By the definition of $Y'_{\rm r}$, we can choose a function $f\colon
     Y'_{\rm
     r}\to Y'$ such that $f(x',z')$ is $(x',z'')$ such that
     $Q^{\zeta}(z')=Q^{\zeta}(z'')$. This means that the following diagram commutes.
      \begin{displaymath}
       \vcenter{
       \xymatrix@R=.8em{
      {Y'+Y'_{\rm r}+Y'_{\rm n}}
      \ar[d]_{[Y',f]+Y'_{\rm n}}
      \ar@{^{(}->}[r]%^-{\iota_{0}}
      &
      {X\times Z}
      \ar[r]^-{X\times Q^{\zeta}}
      &
      {X\times \pPMpa}
      \\
      {Y'+Y'_{\rm n}}
      \ar@{^{(}->}[r]%^-{\iota_{0}}
      &
      {X\times Z}
            \ar[ru]_-{X\times Q^{\zeta}}
       }}
      \end{displaymath}
          By (essentially) applying $F$ to the diagram we obtain the
     following, where $g$ denotes the arrow
     $\tuple{F\pi_{1}, F(Q^{\zeta}\co \pi_{2})}$. The images
     $g[FY_{0}]$ and $g[F(Y'+Y'_{\rm n})]$ are characterized by epi-mono
     factorizations.
     \begin{displaymath}
      \vcenter{\xymatrix@R=.8em@C-.5em{
      &
      {g[FY_{0}]}
      \ar@/^1pc/@{ >->}[rd]
      \ar@/_3pc/@{-->}[ddd]
      \\
      {F(Y'+Y'_{\rm r}+Y'_{\rm n})}
      \ar[d]_{F([Y',f]+Y'_{\rm n})}
      \ar@{^{(}->}[r]%^-{\iota_{0}}
      \ar@/^1pc/@{->>}[ru]
      &
      {F(X\times Z)}
      \ar[r]^-{g}
      &
      {FX\times (F\pPMpa)}
      \\
      {F(Y'+Y'_{\rm n})}
      \ar@{^{(}->}[r]%^-{\iota_{0}}
            \ar@/_1pc/@{->>}[rd]
      &
      {F(X\times Z)}
      \ar[ru]_-{g}
      &
      \\
      &
      {g[F(Y'+Y'_{\rm n})]}
             \ar@/_1pc/@{ >->}[ruu]
      }}
     \end{displaymath}
     The dashed arrow arises as a diagonal fill-in. This proves
     \begin{equation}\label{eq:201507071724}
       \bigl\langle\, F\pi_{1},\, F(Q^{\zeta}\co \pi_{2})\bigr\rangle
      [FY_{0}]
      \;=\;
            \bigl\langle\, F\pi_{1},\, F(Q^{\zeta}\co \pi_{2})\bigr\rangle
      \bigl[F(Y'+Y'_{\rm n})\bigr],
     \end{equation}
     since the other direction $\supseteq$ is straightforward from
     $Y_{0}\supseteq Y'+Y'_{\rm n}$.
     It is  an immediate corollary of~(\ref{eq:201507071724}) that $Y'_{\rm
     n}\neq\emptyset$---otherwise we have $g[FY_{0}]=g[FY']$ that
     contradicts with the assumption that the
     condition~(\ref{eq:201507071341})
     holds with $U=Y', U'=Y_{0}$ but fails with $U=U'=Y'$. It 
     follows too that the
     condition~(\ref{eq:201507071341}) holds with
     $U=Y', U'=Y'+Y'_{\rm n}$ since~(\ref{eq:201507071724}) asserts that
     $Y_{0}$ and $Y'+Y'_{\rm n}$ ``have the same strength'' when it comes
     to the
     condition~(\ref{eq:201507071341}). It is also obvious from the
     definition of $Y'_{\rm n}$---it is defined by excluding ``redundant''
     elements in $Y'_{\rm r}$---that $Y'+Y'_{\rm n}$ satisfies Cond.~2 of
     the set $\mathcal{Y}$. Summarizing, we have shown that $Y'+Y'_{\rm
     n}\in \mathcal{Y}$ is such that $Y'\unlhd Y'+Y'_{\rm n}$, with a
     strict inequality. This contradicts with the maximality of $Y'$. 

     Thus we have shown that the
     condition~(\ref{eq:201507071341}) holds with $U=U'=Y'$.
     Now let us go back to the
    construction of the ``small'' LTMC
     progress measure required in the statement. 
     We define
\begin{displaymath}
  Y:=\bigl\{\,\bigl(x',Q^{\zeta}(z')\bigr)\in X\times \pPMpa
	    \mid
	    (x',z')\in Y'
     \,\bigr\}.
\end{displaymath}     
The functions $r,s$ are defined by
     projections: $r:=\pi_{2}$ and $s:=\pi_{1}$.
The coalgebraic structure $q\colon Y\to FY$ is defined using the above
     fact that the
     condition~(\ref{eq:201507071341}) holds for
     $U=U'=Y'$. Specifically, let $(x',p)\in Y$; then by Cond.~2 of
     $Y'\in \mathcal{Y}$, there exists a unique $z'\in Z$ such that
     $p=Q^{\zeta}(z')$ and $(x',z')\in Y'$. We use the 
     condition~(\ref{eq:201507071341}) (for $U=U'=Y'$) to find $t''\in
     FY'$
      that satisfies the two equalities
     in~(\ref{eq:201507071341}). Finally we define
     \begin{displaymath}
      q(x',p):= F(\id_{X}\times Q^{\zeta})(t'').
     \end{displaymath}
     Note that the types match up: $t''\in FY'\subseteq F(X\times Z)$
     and $F(\id_{X}\times Q^{\zeta})\colon F(X\times Z)\to F(X\times
     \pPMpa)$, and the latter obviously factors through
     $FY\hookrightarrow F(X\times
     \pPMpa)$.

     It is straightforward to check that
     $(\alpha,Y\stackrel{q}{\to} FY,r,s)$ 
     thus obtained indeed constitutes an LTMC progress measure.
     Let us turn to
          Cond.~5(c) of  Def.~\ref{def:nondetExistProgMeas}, 
     and in particular~(\ref{eq:201507051642}), for example.
     Let $(x',p)\in Y$ and $z'\in Z$ be the unique one
     such that
     $p=Q^{\zeta}(z')$ and $(x',z')\in Y'$.
     We have
     \begin{align*}
      &(\text{LHS})
      \\
      &=
      (Q^{\zeta}(z'))_{i}(\alpha_{1},\dotsc,\alpha_{a},\dotsc,\alpha_{k})
      \\
      &\sqsubseteq
       		\PT_{\heartsuit_{\lambda}(u_{j_{1}},\dotsc,u_{j_{n}})}
		%\PT_{\varphi_{i_{a}}}
		(\beta_{1},\dotsc,\beta_{a-1},
		\alpha_{a},\dotsc, \alpha_{k})
      \bigl(\,(FQ^{\zeta}\co \zeta)(z')\,\bigr)
      \\
      &\quad\text{since $Q^{\zeta}$ is an MC progress measure for
      $\zeta\colon Z\to FZ$ (Def.~\ref{def:branchingTimeProgMeas})}
      \\
      &=
       		\PT_{\heartsuit_{\lambda}(u_{j_{1}},\dotsc,u_{j_{n}})}
		%\PT_{\varphi_{i_{a}}}
		(\beta_{1},\dotsc,\beta_{a-1},
		\alpha_{a},\dotsc, \alpha_{k})
      \bigl(\,(F\pi_{2})(q(x',p))\,\bigr)
      \\
      &\quad\text{by def.\ of $q$; see~(\ref{eq:201507071341})}
      \\
      &=
       		\PT_{\heartsuit_{\lambda}(u_{j_{1}},\dotsc,u_{j_{n}})}
		%\PT_{\varphi_{i_{a}}}
		(\beta_{1},\dotsc,\beta_{a-1},
		\alpha_{a},\dotsc, \alpha_{k})
      \bigl(\,(Fr\co q)(x',p)\,\bigr).
     \end{align*}
     for suitable $\beta_{1},\dotsc,\beta_{a-1}$, as required.
     %------------------------------------------------
     For
     Cond.~6, let $z'$ be chosen in the same way as above. Then we have
     \begin{align*}
      &(Fs\co q)(x',p) = (F\pi_{1})(q(x',p))
      \\
      &= F\pi_{1} (q_{0}(x',z'))
      \\
      &\quad\text{by def.\ of $q$; see~(\ref{eq:201507071341})}
      \\
      &\in c(x'),
%      \quad\text{since }
     \end{align*}
     where the last membership is because
     $(\alpha,Y_{0}\stackrel{q_{0}}{\to} FY_{0},r_{0},s_{0})$
     is an LTMC progress measure for $c$.

     It remains to show that the LTMC progress measure that we
     have constructed indeed realizes $\sem{\varphi}_{\zeta}(z)$. Let
     $y:=(x,Q^{\zeta}(z))$ be the element $y\in Y$ required in the
     statement. We have
     \begin{align*}
      &(r(y))_{m}(\alpha,\dotsc,\alpha)
      \\
      &= (Q^{\zeta}(z))_{m}(\alpha,\dotsc,\alpha)
      \quad\text{by definition}
      \\
      &= \sem{\varphi}_{\zeta}(z)
      \quad\text{by the choice of $Q^{\zeta}$, see the proof of Thm.~\ref{thm:correctnessOfNondetExistProgMeasEqSys}.}
     \end{align*}
     This concludes the proof. \myqed
    \end{myproof}

   \subsection{Proof of
   Thm.~\ref{thm:decidabilityOfLinTimeCmuGLModelChecking}}
   	\begin{myproof}
	 We first check if $\varphi$ is satisfiable by $F$-coalgebras. 
	If $\varphi$  not, then obviously the answer to the problem in
	 the statement is false. Assume otherwise.

	 We claim that the maximum ordinal $\alpha$ in
	 Theorem~\ref{thm:smallExistentialProgMeas} can be bounded by
	 $N_{\varphi}$ from Assumption~\ref{asm:smallModel}. Indeed,
       by the small model property
      (Assumption~\ref{asm:smallModel}), we can assume that
      there exists a coalgebra
      $\varepsilon\colon E\to FE$, its state $e\in E$ and an
      MC progress measure $Q'\colon E\to
      \pPM_{\varphi,\alpha'}$ (Def.~\ref{def:branchingTimeProgMeas}) such
      that $Q'(e)(\alpha',\dotsc,\alpha')=\ttrue$ and $|E|\le N_{\varphi}$.
      Moreover, by
      Thm.~\ref{thm:correctnessOfBranchingTimeProgMeas}.1--2, we can
      choose
      the maximum ordinal $\alpha'$ of $Q'$ to be $\alpha'=|E|\le
      N_{\varphi}$. It is then straightforward to adapt the proof of
	 Thm.~\ref{thm:smallExistentialProgMeas} in the following way.
	 \begin{itemize}
	  \item The final coalgebra $\zeta\colon Z\to FZ$ is replaced by
		$\varepsilon\colon E\to FE$. 
	  \item The optimal MC progress measure $Q^{\zeta}$ is replaced by
		the MC progress measure $Q'$ in the above.
	 \end{itemize}
	 Going through the adapted proof  proves the
	 statement of Thm.~\ref{thm:smallExistentialProgMeas} with the
	 ordinal $\alpha$ in it bounded by $\alpha'$, hence by
	 $N_{\varphi}$. 

       Since $X$ is assumed to be finite and so is
	 $N_{\varphi}\in\omega$, we see that the set $X\times
	 \pPM_{\varphi,N_{\varphi}}$ is a finite set. 
	 By the last assumption in Assumption~\ref{asm:smallModel}, 
	 the set $FY$ is finite for any (necessarily finite) subset $Y\in X\times
	 \pPM_{\varphi,N_{\varphi}}$. Therefore there are only finitely
	 many functions $q\colon Y\to FY$. We enumerate all such, for each
      $Y\in X\times
	 \pPM_{\varphi,N_{\varphi}}$, and check if there is any that
	 makes
	 $(\alpha,q,\pi_{2},\pi_{1})$ an LTMC progress measure. If there
	 is, then the answer to  the problem in
	 the statement is true (by
	 Thm.~\ref{thm:correctnessOfNondetExistProgMeasEqSys}.1). If
	 there is none, then the answer is false by the above arguments.
	 \myqed
	\end{myproof}

%\subsection{Proof of }

%\subsection{Proof of }

\end{document}